\documentclass[aps,prd,showpacs,floatfix,
onecolumn,
superscriptaddress,nofootinbib,preprintnumbers, 10pt, notitlepage]{revtex4-1} 

\makeatletter
\def\p@subsection{}
\makeatother

\usepackage{graphicx,amsmath,amssymb,lipsum,bm}

\usepackage[dvipsnames]{xcolor}
\definecolor{xlinkcolor}{rgb}{0.7752941176470588, 0.22078431372549023, 0.2262745098039215}

\usepackage{booktabs} 
\usepackage[normalem]{ulem}
\usepackage[utf8]{inputenc} 
\usepackage{mathtools}
\usepackage{aas_macros}
\usepackage{xspace}

\usepackage{ulem}
\usepackage{diagbox}

\usepackage[dvipsnames]{xcolor}
\usepackage{mathrsfs}

\usepackage[colorlinks=true,citecolor=xlinkcolor,linkcolor=xlinkcolor,urlcolor=xlinkcolor, backref=false,pdfborder={0 0 0}]{hyperref}

\newcommand{\be}{\begin{equation}}
\newcommand{\ee}{\end{equation}}
\newcommand{\beqa}{\begin{eqnarray}}
\newcommand{\eeqa}{\end{eqnarray}}

\newcommand\p{{\bm p}}
\renewcommand\k{{\bm k}}
\newcommand\q{\bm{q}}

\newcommand\g{\gamma}

\renewcommand\r{\rho}
\newcommand\s{\sigma}

\newcommand\F{\mathcal{F}}

\usepackage{xspace}

\newcommand{\bseq}{\begin{subequations}}
\newcommand{\eseq}{\end{subequations}}

\def\ltsima{$\; \buildrel < \over \sim \;$\xspace}
\def\gtsima{$\; \buildrel > \over \sim \;$\xspace}
\def\simlt{\lower.5ex\hbox{\ltsima}}
\def\simgt{\lower.5ex\hbox{\gtsima}}

\newcommand{\Mpch}{\, h^{-1}\mathrm{Mpc}\, }
\newcommand{\hMpc}{\, h\mathrm{Mpc}^{-1}\, }

\newcommand{\kmax}{\, k_{\rm max}\, }
\newcommand{\knl}{\, k_{\rm NL}\, }

\newcommand{\vk}{\bm k}

\newcommand{\vp}{\bm p}

\newcommand{\vq}{\bm q}

\newcommand{\Ids}{\mathcal{I}_{\delta^2}}
\newcommand{\IG}{\mathcal{I}_{\Gal_2}}
\newcommand{\FG}{\mathcal{F}_{\Gal_2}}
\newcommand{\Idsds}{\mathcal{I}_{\delta^2\delta^2}}
\newcommand{\IdsG}{\mathcal{I}_{\delta^2\Gal_2}}
\newcommand{\IGG}{\mathcal{I}_{\Gal_2\Gal_2}}

\def\gsim{\raise0.3ex\hbox{$\;>$\kern-0.75em\raise-1.1ex\hbox{$\sim\;$}}}
\def\lsim{\raise0.3ex\hbox{$\;<$\kern-0.75em\raise-1.1ex\hbox{$\sim\;$}}}

\def\beqn#1{\begin{equation}\label{#1}}
\def\eeqn{\end{equation}}

\def\beqa#1{\begin{eqnarray}\label{#1}}
\def\eeqa{\end{eqnarray}}

\def\kmax{{k_\text{max}}}
\def\hMpc{h{\text{Mpc}}^{-1}}
\def\Mpch{h^{-1}{\text{Mpc}}}

\def\Z2{$\mathcal{Z_2}$}

\def\beq{\begin{eqnarray}}
\def\eeq{\end{eqnarray}}
\let\vec\mathbf

\newcommand{\PL}{P_{\rm lin}}
\newcommand{\Kcal}{\mathcal{K}}
\newcommand{\Lcal}{\mathcal{L}}
\newcommand{\Gal}{\mathcal{G}}

\newcommand{\dL}{\delta_L}
\newcommand{\dg}{\delta_g}
\newcommand{\abs}[1]{\textbf{absorbed}}

\newcommand{\vPsi}{\boldsymbol{\Psi}}
\hypersetup{colorlinks=true,linkcolor=blue}

\newcommand {\ignore}[1]{}

\usepackage{amsmath}              
\usepackage{tikz}
\usetikzlibrary{decorations.markings,arrows.meta,positioning}
 
\tikzset{
  prop/.style ={line width=0.9pt},                       
  vsq/.style  ={draw,line width=0.9pt,fill=white,         
                minimum size=9pt,inner sep=0pt},
  marr/.style ={prop,postaction={decorate},
                decoration={markings,
                  mark=at position 0.5 with
                    {\arrow{Stealth[length=2.6mm,width=2.8mm]}}}},
}
\providecommand{\Pdot}[1]{\fill (#1) circle (2.5pt);}     
\providecommand{\Plin}{$P_{\mathrm{lin}}$}                 
 \providecommand{\bPlin}{$b_1 P_{\mathrm{lin}}$}                 
 \providecommand{\btwoPlin}{$b_1^2 P_{\mathrm{lin}}$}                 

\newcommand{\diagPELE}{
\begin{tikzpicture}
  \node at (-1.7,1.15) {$P_{11}$};
  \coordinate (c) at (0,0);
  \draw[marr] (c) -- (-1.7,0);
  \draw[marr] (c) -- ( 1.7,0);
  \Pdot{c}\node[below=3pt] at (c) {\btwoPlin};
\end{tikzpicture}}
 
\newcommand{\diagPTT}{
\begin{tikzpicture}
  \def\r{1.0}
  \node at (-2.3,1.55) {$P_{22}$};
  \coordinate (T) at (0,\r);\coordinate (B) at (0,-\r);
  \draw[marr] (T) arc[start angle= 90,end angle=180,radius=\r];
  \draw[marr] (T) arc[start angle= 90,end angle=  0,radius=\r];
  \draw[marr] (B) arc[start angle=270,end angle=180,radius=\r];
  \draw[marr] (B) arc[start angle=270,end angle=360,radius=\r];
  \draw[marr] (-\r,0) -- (-2.4,0);
  \draw[marr] ( \r,0) -- ( 2.4,0);
  \node[vsq] at (-\r,0) {};\node[vsq] at ( \r,0) {};
  \Pdot{T}\Pdot{B}
  \node[below=2pt] at (T) {\Plin};\node[above=2pt] at (B) {\Plin};
  \node at (-\r-0.28,-0.33) {$K_2$};
  \node at ( \r+0.28,-0.33) {$K_2$};
\end{tikzpicture}}
 
\newcommand{\diagPOT}{
\begin{tikzpicture}
  \def\rr{0.6}
  \node at (-1.7,1.8) {$P_{13}$};
  \coordinate (V) at (0,0);\coordinate (D) at (1.3,0);\coordinate (T) at (0,2*\rr);
  \draw[marr] (T) arc[start angle=90,end angle=270,radius=\rr];
  \draw[marr] (T) arc[start angle=90,end angle=-90,radius=\rr];
  \draw[marr] (V) -- (-1.6,0);
  \draw[marr] (D) -- (V);
  \draw[marr] (D) -- (2.7,0);
  \node[vsq] at (V) {};
  \Pdot{T}\Pdot{D}
  \node at (0,\rr) {\Plin};
  \node[below=4pt] at (V) {$K_3$};\node[below=4pt] at (D) {\bPlin};
\end{tikzpicture}}
 
\newcommand{\diagPOF}{
\begin{tikzpicture}
  \def\rr{0.6}
  \node at (-1.7,1.8) {$P_{15}$};
  \coordinate (V) at (0,0);\coordinate (D) at (1.4,0);
  \coordinate (T) at (0,2*\rr);\coordinate (Bt) at (0,-2*\rr);
  \draw[marr] (T) arc[start angle=90,end angle=270,radius=\rr];
  \draw[marr] (T) arc[start angle=90,end angle=-90,radius=\rr];
  \draw[marr] (Bt) arc[start angle=270,end angle= 90,radius=\rr];
  \draw[marr] (Bt) arc[start angle=270,end angle=450,radius=\rr];
  \draw[marr] (V) -- (-1.7,0);
  \draw[marr] (D) -- (V);
  \draw[marr] (D) -- (2.8,0);
  \node[vsq] at (V) {};
  \Pdot{T}\Pdot{Bt}\Pdot{D}
  \node at (0, \rr+0.18) {\Plin};\node at (0,-\rr-0.18) {\Plin};
  \node at (0,-0.32) {$K_5$};
  \node[below=4pt] at (D) {\bPlin};
\end{tikzpicture}}
 
\newcommand{\diagPTF}{
\begin{tikzpicture}
  \def\R{1.35}
  \node at (-2.4,1.75) {$P_{24}$};
  \coordinate (L) at (-\R,0);\coordinate (Rr) at (\R,0);
  \coordinate (T) at (0,\R);\coordinate (B) at (0,-\R);\coordinate (M) at (-0.2,0);
  \draw[marr] (T) arc[start angle= 90,end angle=180,radius=\R];
  \draw[marr] (T) arc[start angle= 90,end angle=  0,radius=\R];
  \draw[marr] (B) arc[start angle=270,end angle=180,radius=\R];
  \draw[marr] (B) arc[start angle=270,end angle=360,radius=\R];
  \draw[marr] (M) to[bend left =45] (L);
  \draw[marr] (M) to[bend right=45] (L);
  \draw[marr] (L) -- (-2.5,0);
  \draw[marr] (Rr) -- (2.5,0);
  \node[vsq] at (L) {};\node[vsq] at (Rr) {};
  \Pdot{T}\Pdot{B}\Pdot{M}
  \node[above=2pt] at (T) {\Plin};\node[below=2pt] at (B) {\Plin};
  \node[right=2pt] at (M) {\Plin};
  \node at (-\R-0.28,-0.33) {$K_4$};
  \node at ( \R+0.28,-0.33) {$K_2$};
\end{tikzpicture}}
 
\newcommand{\diagPTHa}{
\begin{tikzpicture}
  \def\R{1.35}
  \node at (-2.5,1.75) {$P_{33,\mathrm{I}}$};
  \coordinate (L) at (-\R,0);\coordinate (Rr) at (\R,0);
  \coordinate (T) at (0,\R);\coordinate (B) at (0,-\R);\coordinate (Mid) at (0,0);
  \draw[marr] (T) arc[start angle= 90,end angle=180,radius=\R];
  \draw[marr] (T) arc[start angle= 90,end angle=  0,radius=\R];
  \draw[marr] (B) arc[start angle=270,end angle=180,radius=\R];
  \draw[marr] (B) arc[start angle=270,end angle=360,radius=\R];
  \draw[marr] (Mid) -- (L);
  \draw[marr] (Mid) -- (Rr);
  \draw[marr] (L) -- (-2.5,0);
  \draw[marr] (Rr) -- (2.5,0);
  \node[vsq] at (L) {};\node[vsq] at (Rr) {};
  \Pdot{T}\Pdot{B}\Pdot{Mid}
  \node[below=2pt] at (T) {\Plin};\node[above=2pt] at (B) {\Plin};
  \node[below=5pt] at (Mid) {\Plin};
  \node at (-\R-0.28,-0.33) {$K_3$};
  \node at ( \R+0.28,-0.33) {$K_3$};
\end{tikzpicture}}
 
\newcommand{\diagPTHb}{
\begin{tikzpicture}
  \def\rr{0.55}\def\s{1.25}
  \node at (-2.5,1.8) {$P_{33,\mathrm{II}}$};
  \coordinate (L) at (-\s,0);\coordinate (Rr) at (\s,0);
  \coordinate (TL) at (-\s,2*\rr);\coordinate (TR) at (\s,2*\rr);\coordinate (D) at (0,0);
  \draw[marr] (TL) arc[start angle=90,end angle=270,radius=\rr];
  \draw[marr] (TL) arc[start angle=90,end angle=-90,radius=\rr];
  \draw[marr] (TR) arc[start angle=90,end angle=270,radius=\rr];
  \draw[marr] (TR) arc[start angle=90,end angle=-90,radius=\rr];
  \draw[marr] (D) -- (L);
  \draw[marr] (D) -- (Rr);
  \draw[marr] (L) -- (-\s-1.3,0);
  \draw[marr] (Rr) -- (\s+1.3,0);
  \node[vsq] at (L) {};\node[vsq] at (Rr) {};
  \Pdot{TL}\Pdot{TR}\Pdot{D}
  \node at (-\s,\rr) {\Plin};\node at ( \s,\rr) {\Plin};
  \node[below=4pt] at (L) {$K_3$};\node[below=4pt] at (Rr) {$K_3$};
  \node[below=4pt] at (D) {\Plin};
\end{tikzpicture}}

\begin{document}

\preprint{MIT-CTP/6054}

\title{Galaxy Power Spectrum at Two-Loop Order: \\
Implications for Weak Lensing Surveys and New Physics
}

\author{Mikhail M. Ivanov}
\email{ivanov99@mit.edu}
\affiliation{Center for Theoretical Physics -- a Leinweber Institute, Massachusetts Institute of Technology, 
Cambridge, MA 02139, USA}
\affiliation{The NSF AI Institute for Artificial Intelligence and Fundamental Interactions, Cambridge, MA 02139, USA}

\begin{abstract} 
\noindent 
We compute the galaxy power spectrum at two-loop order in
cosmological perturbation theory (effective field theory, EFT). 
We derive galaxy bias operators through the fifth order 
and obtain two-loop renormalization conditions 
for their bias coefficients. 
We also derive the relevant higher-derivative and stochastic contributions, and implement IR resummation using 
\textit{time-sliced perturbation theory}. 
We obtain the complete dark-matter and galaxy renormalization conditions 
for the higher-derivative operators at the power spectrum, bispectrum, and trispectrum level.
Having identified the redundant operators, we find that the two-loop galaxy power spectrum 
requires
21 additional
bias, higher derivative and stochastic parameters per galaxy sample relative to the one-loop model. We compare our computation with the galaxy-galaxy and galaxy-matter power spectra from the PT Challenge N-body simulation
at $z=0.61$
and find a per mille-level agreement up to $k=0.85~h$Mpc$^{-1}$.
We show that even with 
conservative priors on all EFT parameters, 
the two-loop model produces an unbiased measurement of the mass fluctuation amplitude $\sigma_8$ with 
three times narrower error-bars
than the linear theory model.
The improvement over the one-loop model
is $\simeq 40\%$.
This suggests significant gains 
in the two-loop EFT analyses
of galaxy clustering and 
galaxy--lensing two-point functions 
(``$2\times2$ pt'') from 
CMB lensing maps and
imaging surveys
like Euclid, LSST, and Roman.
In addition, 
our two-loop computation offers a probe of new physics scenarios 
that modify the shape of the matter power spectrum at wavenumbers 
$(0.4-0.8)~h$Mpc$^{-1}$ such as 
the presence of ultra-light axion dark matter
sub-components 
with masses $m_a\sim 10^{-24}$ eV.
\end{abstract}

\maketitle

\section{Introduction}

Large-scale structure (LSS) of the Universe 
has the potential to deliver new insights into the fundamental 
questions of physics: the nature of dark matter, dark energy, 
and cosmic inflation. The importance of LSS data is emphasized by
a growing number of ongoing and upcoming 
galaxy surveys such as 
DESI~\cite{Aghamousa:2016zmz}, Euclid~\cite{Laureijs:2011gra}, 
LSST~\cite{LSST:2008ijt}, 
and Roman Space Telescope~\cite{Akeson:2019biv}.

Non-linear cosmological perturbation theory is 
an analytic approach to describe the distribution
of galaxies and matter in the low-redshift Universe. 
{The} 
effective field theory (EFT) for LSS~\cite{Baumann:2010tm,Carrasco:2012cv} 
(see~\cite{Ivanov:2022mrd} for review) is a consistent 
formulation of cosmological perturbation theory 
which allows for 
computations of non-linear LSS observables 
whose precision can be 
improved to a desired order in a systematic fashion. 

The key advantages of EFT over popular empirical
or simulation-based models are (i) EFT computations are 
highly {accurate} on quasi-linear scales and (ii) they retain
this accuracy when applied to new physics scenarios, 
which are difficult to explore with simulation-based tools
due to their prohibitive computational cost. These 
advantageous features of EFT are best illustrated by
the EFT-based analyses of redshift-space galaxy clustering
data from BOSS~\cite{BOSS:2016wmc}, eBOSS~\cite{eBOSS:2020yzd}
and DESI~\cite{DESI:2024jis} surveys both in the context
of the standard cosmological model $\Lambda$CDM~\cite{DAmico:2019fhj,Ivanov:2019pdj,Chen:2021wdi,Philcox:2020vvt,Chen:2022jzq,Philcox:2021kcw,Chudaykin:2022nru,Ivanov:2021zmi} and 
its various extensions~\cite{Ivanov:2019hqk,Ivanov:2020ril,Chudaykin:2020ghx,Cabass:2022epm,Cabass:2022ymb,He:2023dbn,He:2023oke,Rogers:2023ezo,Chen:2024vuf,Chudaykin:2025aux,Chudaykin:2025lww,Chudaykin:2025vdh,Chudaykin:2026nls,Ivanov:2026dvl}. 

Real space galaxy power spectrum and galaxy-matter
cross-spectrum EFT computations have been previously used 
in many analyses of projected 
galaxy-galaxy and galaxy-CMB lensing data from  
BOSS, unWISE, and DESI galaxy surveys and Planck+ACT lensing maps~\cite{Chen:2022jzq,Krolewski:2021yqy,Sailer:2024jrx,Qu:2024sfu,
Farren:2024rla,
deBelsunce:2025qku,Maus:2025rvz,Ivanov:2026dvl}.
EFT tools have also been recently applied to galaxy clustering and 
galaxy-galaxy lensing data
from DES and DESI surveys~\cite{DES:2020yyz,DES:2021zxv,Giri:2023ghr,Chen:2024vvk,Chen:2026usz,DAmico:2025zui}. 
For imaging data, the state-of-the-art EFT computations 
are the one-loop order 
for the galaxy power spectrum and galaxy-galaxy lensing
cross-spectrum~\cite{DES:2020yyz,DES:2021zxv,Giri:2023ghr,Chen:2024vvk,Chudaykin:2020aoj}, 
and the two-loop order for the 
matter power spectrum~\cite{Chen:2026usz,Saraivanov:2026sxc} 
relevant for the weak gravitational lensing of galaxies 
(cosmic shear).\footnote{Also see~\cite{Carrasco:2013sva,Blas:2013bpa,Blas:2013aba,Bakx:2025cvu,Bakx:2025jwa,Anastasiou:2025jsy}
for the two-loop matter power spectrum.} 

In this work we 
extend the real space galaxy power spectrum computations
relevant for imaging survey data 
to the two-loop order. Our work builds on previous 
one-loop calculations for the galaxy power spectrum
and bispectrum in real space~\cite{Assassi:2014fva,Eggemeier:2018qae,Eggemeier:2020umu,Philcox:2022frc,Bakx:2025pop}. 
In particular, we will use the 
basis of bias operators first 
used in Ref.~\cite{Eggemeier:2018qae} for the one-loop
galaxy bispectrum.

Our paper is organized as follows.
In Section~\ref{sec:power} we introduce EFT power counting rules
and determine the contributions which we need to compute at the two-loop order. 
In Section~\ref{sec:bias} we derive the galaxy bias operators to the fifth order and study their renormalization. In general, there are 29 distinct operators at this order. In particular, we will
derive the renormalization condition for the third order 
bias operators in the Galileon basis. We will show that many fourth and fifth order bias operators produce redundant contributions to the galaxy power spectrum,
so the actual number of bias parameters will be 17.\footnote{A similar observation was made in Ref.~\cite{Bakx:2025cvu} using a different basis of operators. That work also derived 
the renormalization conditions for the fifth order operators equivalent to ours. 
We thank Mathias Garny for pointing this out.}
We wrap up Section~\ref{sec:bias} by presenting numerical 
results for the relevant bias shapes. 
In Section~\ref{sec:HD} we derive the higher-derivative operators
and discuss the stochastic contributions. 
In Section~\ref{sec:data} we apply our two-loop model to the 
galaxy-galaxy and galaxy-matter power spectrum data from 
the PT Challenge simulation~\cite{Nishimichi:2020tvu}. 
We find an excellent agreement
between our computation and the data for quasi-linear
modes with wavenumbers $k\lesssim 0.85~\hMpc$. 
Section~\ref{sec:implications} focuses on implications 
of our results for weak lensing surveys and emphasizes 
the benefits of the EFT approach to galaxy clustering: even though
EFT models depend on a large number of free parameters, they 
allow one to model significantly smaller scales than 
other modeling approaches, which results in more powerful
cosmological constraints both on $\Lambda$CDM and new physics.
Section~\ref{sec:conclusions} draws conclusions and outlines 
directions of future research. Appendix~\ref{sec:mixed}
contains a proof of a statement that the mixed deterministic-stochastic
contributions do not generate any physical contributions
to the galaxy power spectrum to all orders in perturbation theory.

\section{Power counting}
\label{sec:power}

An important aspect of EFT is power counting, i.e. rules to estimate the size of various corrections. 
These rules allow one to identify all terms relevant at a given 
specified precision of the calculation.
The EFT power counting is the simplest for a power-law Universe~\cite{Pajer:2013jj,Carrasco:2013mua,Ivanov:2022mrd}, whose \textit{linear} matter power spectrum is 
\be 
\PL(k)=\frac{2\pi^2}{k_{\rm NL}^3}
\left(\frac{k}{k_{\rm NL}}\right)^n\,.
\ee 
In our Universe $n\approx -1.5$ around $k\sim 0.2~\hMpc$, 
and $k_{\rm NL}\simeq 0.45 [D(z)]^{-2/(n+3)}~\hMpc$ is the non-linear scale of cosmological perturbation theory, 
where $D(z)$ is the growth factor. 
In what follows we will always assume that the linear 
power spectrum in all expressions
is evaluated at a given fiducial redshift $z$,
and suppress the explicit redshift dependence in $\PL(k)$.

The position space
matter density variance produced by modes in a logarithmic 
interval around $k$ is 
\be 
\Delta^2_{\rm lin}=\frac{k^3}{2\pi^2}\PL(k)=\left(\frac{k}{k_{\rm NL}}\right)^n\,.
\ee 
Non-linear terms in matter cosmological
perturbation theory generate loop 
corrections to the above linear density variance~\cite{Scoccimarro:1996jy,Scoccimarro:1999kp,Bernardeau:2001qr}. 
The non-linear relationship between 
the number density of galaxies and the underlying dark matter field
(galaxy bias~\cite{Desjacques:2016bnm}) produces additional
loop corrections to the galaxy correlation functions.
For these deterministic 
$L$-loop corrections due to either bias or dark matter non-linearities, assuming all bias parameters are 
$\mathcal{O}(1)$ numbers, we can estimate:
\be 
\begin{split}
\Delta^2_{L-\text{loop}}(k)\equiv \frac{k^3}{2\pi^2} P_{L-\text{loop}}(k)\sim \left(\frac{k}{k_{\rm NL}}\right)^{(1+L)(3+n)}\,.
\end{split}
\ee 
EFT requires the presence of higher-derivative terms 
both in the dark matter non-linearities and bias expansion.
These higher derivative terms (also referred to as counterterms) acting on $L$-loop fields scale as 
\be 
\begin{split}
k^{2m}P_{L-\text{loop}}(k) \to \Delta^2_{\mathrm{HD},~mL} \sim \left(\frac{k}{k_{\rm NL}}\right)^{(1+L)(n+3)+2m}\,.
\end{split}
\ee 
Next, the part of the 
galaxy density uncorrelated with cosmological
initial conditions 
(galaxy stochasticity) produces corrections that 
scale as a power-law in $(k/\knl)^2$:
\be 
\Delta^2_{\rm stoch}=\frac{k^3}{\bar n_g} + \frac{k^3}{\bar n_g }\frac{k^2}{k_{\rm NL}^2}+...\,,
\ee 
where $\bar n_g$ is the galaxy number density.
Combining all the above effects we get the following estimate 
for the perturbative non-linear density variance of galaxies:
\be 
\begin{split}
\Delta^2\sim  
&\underbrace{\left(\frac{k}{k_{\rm NL}}\right)^{3+n}}_{\rm tree}+
\underbrace{\left(\frac{k}{k_{\rm NL}}\right)^{2(3+n)}}_{\rm 1-loop}+
\underbrace{\left(\frac{k}{k_{\rm NL}}\right)^{3(3+n)}}_{\rm 2-loop}+
\underbrace{\left(\frac{k}{k_{\rm NL}}\right)^{4(3+n)}}_{\rm 3-loop}+ \\
&+\underbrace{\left(\frac{k}{k_{\rm NL}}\right)^{5+n}}_{k^2\PL-\mathrm{ctr}}+\underbrace{\left(\frac{k}{k_{\rm NL}}\right)^{7+n}}_{k^4\PL-\mathrm{ctr}}+\underbrace{\left(\frac{k}{k_{\rm NL}}\right)^{2n+8}}_{k^2P_{\rm 1-loop}-\mathrm{ctr}}\\
&+\underbrace{\frac{k^3}{\bar n_g}}_{\rm LO~stoch.} + \underbrace{\frac{k^3}{\bar n_g}\frac{k^2}{k_{\rm NL}^2}
}_{\rm NLO~stoch.}+ \underbrace{\frac{k^3}{\bar n_g}\frac{k^4}{k_{\rm NL}^4}
}_{\rm NNLO~stoch.}+...\,.
\end{split}
\ee
For spectroscopic surveys ${\bar n}^{-1}_g\sim b_1^2\PL(\kmax= 0.2~\hMpc)$, and $b_1\sim 1$ is the linear galaxy bias.
For imaging surveys the number density is typically much higher
than $[b_1^2\PL(\kmax= 0.2~\hMpc)]^{-1}$, 
but we will use ${\bar n}^{-1}_g\sim b_1^2\PL(\kmax= 0.2~\hMpc)$
as a conservative estimate.
This produces the following scaling:
\be 
\begin{split}
\Delta^2(\kmax)\sim  
&\underbrace{\left(\frac{\kmax}{k_{\rm NL}}\right)^{1.5}}_{\rm tree}+
\underbrace{\left(\frac{\kmax}{k_{\rm NL}}\right)^{3}}_{\rm 1-loop}+
\underbrace{\left(\frac{\kmax}{k_{\rm NL}}\right)^{4.5}}_{\rm 2-loop}+
\underbrace{\left(\frac{\kmax}{k_{\rm NL}}\right)^{6}}_{\rm 3-loop}\\
&+
\underbrace{\left(\frac{\kmax}{k_{\rm NL}}\right)^{3.5}}_{k^2\PL-\mathrm{ctr}}
+\underbrace{\left(\frac{\kmax}{k_{\rm NL}}\right)^{5.5}}_{k^4\PL-\mathrm{ctr}}
+\underbrace{\left(\frac{\kmax}{k_{\rm NL}}\right)^{5}}_{k^2P_{\rm 1-loop}-\mathrm{ctr}}
\\
&+\underbrace{\left(\frac{\kmax}{k_{\rm NL}}\right)^{1.5}}_{\rm LO~stoch.} + \underbrace{\left(\frac{\kmax}{k_{\rm NL}}\right)^{3.5}
}_{\rm NLO~stoch.}+ \underbrace{\left(\frac{\kmax}{k_{\rm NL}}\right)^{5.5}
}_{\rm NNLO~stoch.}+...\,.
\end{split}
\ee 
The two-loop contributions, defined as all terms 
starting with 
$O(k^{4.5})$ and more relevant than $O(k^6)$-terms, should include: 
mode-coupling two-loop contributions $O(k^{4.5})$,
$k^4\PL$-counterterms 
$O(k^{5.5})$, $k^2P_{\rm 1-loop}$-like counterterms $O(k^{5})$,
and the NNLO stochastic terms $O(k^{5.5})$. 
In what follows we will compute the relevant contributions one by one.

\section{Two-loop galaxy bias expansion}
\label{sec:bias}

In this Section we handle the bias expansion 
for the galaxy power spectrum at the two-loop order.
In EFT the galaxy overdensity is expanded as 
\be 
\dg(\vk) = \sum_a b_a\,\mathcal{O}_a(\vk)~\,,
\ee
where each operator $\mathcal{O}_a$ 
is built from the underlying degrees of freedom.
At lowest orders these are 
the density and velocity scalar potentials.
A standard practice is to use the Eulerian framework for these terms, 
i.e. to formulate the bias expansion in terms of 
Eulerian operators that depend on the non-linear density
and velocity fields, and then expand them perturbatively
over the linear matter density field using the 
Standard Perturbation Theory expansion. At lowest orders this produces 
results equivalent to Lagrangian perturbation theory
when the displacement is perturbatively expanded. {However, starting from
the fifth order, the Lagrangian bias expansion
generates operators that cannot be written as local functions
of the Eulerian density and velocity potentials evaluated at the same time.
In the Eulerian language these correspond to non-local-in-time operators,
i.e. convective time derivatives of the tidal tensor~\cite{Mirbabayi:2014zca};
once such operators are included, the Eulerian and Lagrangian expansions are equivalent.}
In what follows,
following~\cite{Eggemeier:2018qae} we will
refer to
the Eulerian bias operators as ``local evolution'' (LE) operators,
and to the Lagrangian bias operators as ``non-local evolution'' (NLE) operators.

While there are efficient ways to generate the complete sets of operators
both in Eulerian and Lagrangian frameworks~\cite{Desjacques:2016bnm},
in what follows we will adopt a hybrid approach which uses the Eulerian
fields for the LE operators and the Lagrangian fields for the NLE 
operators. While this approach creates some asymmetry 
in the treatment of bias operators, it has a great advantage of 
allowing us to connect our two-loop power 
spectrum results with the commonly used Eulerian 
one-loop power spectrum and bispectrum
computations. In particular, this way our basis of bias operators
will easily map onto the one used in standard Eulerian 
codes such as \texttt{CLASS-PT}~\cite{Chudaykin:2020aoj}.

In Lagrangian perturbation theory (LPT), the displacement field
$\vPsi$ is expanded order by order in linear density field $\delta_L$:
\begin{equation}
    \vPsi(\q)= \vPsi^{(1)}(\q) + \vPsi^{(2)}(\q) + \vPsi^{(3)}(\q) + \vPsi^{(4)}(\q) + \cdots
\end{equation}
and each $\vPsi^{(n)}$ can be obtained by iterating the equation
of motion for this field. Using the scalar-vector decomposition, 
the displacement field at a given perturbative order $(n)$
can be written as
\be 
\label{eq:Psin_vec_scal}
\Psi_i^{(n)}=-\nabla_i \varphi^{(n)}+(\nabla\times \Vec{A}^{(n)})_i\,.
\ee 
In what follows we adopt a convention that all derivatives 
acting on LPT potentials are with respect to the Lagrangian
coordinate $\q$, i.e. $\nabla_i\varphi^{(n)}\equiv\partial \varphi^{(n)}/\partial q_i$.
An economical way to summarize the statement about the time 
evolution of the bias expansion is to
use the deformation tensor $\nabla_i\Psi_j$ to build the bias expansion~\cite{Mirbabayi:2014zca,Desjacques:2016bnm}, which we adopt in our treatment of NLE terms.

The operators in the bias expansion
should satisfy symmetries of the problem: the equivalence 
principle and the rotational symmetry. The first one dictates
that the galaxy bias 
expansion should be built from operators that have at least two 
spatial 
derivatives acting on the relevant potentials, which is naturally 
accounted for when using the deformation tensor. 
Rotational invariance
implies that such operators should be Euclidean scalars.

\subsection{Recap of the one-loop galaxy power spectrum}

\begin{figure}
  \centering
  \resizebox{\textwidth}{!}{%
  \begin{tikzpicture}
    \node (a) at (0,0)      {\diagPELE};
    \node (b) at (5.4,0)    {\diagPTT};
    \node (c) at (10.8,0)   {\diagPOT};
  \end{tikzpicture}}
  \caption{Diagrammatic representation of the one-loop contributions to the galaxy
  power spectrum in standard perturbation theory. Filled dots denote the
  linear power spectrum $P_{\mathrm{lin}}$ (times the appropriate power of $b_1$) and open squares the galaxy kernels
  $K_n$.}
  \label{fig:spt-diagrams-1L}
\end{figure}
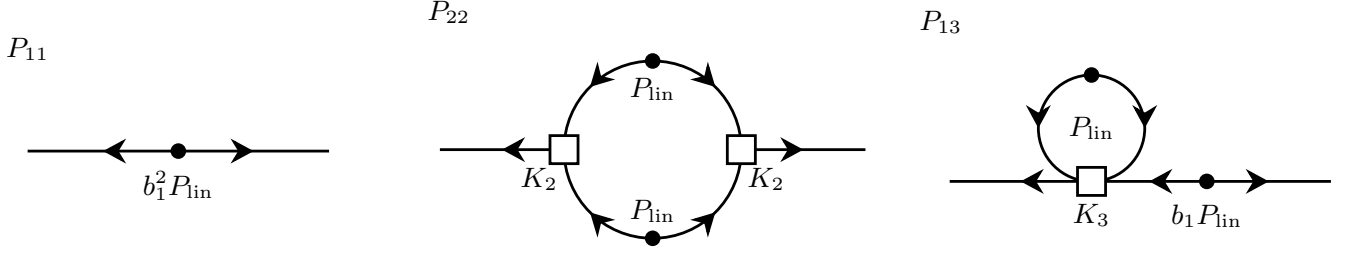

Let us start by listing unique bias operators
constructed using these rules. Up to the fourth order,
these operators have been studied in detail before. Let us review 
these results first. 
At cubic order one has~\cite{Assassi:2014fva,Ivanov:2018gjr,Ivanov:2019pdj,Eggemeier:2018qae}: 
\begin{equation}\label{eq:cubic}
    \delta_g |_{O(\delta^3)}= b_1\delta + \frac{b_2}{2}\delta^2 + \gamma_2\,\Gal_2 + \frac{b_3}{6}\delta^3
    + \gamma_2^\times\delta\,\Gal_2 + \gamma_3\,\Gal_3 + \gamma_{21}\,\Gal_2(\varphi_2,\varphi_1)\,,
\end{equation}
where $\Gal_2$ and $\Gal_3$ are Galileon operators 
$\Gal_2 \equiv (\nabla_i\nabla_j\Phi_v)^2 - (\nabla^2\Phi_v)^2$, $\Gal_3 \equiv 6\det(\nabla_i\nabla_j\Phi_v)$. At the first non-trivial order (i.e. 2nd and 3rd for $\Gal_2$ and $\Gal_3$, respectively), they are expressed 
as operators acting on the velocity potentials, which are equivalent 
to the 1st order scalar LPT displacement potential
\be 
\Psi^{(1)}_i = \nabla_i\varphi_1\,.
\ee 
Using the linear density field $\delta_L$ we get the usual 
Zel'dovich expression $\delta_L=-\nabla_i \Psi^{(1)}_i=-\nabla^2\varphi_1$.
At the cubic order 
it is possible to build a Galileon operator from the 2nd order LPT potential $\varphi_2$ ($\Psi^{(2)}_i = \nabla_i \varphi_2$):
\be 
\Gal_2(\varphi_2,\varphi_1)\equiv (\nabla_i\nabla_j\varphi_2)(\nabla_i\nabla_j\varphi_1) - (\nabla^2\varphi_2)(\nabla^2\varphi_1)\,,
\ee 
where $\nabla^2\varphi_2=\nabla_i \Psi^{(2)}_i=-\Gal_2(\varphi_1)$.

All in all, the cubic expansion \eqref{eq:cubic} contains six local 
evolution (LE) operators built from velocity or density potentials
\be 
\mathcal{O}^{\rm LE}_a=\{\delta,\,\delta^2,\,\delta\Gal_2,\,\delta^3,\,
\Gal_2(\Phi_v),\,\Gal_3(\Phi_v)\}\,,
\ee 
of which $\delta\Gal_2$ is the mixed LE product operator. Plus there is a
non-local evolution (NLE) operator $\Gal_2(\varphi_2,\varphi_1)$ built from
the 2nd order displacement distortion tensor. This operator can be 
represented as Galileon operator that captures the difference between the velocity
and density potentials. 
Subjecting all non-linear density and velocity fields to the SPT expansion
we get
\begin{equation}
    \mathcal{O}_a(\vk) = \sum_{n=n_a}^{\infty} \int_{\vq_1\cdots\vq_n}(2\pi)^3\delta_D(\vk{-}\vq_{1\cdots n})\,K_a^{(n)}(\vq_1,\ldots,\vq_n)\,\dL(\vq_1)\cdots\dL(\vq_n)\,,
\end{equation}
where $\vq_{1\cdots n}=\vq_1+...+\vq_n$, $\delta_L$ is the linear density field, $n_a$ is the original (Eulerian) order of the operator, and 
the 2nd and 3rd galaxy bias kernels are given by
\be 
\begin{split}
& K_{2}(\vq_1,\vq_2)=b_1 F_2(\vq_1,\vq_2)+\frac{b_2}{2}+\g_2\Kcal(\vq_1,\vq_2)\,,\\
&K_{3}(\vq_1,\vq_2,\vq_3)=b_1 F_3(\vq_1,\vq_2,\vq_3)+
{b_2}\frac{1}{3}\big[F_2(\vq_1,\vq_2)+F_2(\vq_1,\vq_3)+F_2(\vq_2,\vq_3)\big]\\
&+{\frac{2\g_2}{3}}\big[\Kcal(\vq_1,\vq_{23})G_2(\vq_2,\vq_3)+\text{2 cyc.}\big]
+
\frac{b_3}{6}+
\gamma_2^\times \frac{1}{3}\big[\Kcal(\vq_1,\vq_2)+\Kcal(\vq_1,\vq_3)+\Kcal(\vq_2,\vq_3)\big]\\
&+
\gamma_3 \Lcal(\vq_1,\vq_2,\vq_3)+\gamma_{21}
\frac{1}{3}\big[\Kcal(\vq_1,\vq_{23})\Kcal(\vq_2,\vq_3)+\text{2 cyc.}\big]\,,
\end{split}
\ee 
where we have defined the Fourier kernels:
\begin{align}
    \Kcal(\vq_1,\vq_2) &= \frac{(\vq_1{\cdot}\vq_2)^2}{q_1^2\,q_2^2} - 1 = \mu_{12}^2 - 1 \,,\label{eq:Kdef}\\
    \Lcal(\vq_1,\vq_2,\vq_3) &= 1 - \mu_{12}^2 - \mu_{23}^2 - \mu_{31}^2 + 2\mu_{12}\mu_{23}\mu_{31} \,,\label{eq:Ldef}
\end{align}
with $\mu_{ij} = \hat\vq_i\cdot\hat\vq_j$. 
The above kernels have the following important properties:
\be 
\Kcal(\vq,-\vq) = 0\,,\quad \Lcal(\mathbf{a},\mathbf{b},-\mathbf{a}{-}\mathbf{b}) = 0\,.
\ee 
The key effect of the SPT expansion is to displace (or evolve) the 
local Lagrangian fields by the LPT displacements. For this reason in what follows
we will call the operators stemming from the SPT expansion ``evolved.''
Schematically the cubic kernel 
can be written as 
\be 
K^{(3)} = K^{(3)}_{\rm LE}[b_3,\gamma_2^\times,\gamma_3] + K^{(3)}_{\rm NLE}[\gamma_{21}] + K^{(3)}_{\rm evolved}[b_1,b_2,\gamma_2]\,,
\ee 
where $K^{(3)}_{\rm LE/NLE}$ are the kernels of LE/NLE
cubic operators in which all composite fields are taken to be linear,
while $K^{(3)}_{\rm evolved}$ represents the lower-order 
bias operators in which the composite fields are evolved using 
the SPT expansion. In what follows we will use 
\be 
K_{\gamma_{21}}^{(3)}(\vq_c,\vq_d,\vq_e) = 
\frac{1}{3}\sum_{\text{3 cyc}}
    \Kcal(\vq_c,\vq_{de})\,\Kcal(\vq_d,\vq_e)\,,
    \ee 
where $\sum_{\text{3 cyc}}$ denotes a sum over three 
distinct momenta configurations obtained by a cyclic permutation.
In what follows we suppress ``cyc'' in this and analogous sums to simplify the notation.
The deterministic part of the one-loop galaxy power spectrum is given by
\be 
\begin{split}
P^{\rm tot}_{gg}(k)&=b_1^2 \PL(k)+P_{13}(k)+P_{22}(k)\\
&=b_1^2 \PL(k)+6\PL(k)\int_{\q} K_{3}(\vk,-\vq,\vq)\PL(q)+2\int_{\q}[K_{2}(\vk-\vq,\vq)]{^2}\PL(|\vk-\vq|)\PL(q)\,.
\end{split}
\ee 
Note that following~\cite{Scoccimarro:1995if} we absorb all symmetry
factors in the definition of $P_{nm}$, i.e. our $P_{13}$ includes
the factor of two.
Feynman diagrams representing these corrections are displayed 
in Fig.~\ref{fig:spt-diagrams-1L}.
The expressions for the loop corrections are well known in the literature.
Let us highlight a couple of points that become important moving to higher orders. First of all, the $b_2^2$ term inside $P_{22}$ above has a constant 
power spectrum at $k\to 0$. This can be interpreted as a shot noise
renormalization. In practice, the $k=0$ limit of this integral is often subtracted from the integrand. We will follow this convention in our work. 
Second, the $P_{13}$ contributions from the shifted (evolved) $\delta^2$ and from all the LE operators are trivial. Specifically, focusing on these terms we have:
\be 
P_{gg}\supset b_1^2 \PL(k)+ 2b_1\PL(k)\left(\frac{b_3}{2} + \frac{34 b_2}{21} - \frac{4\gamma_2^\times}{3}\right)\int_{\q}\PL(q)\,.
\ee  
The contribution from $\Gal_3$ vanishes identically at the one-loop order.
Importantly the above expression features a UV-divergent mass variance integral
$\sigma^2\equiv \int_{\q}\PL(q)$. This divergence is eliminated by 
the linear bias renormalization,
\be 
b_1^2 \PL(k)  + 2b_1\PL(k)\left(\frac{b_3}{2} + \frac{34 b_2}{21} - \frac{4\gamma_2^\times}{3}\right)\sigma^2\equiv (b_1+\delta b_1 \sigma^2)^2
\PL(k)
\equiv [b^{(R)}_1]^2\PL(k) + O(\sigma^4)\,,\ee
where $b_1^{(R)}$ is the finite renormalized bias, 
which should be contrasted with the infinite bare bias $b_1$ which absorbs
the divergent loop contribution $\delta b_1$:
\be 
\label{eq:1l_ren_cond}
\delta b_1 \equiv \frac{b_3}{2} + \frac{34 b_2}{21} - \frac{4\gamma_2^\times}{3}\,.
\ee 
The above equation is a $b_1$ renormalization condition to order 
$\sigma^2$, assuming in what follows 
that $\delta b_{\mathcal{O}}\sigma^2$ is a loop correction that has to be 
exactly cancelled by the bare bias parameter $\delta b_{\mathcal{O}}$ to yield 
the renormalized bias parameter $b^{(R)}_{\mathcal{O}}$ 
\be 
b_{\mathcal{O}} + \delta b_{\mathcal{O}} \sigma^2=  b^{(R)}_{\mathcal{O}}\Big|_{\sigma^2}\,.
\ee 
Importantly, the LE operators and the shifted $\delta^2$
do not have finite, calculable loop corrections. This is a consequence
of using the Galileon bias. In contrast, the NLE operator $\gamma_{21}$ has a 
convergent finite contribution. The reason this contribution does not vanish
is that its complicated momentum structure implies the terms like $1/|\vk-\vq|^2$
in the integrand, which cannot be eliminated by angular integration.
Similar terms stem from the shifted $\gamma_2$ kernel.

\subsection{Fourth order bias}

Moving to the fourth order we have to take into account that the LPT
displacement ceases to be purely potential at the cubic order, see eq.~\eqref{eq:Psin_vec_scal}. 
Explicitly, one has~\cite{Eggemeier:2018qae,Matsubara:2007wj,Matsubara:2015ipa}:
\be 
\vPsi^{(3)}(\q,\tau)=D_3^{(a)}\vPsi^{(3a)}+{D_3^{(b)}\vPsi^{(3b)}}
+D_3^{(c)}\vPsi^{(3c)}\,,
\ee 
where (a,b) are scalar components,
whilst (c) is the vector component $\vPsi^{(3c)} = \nabla\times\Vec{A}_3$;
the LPT solution for the growth factors in the EdS approximation reads
$D_3^{(a)}/D^3 = -1/18$, $D_3^{(b)}/D^3 = 5/42$, $D_3^{(c)}/D^3 =
-1/14$.
The scalar and vector components above are sourced by the low order scalar potentials: 
\be 
\begin{split} 
& \nabla\cdot \vPsi^{(3a)} =\nabla^2\varphi_3^{(a)} = -\Gal_3(\varphi_1) \,,\\
& \nabla\cdot \vPsi^{(3b)} =\nabla^2\varphi_3^{(b)} = -\Gal_2(\varphi_2,\varphi_1)\,, \\
& \nabla \times \vPsi^{(3c)} = -\nabla^2\Vec{A}_3=\hat{\Vec{e}}_i\epsilon_{ijk}
(\nabla_{jl}\varphi_1)(\nabla_{kl}\varphi_2)\,.
\end{split}
\ee 
We define now the Lagrangian kernel in Fourier space:
\be
\vPsi^{(n)}=i D^n \int_{\q_1...\q_{n}}(2\pi)^3\delta_D^{(3)}
(\k-\q_{1...n})\Vec{L}_n(\q_1,...,\q_{n})
\delta_L(\q_1)...\delta_L(\q_n)\,,
\ee 
which can be decomposed into the scalar divergence 
and curl parts $S_n\equiv \k\cdot \Vec{L}_n$, 
$\Vec{T}_n=-\k\times \Vec{L}_n$ where for $n=3$ we have
\be 
S_3(\vq_j,\vq_k,\vq_l) \equiv
   \frac{1}{18}
   \left(
   -\Lcal(\vq_j,\vq_k,\vq_l)
+ \frac{15}{7}K_{\gamma_{21}}^{(3)}(\vq_j,\vq_k,\vq_l)
\right)
\,,
\ee 
and  
\be\label{eq:T3def}
\Vec{T}_3(\vq_1,\vq_2,\vq_3) = -\frac{1}{42}\Big[
\Vec{W}(\vq_1,\vq_{23})\,\Kcal(\vq_2,\vq_3)
+\Vec{W}(\vq_2,\vq_{13})\,\Kcal(\vq_1,\vq_3)
+\Vec{W}(\vq_3,\vq_{12})\,\Kcal(\vq_1,\vq_2)\Big]\,,
\ee
where we used
\begin{equation}\label{eq:Wdef}
    \Vec{W}(\vk_1,\vk_2) = \frac{(\vk_1\times\vk_2)(\vk_1\cdot\vk_2)}{k_1^2\,k_2^2}\,.
\end{equation}
While the vector potential does not directly contribute
to the cubic LPT solution for $\delta$, it generates non-trivial corrections
at higher orders through LPT recursion relations~\cite{Matsubara:2007wj,Matsubara:2015ipa}. 

We introduce now Galileon operators built 
from higher order distortion
tensors, i.e. starting from the cubic order we use the notation
($M^{(n)}_{ij}\equiv \nabla_j \Psi_i^{(n)}$)
\be 
\begin{split}
\Gal_2(\vPsi^{(n)},\vPsi^{(m)})&
\equiv 
\text{Tr}(M^{(n)}M^{(m)})-\text{Tr}(M^{(m)})
\text{Tr}(M^{(n)})
=(\nabla_i\Psi_j^{(n)})(\nabla_j\Psi_i^{(m)}) - (\nabla_i\Psi_i^{(n)})(\nabla_j\Psi_j^{(m)})\,,\\
-\Gal_3(\vPsi^{(l)},\vPsi^{(m)},\vPsi^{(n)}) 
&\equiv \text{Tr}(M^{(l)})\text{Tr}(M^{(m)})
\text{Tr}(M^{(n)})+\text{Tr}(M^{(l)}M^{(m)}M^{(n)})
+\text{Tr}(M^{(l)}M^{(n)}M^{(m)})\\
&\quad \quad \quad -\sum_{\rm cyc} \text{Tr}(M^{(l)})
\text{Tr}(M^{(m)}M^{(n)})\,.
\end{split}
\ee 
Since $\Vec{A}^{(1)}=\Vec{A}^{(2)}=0$, this distinction was
not important at the one-loop order.
At the 4th order we get 8 new contributions from four LE and four new NLE operators:
\begin{equation}
    \mathcal{O}_4 = \Big\{\;\underbrace{\frac{\delta^4}{24}\,,\;\frac{\delta^2\Gal_2}{2}\,,\;\delta\Gal_3\,,\;\frac{\Gal_2^2}{2}}_{\text{4 LE}}\,,\quad\underbrace{\delta\,\Gal_2(\varphi_2,\varphi_1)\,,\;\Gal_3(\varphi_2,\varphi_1,\varphi_1)\,,\;\Gal_2(\varphi_2,\varphi_2)\,,\;\Gal_2(\vPsi^{(3)},\vPsi^{(1)})}_{\text{4 NLE}}\;\Big\}\,.
\end{equation}
where the last operator is given by
\begin{equation}\label{eq:gamma31_ESS}
\Gal_2(\vPsi^{(3)},\vPsi^{(1)})\equiv
 D^4\left[    {-\frac{1}{18}}\,\Gal_2(\varphi_3^{(a)},\varphi_1)
    + \frac{5}{42}\,\Gal_2(\varphi_3^{(b)},\varphi_1)
    {-\frac{1}{14}}\,\nabla_i(\nabla\times \Vec{A}_3)_j\;\nabla_{ij}\,\varphi_1\right]\,,
\end{equation}
where we have explicitly factored out an overall growth factor $D^4$
and the two sign flips relative to~\cite{Eggemeier:2018qae} follow from
the growth factors above.
The rightmost transverse term is built from the pseudovector
$\nabla\times\Vec{A}_3$, but it enters the operator through a scalar
contraction of two $\Vec{W}$ vectors, cf.~eq.~\eqref{eq:WdotW}: it is
even under the parity reflection $\vq_i\to-\vq_i$ of the loop
momenta and thus it does not vanish upon angular integration and is retained in what follows.  
All together, the above fourth order operators produce the following 
symmetrized 
Fourier space kernel:
\begin{align}\label{eq:Gamma4_full}
    \tilde{K}_{4}(\vq_1,\vq_2,\vq_3,\vq_4) =
    &\quad \frac{b_4}{24}
    + {\frac{\gamma_2^{\times\times}}{12}}\sum_{6}\!\Kcal(\vq_i,\vq_j)
    + \frac{\gamma_3^\times}{4}\sum_{4}\!\Lcal(\vq_j,\vq_k,\vq_l) \nonumber
    + {\frac{\gamma_2^{{sq}}}{6}}\sum_{3}\!\Kcal(\vq_a,\vq_b)\,\Kcal(\vq_c,\vq_d)
    \nonumber\\[6pt]
    &\quad + \frac{\gamma_{21}^\times}{12}\sum_{12}\!
    \Kcal(\vq_i,\vq_{jk})\,\Kcal(\vq_j,\vq_k)
  + \frac{\gamma_{211}}{6}\sum_{6}\!
    \Lcal(\vq_a,\vq_b,\vq_{cd})\,\Kcal(\vq_c,\vq_d)
    \nonumber\\
    &\quad + \frac{\gamma_{22}}{3}\sum_{3}\!
    \Kcal(\vq_{ab},\vq_{cd})\,\Kcal(\vq_a,\vq_b)\,\Kcal(\vq_c,\vq_d)\\
 &\quad  + {\frac{\gamma_{31}}{4}\sum_{4}\!
    \Big[\Kcal(\vq_i,\vq_{jkl})\,S_3(\vq_j,\vq_k,\vq_l)
    +\Vec{W}(\vq_i,\vq_{jkl})\cdot\Vec{T}_3(\vq_j,\vq_k,\vq_l)\Big]}\,, 
\end{align} 
and the sums $\sum_n$  run over n  distinct 
combinations of four momenta,
$\vq_{ij} = \vq_i{+}\vq_j$, $\vq_{ijk} = \vq_i{+}\vq_j{+}\vq_k$.
Note that the above kernel does not 
account for SPT evolution or advection. We will 
discuss these effects shortly.

\subsection{Fifth order bias}

At 5th order we find 5 LE operators
\begin{equation}
    \mathcal{O}_5^{\mathrm{LE}} = \Big\{\;\frac{\delta^5}{5!}\,,\quad \frac{\delta^3\Gal_2}{6}\,,\quad \frac{\delta^2\Gal_3}{2}\,,\quad \frac{\delta\Gal_2^2}{2}\,,\quad \Gal_2\Gal_3\;\Big\}\,,
\end{equation}
5 NLE product-type operators,
\begin{equation}
    \mathcal{O}_5^{\mathrm{NLE,prod}} = \Big\{\;\frac{\delta^2_L}{2}\Gal_2(\varphi_2,\varphi_1)\,,\; \Gal_2(\Phi_v)\Gal_2(\varphi_2,\varphi_1)\,,\; \delta_L\Gal_3(\varphi_2,\varphi_1,\varphi_1)\,,\; \delta_L\Gal_2(\varphi_2,\varphi_2)\,,\; \delta_L\Gal_2(\vPsi^{(3)},\vPsi^{(1)})\;\Big\}\,,
\end{equation}
where 
$\delta_L$ explicitly reflects the 
Lagrangian nature of these operators,
and 4 genuinely new NLE operators
\begin{equation}
    \mathcal{O}_5^{\mathrm{NLE,new}} = \Big\{\;\Gal_2(\vPsi^{(3)},\vPsi^{(2)})\,,\quad \Gal_3(\varphi_2,\varphi_2,\varphi_1)\,,\quad \Gal_3(\vPsi^{(3)},\vPsi^{(1)},\vPsi^{(1)})\,,\quad \Gal_2(\vPsi^{(4)},\vPsi^{(1)})\;\Big\}~\,.
\end{equation}
$\Gal_2(\vPsi^{(3)},\vPsi^{(2)})$ and 
$\Gal_3(\vPsi^{(3)},\vPsi^{(1)},\vPsi^{(1)})$
are simple generalizations 
of the quadratic and cubic Galileon matrices
which depend now on the full cubic potential. 
The parts depending on the transverse potential $\Vec{A}_3$
enter these operators through scalar contractions of two $\Vec{W}$
vectors and are parity-\emph{even}, exactly as for $\gamma_{31}$ above.
The last term above has the following explicit expression:
\begin{align}\label{eq:gamma41_master}
    \Gal_2(\vPsi^{(4)},\vPsi^{(1)}) &=
    D^5\Big[{\frac{7}{198}}\,\Gal_2(\varphi_4^{(1,3a)},\varphi_1)
    {- \frac{5}{66}}\,\Gal_2(\varphi_4^{(1,3b)},\varphi_1)
    {- \frac{51}{4312}}\,\Gal_2(\varphi_4^{(2,2)},\varphi_1)\nonumber\\
    &\quad {- \frac{13}{308}}\,\Gal_2(\varphi_4^{(2,1,1)},\varphi_1)
    {+ \frac{1}{22}}\,\Gal_2(\varphi_4^{(1,3c)},\varphi_1)
    \nonumber\\
    &\quad {- \frac{1}{36}}\,\mathcal{D}[\vec{A}_4^{(1,3a)}]
    {+ \frac{5}{84}}\,\mathcal{D}[\vec{A}_4^{(1,3b)}]
    {- \frac{1}{28}}\,\mathcal{D}[\vec{A}_4^{(1,A_3)}]\Big]\,,
\end{align}
where we have introduced the following notation for the deformation
tensor built from the transverse displacement,
\begin{equation}\label{eq:D_A4}
    \mathcal{D}[\Vec{A}_4^{(X)}] \equiv
    \Psi_{i,j}^{(4,X)}\;\varphi_{1,ij}
    = (\nabla\times\Vec{A}_4^{(X)})_{i,j}\;\varphi_{1,ij}\,.
\end{equation}
The above Galileon operator $\gamma_{41}$ is a straightforward
generalization of $\gamma_{31}$. Specifically, the scalar part reads
\begin{align}
    &\nabla^2\varphi_4^{(1,3a)} = -\Gal_2(\varphi_1,\varphi_3^{(a)})\,,
\\
    &\nabla^2\varphi_4^{(1,3b)} = -\Gal_2(\varphi_1,\varphi_3^{(b)})\,,
\\
    &\nabla^2\varphi_4^{(2,2)} = -\Gal_2(\varphi_2,\varphi_2)\,,
\\
    &\nabla^2\varphi_4^{(2,1,1)} = \Gal_3(\varphi_2,\varphi_1,\varphi_1)\,,
\\
    &\nabla^2\varphi_4^{(1,3c)} = -(\nabla\times\Vec{A}_3)_{i,j}\;\varphi_{1,ij}\,.
\end{align}
Importantly, we note the appearance of a 
scalar sub-potential sourced by the transverse vector
$\Vec{A}_3$. As for the vector potentials, their explicit
contributions generated by the LPT recursion relations take the following form:
\begin{alignat}{3}\label{eq:Apot}
 & \nabla^2 A_{4,i}^{(1,3a)} = -\epsilon_{ijk}\,(\nabla_{jl}\varphi_1)(\nabla_{kl}\varphi_3^{(a)})\\
  & \nabla^2 A_{4,i}^{(1,3b)} = -\epsilon_{ijk}\,(\nabla_{jl}\varphi_1)(\nabla_{kl}\varphi_3^{(b)})\,,\\
   &  \nabla^2 A_{4,i}^{(1,A_3)} = -\epsilon_{ijk}\,(\nabla_{jl}\varphi_1)\,
    \nabla_k(\nabla\times\Vec{A}_3)_l\,,\\
   &  \nabla^2 A_{4,i}^{(2,1,1)} =
    -\epsilon_{ijk}\,(\nabla_{jl}\varphi_2)\,(\nabla_{lm}\varphi_1)\,(\nabla_{km}\varphi_1)\,.
\end{alignat}
The transverse contribution sourced by two 
2nd order scalar potentials vanishes identically 
$\epsilon_{ijk}(\nabla_{jl}\varphi_2)(\nabla_{kl}\varphi_2)=0$.
The non-trivial scalar part stems from the following LPT kernel:
\begin{equation}\label{eq:S4decomp}
    S_4^{\text{sym}}(\vq_1,\ldots,\vq_4) = -\frac{28}{33}\,\mathcal{S}_{(1,3a)}
    + \frac{20}{11}\,\mathcal{S}_{(1,3b)} + \frac{153}{539}\,\mathcal{S}_{(2,2)}
   -\frac{78}{77}\,\mathcal{S}_{(2,1,1)} {+\frac{12}{11}}\,\mathcal{S}_{(1,3c)}\,,
\end{equation}
where the scalar pieces are degenerate with the 
other 4th order operators ($\vq_{\hat{i}}=\sum_{j\neq i}\vq_j$):
\begin{align}
    \mathcal{S}_{(1,3a)} &= \frac{1}{4}\sum_{i=1}^4
    \Kcal(\vq_i,\vq_{\hat{i}})\;\Lcal(\vq_j,\vq_l,\vq_k)\,,
\quad \quad \quad      \mathcal{S}_{(1,3b)} = \frac{1}{4}\sum_{i=1}^4
    \Kcal(\vq_i,\vq_{\hat{i}})\;K_{\gamma_{21}}^{(3)}(\vq_j,\vq_l,\vq_k)\,,
    \label{eq:S13b}\\
    \mathcal{S}_{(2,2)} &= K_{\gamma_{22}}^{(4)}(\vq_1,\vq_2,\vq_3,\vq_4)\,,
\quad \quad \quad     \mathcal{S}_{(2,1,1)} = K_{\gamma_{211}}^{(4)}(\vq_1,\vq_2,\vq_3,\vq_4)\,,
    \label{eq:S211}
\end{align}
while the vector potential contribution is 
\begin{equation}\label{eq:S13c}
    \mathcal{S}_{(1,3c)} = -\frac{1}{4!}
    \sum_{\sigma\in S_4}
    \big[\Vec{W}(\vq_{\sigma_1},\vq_{\sigma_{234}})
    \cdot\Vec{W}(\vq_{\sigma_2},\vq_{\sigma_{34}})\big]\;
    \Kcal(\vq_{\sigma_3},\vq_{\sigma_4})\,.
\end{equation}
The above dot product evaluates to:
\begin{equation}\label{eq:WdotW}
    \Vec{W}(\vk_a,\vk_b)\cdot\Vec{W}(\vk_c,\vk_d) =
    \frac{\big[(\vk_a\cdot\vk_c)(\vk_b\cdot\vk_d)
    - (\vk_a\cdot\vk_d)(\vk_b\cdot\vk_c)\big]
    (\vk_a\cdot\vk_b)(\vk_c\cdot\vk_d)}{k_a^2\,k_b^2\,k_c^2\,k_d^2}\,,
\end{equation}
which gives rise to a new momentum configuration
that has not appeared from low order scalar potentials.

The full symmetrized galaxy kernel at 5th order is
\begin{align}\label{eq:K5_full}
  &  \tilde{K}_{5}(\vq_1,\ldots,\vq_5) =
    \frac{b_5}{120}
    + \frac{\gamma_2^{\times\times\times}}{6\cdot 10}\sum_{10}\Kcal(\vq_a,\vq_b)
    + \frac{\gamma_3^{\times\times}}{2\cdot 10}\sum_{10}\Lcal(\vq_c,\vq_d,\vq_e)
+   \frac{\gamma_2^{{sq}\times}}{2\cdot 15}\sum_{15}
    \Kcal(\vq_a,\vq_b)\,\Kcal(\vq_c,\vq_d)
   \nonumber\\
    &\quad +    \frac{b_{\Gal_2\Gal_3}}{10}\sum_{10}
    \Kcal(\vq_a,\vq_b)\,\Lcal(\vq_c,\vq_d,\vq_e)
+ \frac{\gamma_{21}^{\times\times}}{20}\sum_{10}
    K_{\gamma_{21}}^{(3)}(\vq_c,\vq_d,\vq_e)
    + \frac{\gamma_{21,2}}{10}\sum_{10}
    \Kcal(\vq_a,\vq_b)\,K_{\gamma_{21}}^{(3)}(\vq_c,\vq_d,\vq_e)
    \nonumber\\
    &\quad + \frac{\gamma_{211}^\times}{5}\sum_{5}
    K_{\gamma_{211}}^{(4)}(\hat\vq_i)
    + \frac{\gamma_{22}^\times}{5}\sum_{5}
    K_{\gamma_{22}}^{(4)}(\hat\vq_i)
    + \frac{\gamma_{31}^\times}{5}\sum_{5}
    K_{\gamma_{31}}^{(4)}(\hat\vq_i)
   \nonumber\\
    &\quad + {\frac{\gamma_{23}}{10}\sum_{10}
    \Kcal(\vq_a,\vq_b)\Big[\Kcal(\vq_{ab},\vq_{cde})\,
    S_3^{\mathrm{sym}}(\vq_c,\vq_d,\vq_e)
    +\Vec{W}(\vq_{ab},\vq_{cde})\cdot\Vec{T}_3(\vq_c,\vq_d,\vq_e)\Big]}
    \nonumber\\
    &\quad + \frac{\gamma_{221}}{15}\sum_{15}
    \Lcal(\vq_{ab},\vq_{cd},\vq_e)\,\Kcal(\vq_a,\vq_b)\,\Kcal(\vq_c,\vq_d)
  \nonumber\\
    &\quad  + {\frac{\gamma_{311}}{10}\sum_{10}
    \Big[\Lcal(\vq_{abc},\vq_d,\vq_e)\,S_3^{\mathrm{sym}}(\vq_a,\vq_b,\vq_c)
    + X_{311}(\vq_a,\vq_b,\vq_c;\vq_d,\vq_e)\Big]}
 \nonumber\\
    &\quad  + {\frac{\gamma_{41}}{5}\sum_{5}
    \Big[\Kcal(\vq_i,\vq_{\hat{i}})\,S_4^{\mathrm{sym}}(\hat\vq_i)
    +\Vec{W}(\vq_i,\vq_{\hat i})\cdot\Vec{T}_4^{\mathrm{sym}}(\hat\vq_i)\Big]}\,,
\end{align}
where $S_3^{\mathrm{sym}}$ is the symmetrized 
3rd order LPT kernel, $S_4^{\mathrm{sym}}(\hat\vq_i)$ is the fully symmetrized scalar
LPT kernel of $\varphi_4$ evaluated on the 4 momenta complementary to $\vq_i$,
$\hat\vq_i$ denotes the complement $\{\vq_1,\ldots,\vq_5\}\setminus\{\vq_i\}$.
The transverse companions are built from the curl kernels: the
$\gamma_{311}$ term contracts the Fourier kernel of the transverse
displacement itself,
$\Vec{{V}}_3(\vq_a,\vq_b,\vq_c) \equiv
\vq_{abc}\times\Vec{T}_3(\vq_a,\vq_b,\vq_c)/q_{abc}^2$ against the two
$\vPsi^{(1)}$ terms inside $\Gal_3$,
\be\label{eq:X311def}
X_{311}(\vq_a,\vq_b,\vq_c;\vq_d,\vq_e) =
\sum_{d\leftrightarrow e}
\left[
\frac{(\vq_{abc}\!\cdot\!\vq_d)(\vq_d\!\cdot\!\vq_e)
(\vq_e\!\cdot\!\Vec{{V}}_3)}{q_d^2\,q_e^2}
-\frac{(\vq_{abc}\!\cdot\!\vq_d)(\vq_d\!\cdot\!\Vec{{V}}_3)}{q_d^2}
\right]\,,
\ee
and $\Vec{T}_4^{\mathrm{sym}} \equiv -24\,\Vec{T}_4$ carries the same
$-24$ normalization as $S_4^{\mathrm{sym}}$, with the explicit
decomposition over the vector potentials of
eq.~\eqref{eq:gamma41_master},
\be\label{eq:T4def}
\Vec{T}_4^{\mathrm{sym}}(\vq_1,\ldots,\vq_4) = -\,\k\times\Bigl(\k\times
\Bigl[\tfrac{2}{3}\,\tilde{\vec{A}}_4^{(1,3a)}
-\tfrac{10}{7}\,\tilde{\vec{A}}_4^{(1,3b)}
+\tfrac{6}{7}\,\tilde{\vec{A}}_4^{(1,A_3)}\Bigr]\Bigr)\,,
\ee
with $\k \equiv \vq_{1234}$ and $\tilde{\vec{A}}_4^{(X)}$ denoting the symmetrized Fourier kernels of the unit
vector potentials $\vec{A}_4^{(X)}$ defined by Eq.~\eqref{eq:Apot}.
The $\gamma_{31}^\times$ term and the advected $\gamma_{31}$ 
are built from $K^{(4)}_{\gamma_{31}}$ and are
completed automatically by the bracket of eq.~\eqref{eq:Gamma4_full}.

\subsection{Evolution and advection of operators}

The LPT operators we have used are subject to advection, 
which produces higher order contributions. Let us remind
the reader that 
in our approach all LE operators built from non-linear 
Eulerian velocity or density 
potentials encode the evolution of cosmological fields 
from the SPT expansion for the density and velocity 
fields from the non-linear perfect fluid equations. 
This is implemented by plugging the SPT formulas 
$\delta=\delta_L+\int F_2 \delta_L^2 + ...$
and $\theta=\delta_L+\int G_2 \delta_L^2 + ...$
in the relevant LE operators.

For the NLE Lagrangian operators, we implement an explicit LPT advection.
The new contributions are generated by the displacement field 
acting on an LPT operator $\mathcal{O}$ as follows:
\begin{equation}\label{eq:advection}
    \mathcal{O} \;\to\;
   \mathcal{O}+ \nabla_i\!\left(\mathcal{O}\;\frac{\nabla_i\delta_L}{\nabla^2}\right)
    + \frac{1}{2}\,\nabla_i\nabla_j\!\left(\mathcal{O}\;
    \frac{\nabla_i\delta_L}{\nabla^2}\;\frac{\nabla_j\delta_L}{\nabla^2}\right)
    - \frac{3}{14}\nabla_i\!\left(\mathcal{O}\;\frac{\nabla_i \Gal_2(\Phi_v)}{\nabla^2}\right)
    + \cdots\,,
\end{equation}
where we kept only the terms through 2LPT order 
relevant for the two-loop galaxy power spectrum
computation.
At fourth order, this gives the following final galaxy bias kernel:
\begin{align}\label{eq:K4_full}
 &   K^{(4)}_g = \tilde{K}^{(4)} + b_1 F_4
    + \frac{b_2}{4}\sum_{4}F_3(\hat\vq_i)
    + \frac{b_2}{6}\sum_{3}F_2(\vq_a,\vq_b)F_2(\vq_c,\vq_d)
    + {\frac{\gamma_2}{2}}\sum_{4}\Kcal(\vq_i,\vq_{\hat{i}})G_3(\hat\vq_i)
        \nonumber\\
    &\quad
    + \frac{\gamma_2}{3}\sum_{3}\Kcal(\vq_{ab},\vq_{cd})G_2(\vq_a,\vq_b)G_2(\vq_c,\vq_d)
    + {\frac{b_3}{12}}\sum_{6}F_2(\vq_i,\vq_j)
    + \frac{\gamma_2^\times}{12}\sum_{12}
    \Big[\Kcal(\vq_k,\vq_l)F_2(\vq_i,\vq_j)
    + 2\Kcal(\vq_{ij},\vq_k)G_2(\vq_i,\vq_j)\Big]
    \nonumber\\
    &\quad
    + {\frac{\gamma_3}{2}}\sum_{6}\Lcal(\vq_{ij},\vq_k,\vq_l)G_2(\vq_i,\vq_j)
    + \frac{\gamma_{21}}{4}\sum_{4}
    \frac{\vk\cdot\vq_i}{q_i^2}\,K_{\gamma_{21}}^{(3)}(\hat\vq_i)\,,
\end{align}
where $\tilde{K}_{4}$ is the leading 4th-order kernel
Eq.~(\ref{eq:Gamma4_full}), and the remaining terms evolve or advect
lower-order operators.
The evolved 5th order kernel reads:
\begin{align}\label{eq:K5_complete}
   & K_{5} = \tilde{K}_{5}+ 
   b_1\,F_5
 + \frac{b_2}{5}\sum_{5}F_4(\hat\vq_i)
    + \frac{b_2}{10}\sum_{10}F_2(\vq_a,\vq_b)\,F_3(\vq_c,\vq_d,\vq_e)
   + {\frac{2\gamma_2}{5}}\sum_{5}\Kcal(\vq_i,\vq_{\hat{i}})\,G_4(\hat\vq_i)  \nonumber\\
    &\quad 
    + {\frac{\gamma_2}{5}}\sum_{10}\Kcal(\vq_{ab},\vq_{cde})\,
    G_2(\vq_a,\vq_b)\,G_3(\vq_c,\vq_d,\vq_e)
 + {\frac{b_3}{20}}\sum_{10}F_3(\vq_c,\vq_d,\vq_e)
    + {\frac{b_3}{30}}\sum_{15}F_2(\vq_a,\vq_b)\,F_2(\vq_c,\vq_d)
    \nonumber\\
    &\quad + \frac{\gamma_2^\times}{10}\sum_{10}
    \Big[\Kcal(\vq_d,\vq_e)\,F_3(\vq_a,\vq_b,\vq_c)
    + 2\,\Kcal(\vq_{abc},\vq_d)\,G_3(\vq_a,\vq_b,\vq_c)\Big]
    \nonumber\\
    &\quad + {\frac{\gamma_2^\times}{15}\sum_{30}
    \Kcal(\vq_{cd},\vq_e)\,F_2(\vq_a,\vq_b)\,G_2(\vq_c,\vq_d)
    + \frac{\gamma_2^\times}{15}\sum_{15}
    \Kcal(\vq_{ab},\vq_{cd})\,G_2(\vq_a,\vq_b)\,G_2(\vq_c,\vq_d)}
    \nonumber\\
    &\quad + {\frac{3\gamma_3}{10}}\sum_{10}
    \Lcal(\vq_{abc},\vq_d,\vq_e)\,G_3(\vq_a,\vq_b,\vq_c)
    + {\frac{\gamma_3}{5}}\sum_{15}
    \Lcal(\vq_{ab},\vq_{cd},\vq_e)\,G_2(\vq_a,\vq_b)\,G_2(\vq_c,\vq_d) +
  {\frac{b_4}{60}}\sum_{10}F_2(\vq_i,\vq_j)
    \nonumber\\[3pt]
    &\quad  
    + {\frac{\gamma_2^{\times\times}}{30}\sum_{30}
    \Kcal(\vq_k,\vq_l)
    \,F_2(\vq_i,\vq_j)
    + \frac{\gamma_2^{\times\times}}{30}\sum_{30}
    \Kcal(\vq_{ij},\vq_k)\,G_2(\vq_i,\vq_j)}
     \nonumber\\[3pt]
    &\quad  
   + \frac{\gamma_3^\times}{10}\sum_{10}
    \Big[
    {\Lcal(\vq_c,\vq_d,\vq_e)
    \,F_2(\vq_a,\vq_b)}
    + 3\,\Lcal(\vq_{ij},\vq_k,\vq_l)\,G_2(\vq_i,\vq_j)\Big]
    \nonumber\\
    &\quad + {\frac{\gamma_2^{{sq}}}{15}\sum_{30}}
    \Kcal(\vq_k,\vq_l)
    \,\Kcal(\vq_{ij},\vq_m)\,G_2(\vq_i,\vq_j)
 + \frac{\gamma_{21}^\times}{5}\sum_{5}
    \frac{\vk\cdot\vq_i}{q_i^2}\,K_{\gamma_{21}^\times}^{(4)}(\hat\vq_i)
    +
    \frac{\gamma_{211}}{5}\sum_{5}\!
    \frac{\vk\cdot\vq_i}{q_i^2}\;
    K_{\gamma_{211}}^{(4)}(\hat\vq_i)
 \nonumber\\
   &\quad 
     +
    \frac{\gamma_{22}}{5}\sum_{5}\!
    \frac{\vk\cdot\vq_i}{q_i^2}\;
    K_{\gamma_{22}}^{(4)}(\hat\vq_i) +
    \frac{\gamma_{31}}{5}\sum_{5}\!
    \frac{\vk\cdot\vq_i}{q_i^2}\;
    K_{\gamma_{31}}^{(4)}(\hat\vq_i) 
     \nonumber\\
   &\quad 
   +
     \frac{\gamma_{21}}{20}\sum_{10}\!
    \frac{(\vk\cdot\vq_a)(\vk\cdot\vq_b)}{q_a^2\,q_b^2}  K_{\gamma_{21}}^{(3)}(\vq_c,\vq_d,\vq_e)
  - \frac{3\,\gamma_{21}}{140}\sum_{10}
    \frac{\vk\cdot\vq_{ab}}{q_{ab}^2}\,\Kcal(\vq_a,\vq_b)\,
    K_{\gamma_{21}}^{(3)}(\vq_c,\vq_d,\vq_e)\,.
\end{align}

To summarize, fifth order bias expansion
involves 29 unique bias operators: 
one at linear order, 
2 at quadratic order,
4 at cubic order,
8 at fourth order,
14 at fifth order.
We have explicitly verified that 
all operators at a given order are 
linearly independent.
The fluid flow evolution
induces additional operators
with fixed coefficients in front of them,
which are determined by Galilean invariance.

\subsection{2-loop diagrams}

\begin{figure}
  \centering
  \resizebox{\textwidth}{!}{%
  \begin{tikzpicture}
    \node (d) at (0,-4.2)   {\diagPOF};
    \node (e) at (5.4,-4.2) {\diagPTF};
    \node (f) at (10.8,-4.2){\diagPTHa};
    \node (g) at (0,-8.4)   {\diagPTHb};
  \end{tikzpicture}}
  \caption{Diagrammatic representation of the two-loop bias contributions to the galaxy
  power spectrum. }
  \label{fig:spt-diagrams-2L}
\end{figure}
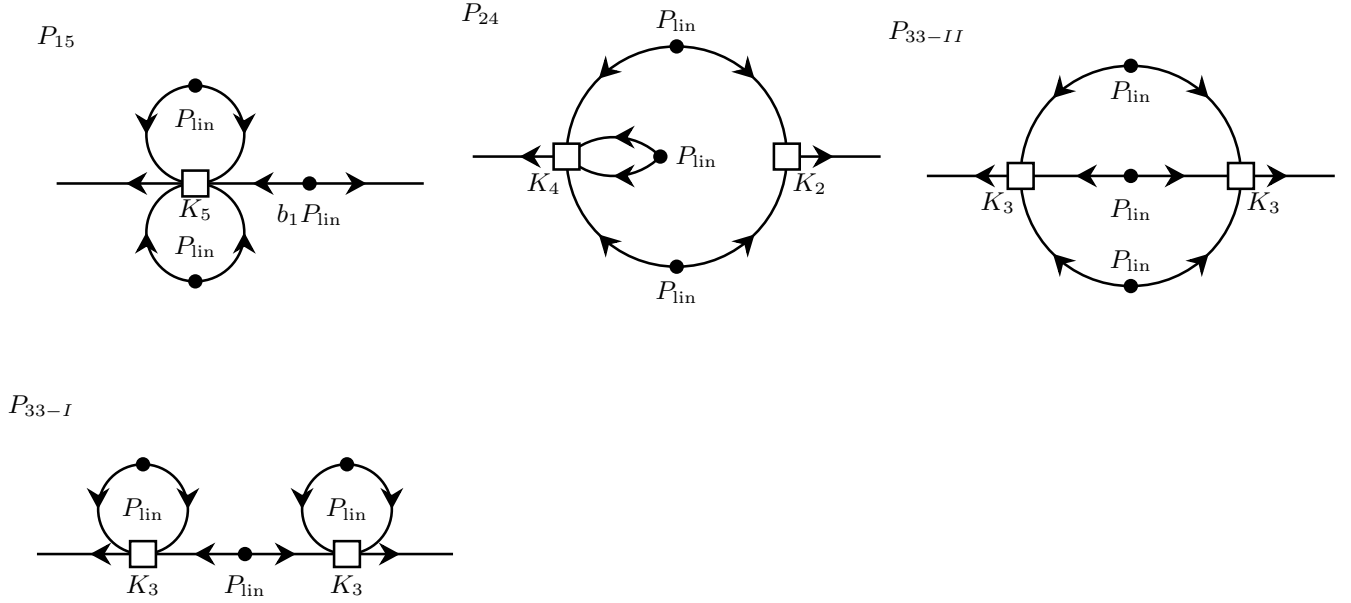

The total (``bare'') two-loop galaxy power spectrum 
from non-linear galaxy bias has the 
following form: 
\begin{equation}\label{eq:P2loop}
    P_{gg}^{\text{2-loop}}(k) =
    P_{15}(k) + P_{24}(k)
    + P_{33,\mathrm{I}}(k) + P_{33,\mathrm{II}}(k)\,,
\end{equation}
where the individual building blocks are given by:
\begin{align}
    P_{24}(k) &= 2\int_{\vp} K_2(\vk{-}\vp,\vp)\;
    \underbrace{
    \Big[12\int_{\vq} K_4(\vk{-}\vp,\vp,\vq,-\vq)\,\PL(q)\Big]
    }_{\mathcal{I}^{(4)}(\vk{-}\vp,\vp)}\;
    \PL(|\vk{-}\vp|)\,\PL(p) \label{eq:P24} \\[6pt]
  P_{33,\mathrm{I}}(k) &= 6\int_{\vp,\vq} K_3(\vk{-}\vp{-}\vq,\vp,\vq)\,K_3(\vk{-}\vp{-}\vq,\vp,\vq)\,\PL(|\vk{-}\vp{-}\vq|)\,\PL(p)\,\PL(q) \label{eq:P33I}\\[6pt]
    P_{33,\mathrm{II}}(k) &= 
    9\,\PL(k)\;[\sigma_{13}(k)]^2
    \,,
    \qquad \sigma_{13}(k) = \int_{\vp} K_3(\vk,\vp,-\vp)\,\PL(p) \label{eq:P33II}\\[6pt]
    P_{15}(k) &= 30\,b_1\,\PL(k)\int_{\vp,\vq} K_5(\vk,\vp,-\vp,\vq,-\vq)\,\PL(p)\,\PL(q) \,.
    \label{eq:P15}
\end{align}
$P_{33,\mathrm{II}}$ factorizes into a product of one-loop integrals
$\sigma_{13}^2$. $P_{24}$ has a factorizable inner loop
$\mathcal{I}^{(4)}$ whose complexity
is similar to that of the $B_{411}$ 
one-loop bispectrum diagram;
but the outer integration over $\vp$ couples
it to $K_2$ and $\PL$. 
Finally, $P_{15}$ and $P_{33,\mathrm{I}}$
are new two-dimensional integrals that cannot be reduced to 
the one-loop power spectrum or bispectrum integrals.
The corresponding Feynman graphs 
are shown in Fig.~\ref{fig:spt-diagrams-2L}.

\subsection{Bias Renormalization and Redundant Operators}

The two-loop power spectrum contains 
divergent contributions, which have to be 
renormalized by the low order bias parameters.
In this Section we will show how this happens
and  explicitly derive new renormalization conditions
relevant at the two-loop order.
In addition, the systematic renormalization
of bias allows us to remove some of the 
contributions which appear trivial, or
redundant at the two-loop order.

The two-loop redundancies and renormalization
conditions can be seen as a natural 
generalization of the one-loop renormalization
conditions. Recall that at one-loop order 
the LE cubic operators with coefficients 
$b_3,\gamma_3$, and $\gamma_2$ (from the SPT evolution) are redundant,
i.e. their $P_{13}$ diagrams produce a pure renormalization
of $b_1$, c.f. Eq.~\eqref{eq:1l_ren_cond}.
These operators do not have any finite contributions.
The situation is somewhat more interesting at the two-loop order. There we also have some purely redundant operators, but in addition to them
we have non-redundant operators that generate 
divergences in the UV. The renormalization of these
contributions is more subtle as it has to preserve
the finite parts. The second important 
point is the presence of two UV domains corresponding
to UV limits $p\gg k$ and $q\gg k$, where $p,q$
are loop momenta and $k$ is the ``on-shell'' external momentum. This gives rise
to two distinct limits: double-hard $p\sim q\gg k$ and 
single-hard, where only one of the loop momenta
is hard, while the other is allowed to take 
arbitrary values. We will develop a renormalization
procedure that removes UV singularities present 
both in single-hard and double-hard limits.

If we want to be consistent 
with the one-loop results coded up in standard software, 
like the \texttt{CLASS-PT} code, we have to use a scheme that 
produces convergent contributions in the limit $\Lambda\to \infty$.
\texttt{CLASS-PT} is based on FFTLog, which implements 
dimensional regularization. In this scheme all divergent 
integrals are set to zero, and all convergent integrals
are evaluated up to $\Lambda = \infty$. If the two-loop
integrals are evaluated up to a finite cutoff $\Lambda_{\rm 2-loop}$, the renormalization
conditions for the low order bias parameters at one and two-loop orders
will be different, in which case one cannot easily relate the one-loop
and two-loop results. One approach is to derive the appropriate 
renormalization group equations for bias parameters~\cite{Bakx:2025cvu},
and evolve the two-loop bias parameters to $\Lambda=\infty$, or equivalently evolve the one-loop bias parameters to $\Lambda_{\rm 2-loop}$.  
Alternatively, one can remove all the divergent contributions 
at the level of the two-loop integrands and then evaluate the remaining 
finite pieces over the entire loop momentum range. This way the bias 
parameters at one-loop and two-loop order will be the same by construction.
We will adopt this approach in our work. Specifically, we shall use
the Bogoliubov-Parasiuk-Hepp-Zimmermann (BPHZ) renormalization scheme~\cite{Bogoliubov:1957gp,Hepp:1966eg} 
and systematically subtract divergences directly in momentum space integrals.
Let us discuss now the renormalization of relevant loop contributions one-by-one.

\textbf{$P_{33,\mathrm{II}}$ renormalization.}
We start by considering the renormalization
of $P_{33,\mathrm{II}}$. This diagram has a very simple topology: 
a product of two one-loop diagrams. As a result, the UV-divergences are trivially factorized: the 
double-hard domain is simply a product of two 
single-hard loop domains from the relevant 
loop momenta. Each individual one-loop diagram 
is renormalized by the one-loop renormalization
condition~\eqref{eq:1l_ren_cond}. Let us demonstrate 
that this condition holds for $P_{33,\mathrm{II}}$ as well.
Focusing on the UV limits of the relevant diagrams we get:
\be 
\begin{split}
P_{33,\mathrm{II}}(k)& = 9\PL(k)\left[\frac{P_{13}}{6b_1\PL(k)}\right]^2=9\PL(k)\left[\frac{1}{3}\left(\frac{b_3}{2} + \frac{34 b_2}{21} - \frac{4\gamma_2^\times}{3}\right)\sigma^2+\sigma_{13}^{\mathrm{ren}}\right]^2\\
&
= 9\,\PL(k)\;\Big[\sigma_{13}^{\mathrm{ren}}
    + \frac{\delta b_1}{3}\sigma^2\,\Big]^2
   = 9\,\PL(k)\;\Big[
    \big(\sigma_{13}^{\mathrm{ren}}\big)^2
    + \frac{2\,\delta b_1}{3}\sigma^2\,
    \;\sigma_{13}^{\mathrm{ren}}
    + \frac{(\delta b_1)^2\sigma^4}{9}\,
    \Big]\,.
    {\label{eq:P33II_expanded}}
\end{split}
\ee 
In the above formula $ O(\delta b_1\sigma^2)$,$ O(\delta b_1^2\sigma^4)$
are the single-hard and double-hard 
contributions, respectively. 
The above loop is renormalized by the $b_1$ counterterm contribution with the one-loop renormalization condition. To see this we add all the relevant pieces together:
\be 
\begin{split}
&P_{33,{\mathrm{II}}}+
P_{13} + P_{11}
 \\
  &  = 9\PL(k)(\sigma_{13}^{\mathrm{ren}}\big)^2
    + 6\delta b_1\sigma^2\sigma_{13}^{\mathrm{ren}}(k)\PL(k)
    + (\delta b_1\sigma^2)^2 \PL(k)+  6 b_1
    \left(\sigma_{13}^{\mathrm{ren}}(k)+\frac{\delta b_1}{3}\sigma^2\right)\PL(k)
+b_1^2 \PL(k)\\
  & 
= 
(b_1+\delta b_1 \sigma^2)^2 \PL(k)
+{6}(b_1+\delta b_1 \sigma^2) \PL(k)\sigma^{\rm ren}_{13}+9\PL(k)\big(\sigma_{13}^{\mathrm{ren}}\big)^2\\
&=(b_1^{[R]})^2 \PL(k)+{6b_1^{[R]}\PL(k)\sigma_{13}^{\mathrm{ren}}(k)}+9\PL(k)\big(\sigma_{13}^{\mathrm{ren}}\big)^2\,,
\end{split}
\ee 
where we emphasize that the above equation 
is exact, i.e. $b_1$
absorbs all the divergences through $O(\sigma^4)$
from the $P_{33,\mathrm{II}}$ diagram. Hence, 
the $b_1$ renormalization to order $O(\sigma^2)$
is systematic in the Bogoliubov sense:
a complex 1-particle-reducible diagram $P_{33,{\mathrm{II}}}$
with nested divergences is renormalized
by the same counterterm as the corresponding
1-particle-irreducible sub-diagrams.

Let us note that the other loop integral
$P_{33,\mathrm{I}}$ does not produce any non-trivial 
renormalization conditions for bias operators.
In contrast to $P_{33,\mathrm{II}}$, $P_{33,\mathrm{I}}$ 
produces a Taylor expansion in $k^2$ in the UV limit.
These UV-sensitive contributions 
are
renormalized by the stochastic counterterms.

\textbf{$P_{24}$ renormalization.}
To study the renormalization 
of the $P_{24}$ term it is convenient to rewrite it as
\be     
P_{24}(k) = 2\int_{\vp} K_2(\vk{-}\vp,\vp)\;
    {\mathcal{I}^{(4)}(\vk{-}\vp,\vp)}\;
    \PL(|\vk{-}\vp|)\,\PL(p) \,,
    \ee 
and consider the loop integral of two momentum variables
\be 
   \mathcal{I}^{(4)}(\vk_1,\vk_2) = 12
   \int_{\vq}
    K_4(\vk_1,\vk_2,\vq,-\vq)\;\PL(q)\,.
\ee
This integral is identical to the 
$B_{411}$ one-loop bispectrum contribution,
which is renormalized by $b_2$, $\gamma_2$,
and the one-loop renormalization condition for $b_1$~\cite{Eggemeier:2018qae,Philcox:2022frc}.
The relevant counterterm can be written as
\be 
P_{24}^{\rm ctr.}=4\int_{\vp} \delta K_2(\vk-\vp,\vp)
 K_2(-\vk+\vp,-\vp)\PL(p)\PL(|\vk-\vp|)\,,
\ee 
from which we establish $2\delta K_2(\vk_1,\vk_2)=\mathcal{I}^{(4)} \Big|_{\rm hard}$.
One can find that the hard limit projects onto the 2nd-order basis:
\begin{equation}\label{eq:hard24}
    6\langle K_4 \PL(q)\rangle_{\hat\vq}\bigg|_{\text{hard}}
   =\delta K_2(\k_1,\k_2) = \left[\delta b_1\,F_2(\vk_1,\vk_2) + \frac{\delta b_2}{2} + \delta \gamma_2\,\Kcal(\vk_1,\vk_2)\right]\sigma^2\,,
\end{equation}
with 
\begin{subequations}
\begin{align}
\delta b_1
    &= \frac{b_3}{2} + \frac{34 b_2}{21} - \frac{4\gamma_2^\times}{3}\,,
    \label{eq:db1_P24}\\[6pt]
    \delta b_2
    &= \frac{b_4}{2} - \frac{8\gamma_2^{\times\times}}{3}
    + \frac{32\gamma_{21}^\times}{15} + \frac{32\gamma_2^{sq}}{15}
    + \frac{68 b_3}{21} + \frac{8126 b_2}{2205}
    - \frac{208\gamma_2^\times}{35}
    \label{eq:db2_P24}\,,\\[6pt]
\delta\gamma_2    
    &= \frac{\gamma_2^{\times\times}}{2} - \frac{2\gamma_{21}^\times}{5}
    - \gamma_3^\times + \frac{4\gamma_2^{sq}}{15}
    + \frac{127 b_2}{2205} + \frac{92\gamma_2^\times}{105}\,,
    \label{eq:dg2_P24}
\end{align}
\end{subequations}
which reproduces the known 1-loop bispectrum renormalization condition.
The fact that a single renormalization condition
for $\delta b_1$ cancels divergences
in $P_{13}$, $P_{33,{\mathrm{II}}}$, and $P_{24}$ is a non-trivial 
consistency test of the bias renormalization program.
The remaining sub-leading divergences in $P_{24}$ due to 
the hard limit of $p$ are much softer
than the leading ones.
The corresponding integrals actually converge 
for the $\Lambda$CDM power spectrum, 
and their UV sensitivity is renormalized
by stochastic counterterms.

Importantly, 
all four LE fourth order operators 
are fully redundant, which is consistent
with the one-loop bispectrum result. 
This leaves us with only 4 non-trivially contributing
fourth order bias operators, all of which are NLE operators.

\textbf{$P_{15}$ renormalization.}
The analysis of $P_{15}$ renormalization
is complicated by the presence of nested
divergences.
To deal with them, we use the BPHZ forest formula
for the renormalized integral $\sigma^{\rm ren}_{15}(k)$:
\be
\sigma^{\rm ren}_{15}(k)\equiv 30\int_{\q,\p}[K_g^{(5)}-K_g^{(5)}\Big|_{\rm p-hard}-K_g^{(5)}\Big|_{\rm q-hard}
+K_g^{(5)}\Big|_{\rm double-hard}]\PL(q)\PL(p)\,.
\ee 
The above 
subtraction of the double-hard limit is needed in 
order to avoid double-counting as each single-hard 
limit above contains the domain where both
loop 
momenta are hard. 
By $\q\leftrightarrow\p$ symmetry, the two single-hard limits are equal, and hence it is sufficient to obtain
renormalization conditions in one single-hard domain only. The single-hard limit of
$P_{15}$ produces non-trivial renormalization conditions 
for linear, quadratic, and 
cubic bias operators at $O(\sigma^2)$, while the double-hard limit renormalizes
$b_1$ to order $\sigma^4$. 

Let us focus on the renormalization
of single-hard limits first. 
The relevant integral is equivalent to 
a one-loop correction to the connected four-point function stemming from the $K_5$ kernel,
called $T_{5111}$ in the SPT nomenclature.
By definition of renormalized bias 
$\delta K_3+K_3$ is finite.
Then consistency 
requires that 
$P_{13}^{ K_3+\delta K_3}+P_{15}$
is finite in the single-hard limit, which
implies
\be 
\begin{split}
3\int_{\vq}\delta K_3(\k,-\vq,\vq)\PL(q) & = 15\int_{\p\vq}K_5(\k,-\vq,\vq,\vp,-\vp)\PL(q)\PL(p)\Big|_{\rm p,q,~single-hard}\\
&=30\int_{\p\vq}K_5(\k,-\vq,\vq,\vp,-\vp)\PL(q)\PL(p)\Big|_{\rm q,~single-hard}\,,
\end{split}
\ee 
where we took into account the $\q\leftrightarrow\p$ symmetry.
The above renormalization condition can be 
derived more formally using the propagator framework
of~\cite{Eggemeier:2018qae}.
The hard limit of the relevant one-loop integral
\be 
\label{eq:P15renorm}
\delta K_3(\vk_1,\vk_2,\vk_3) = 10 \int_{\q}
    K_5(\vk_1,\vk_2,\vk_3,\vq,-\vq) \PL(q)\Big|_{\rm q,~hard}
\ee 
can be projected onto the one-loop operator basis,
\begin{equation}\label{eq:dGamma3_basis}
    \delta K_g^{(3)} = \left[\delta b_1\,F_3 + \frac{\delta b_2}{3}\sum_3 F_2
    + \delta\gamma_2\,K_{\mathcal{G}_2}^{(3)} + \frac{\delta b_3}{6}
    + \frac{\delta\gamma_2^\times}{3}\sum_3\Kcal + \delta\gamma_3\,\Lcal
    + \delta\gamma_{21}\,K_{\gamma_{21}}\right]\sigma^2\,.
\end{equation}
A straightforward computation of eq.~\eqref{eq:P15renorm} recovers
Eqs.~(\ref{eq:db1_P24},\ref{eq:db2_P24},\ref{eq:dg2_P24})
for linear and quadratic bias coefficients and also produces 
new renormalization conditions on the cubic bias parameters:
\begin{align}
    \delta b_3
    &= \frac{b_5}{2} - 4\gamma_2^{\times\times\times}
    + \frac{32\gamma_2^{sq\times}}{5} + \frac{32\gamma_{21}^{\times\times}}{5}
    {{- \frac{144\gamma_{31}^\times}{245}}} {- \frac{384\gamma_{21,2}}{35}}
  + \frac{34 b_4}{7} - \frac{624\gamma_2^{\times\times}}{35}
    + \frac{2304\gamma_2^{sq}}{245} - \frac{32\gamma_{21}^\times}{5}
    \nonumber\\
    &\quad + \frac{8126 b_3}{735}
    - \frac{16504\gamma_2^\times}{1715} + \frac{286004 b_2}{56595} \,,  \label{eq:db3_P15}\\[8pt]
    \delta\gamma_2^\times
    &= \frac{\gamma_2^{\times\times\times}}{2}
    - \gamma_3^{\times\times} - \frac{16\gamma_2^{sq\times}}{15}
    - \frac{2\gamma_{21}^{\times\times}}{5}
    + \frac{8b_{\Gal_2{\cdot}\Gal_3}}{5}
    + \frac{32\gamma_{22}^\times}{15}
    {{- \frac{8\gamma_{31}^\times}{735}}}
    + \frac{8\gamma_{211}^\times}{15}
    {+ \frac{16\gamma_{21,2}}{105}}
    \nonumber\\
    &\quad + \frac{262\gamma_2^{\times\times}}{105}
    - \frac{78\gamma_3^\times}{35}
    - \frac{152\gamma_2^{sq}}{735}
    - \frac{64\gamma_{21}^\times}{105} + \frac{127 b_3}{2205}
    + \frac{12877\gamma_2^\times}{5145} + \frac{7229 b_2}{33957}\,,
    \label{eq:dgx2_P15}\\[8pt]
    \delta\gamma_3
    &= \frac{\gamma_3^{\times\times}}{2}
    + \frac{4b_{\Gal_2{\cdot}\Gal_3}}{5}
    + \frac{4\gamma_{211}^\times}{15}
    {{+ \frac{131\gamma_{31}^\times}{6615}}}
    {+ \frac{16\gamma_{21,2}}{105}}
  + \frac{66\gamma_3^\times}{35}
    + \frac{296\gamma_2^{sq}}{735}
    + \frac{2\gamma_{21}^\times}{5}
  + \frac{3433\gamma_2^\times}{9261} - \frac{421 b_2}{101871} \,,
    \label{eq:dg3_P15}\\[8pt]
    \delta\gamma_{21}
    &= \frac{1}{2}\gamma_{21}^{\times\times}
    + \frac{8\gamma_{22}^\times}{15}
    {{- \frac{31\gamma_{31}^\times}{630}}}
    - \frac{2\gamma_{211}^\times}{3}
    + \frac{4\gamma_{21,2}}{15}
    - \frac{12\gamma_{21}^\times}{35}
    + \frac{142\gamma_2^\times}{735} + \frac{1262 b_2}{33957}\,.
    \label{eq:dg21_P15}
\end{align}
The conditions above are written for the full kernels,
including the transverse (curl) parts.  
The latter affect only the renormalization conditions from the $\gamma_{31}^\times$
operator.\footnote{When
matching an implementation that drops the curl of $\vPsi^{(3)}$, the four
$\gamma_{31}^\times$ coefficients above instead read 
\be\label{eq:g31x_noT}
\delta b_3\big|_{\gamma_{31}^\times} = -\tfrac{32}{49}\,,\qquad
\delta \gamma_2^\times\big|_{\gamma_{31}^\times} = -\tfrac{26}{735}\,,\qquad
\delta \gamma_3\big|_{\gamma_{31}^\times} = +\tfrac{158}{6615}\,,\qquad
\delta \gamma_{21}\big|_{\gamma_{31}^\times} = -\tfrac{4}{63}\,.
\ee}
The recovery of the renormalization conditions
for 
$\delta b_1,\delta b_2$ and $\gamma_2$
is a non-trivial test of the 5th order operators. 
The final double-hard contribution reads
\begin{equation}\label{eq:db1_s4}
    \delta b_1\big|_{\sigma^4} = -\frac{3}{2}
    \int_{\vp: p\gg k}
    \delta K_3(\k,-\vp,\vp)\PL(p)
    = \left(-\frac{\delta b_3}{4} + \frac{2\delta\gamma_2^\times}{3}
    - \frac{17\delta b_2}{21}\right)\sigma^4\,,
\end{equation}
where the 
minus sign reflects the BPHZ subtraction,
whilst $\delta b_3,\delta b_2$ and $\delta \gamma_2^\times$ are 
the divergent contributions given above. 
The factor $\tfrac12$ inside the overall
$-\tfrac32 = 3\times(-\tfrac12)$ removes the double-counting introduced
by the fact that our $\delta b_3$, $\delta b_2$, and
$\delta\gamma_2^\times$ were obtained by summing two single-hard limits.\footnote{The factor of $3$ originates from the Wick pairing
factor of $6$ in the $P_{13}$-like tadpole integral divided by two from the leading order expansion of $(b_1^{(R)})^2=b^2_1+2b_1\delta b_1+...$.}
Note that the above renormalization conditions
are equivalent to
the ones from Eq.~(4.11) of \cite{Bakx:2025cvu},
which were derived using a different basis 
of EFT operators. Also note that our 
treatment of the single/double-hard limits
and the universality checks parallel those of~\cite{Bakx:2025cvu}.

Note that the five fifth order LE operators,
four evolved 4th order LE operators
and three NLE fifth order product operators $\gamma_{21}^{\times\times}$, 
$\gamma_{21,2}$, $\gamma_{211}^{\times}$
are completely redundant, i.e. they 
do not produce any finite $P_{15}$ contributions.
This leaves us with only 17 free bias parameters
in the two-loop galaxy power spectrum. The situation may be different 
if we consider the trispectrum or go beyond the two-loop power
spectrum order, where all the fifth order operators are expected to be
non-redundant in general.

Finally, let us mention that the sub-leading UV limits 
of the $P_{15}$ integral are proportional to $k^2\PL$ and $k^4\PL$. 
The corresponding terms
are renormalized by the higher-derivative operators, which will be discussed
in detail shortly. 

\subsection{IR-resummation}

The perturbative in $\delta_L$ expansion that we have used 
so far does not converge well for the baryon acoustic 
oscillations (BAO) in the 
linear matter power spectrum. The convergence can be significantly
improved by resumming the displacement contributions enhanced
in the IR domains. This procedure of improving the modeling of the 
non-linear evolution of the BAO is known as IR resummation~\cite{Crocce:2005xy,Crocce:2007dt,Senatore:2014via,Baldauf:2015xfa,Vlah:2015zda,Blas:2015qsi}.
In this work we account for it using the formalism of time-sliced
perturbation theory (TSPT)~\cite{Blas:2015qsi,Blas:2016sfa,Ivanov:2018gjr,Vasudevan:2019ewf}. Specifically, we apply the ``wiggly-smooth''
decomposition to the linear power spectrum $\PL=P_s+P_w$ and resum the IR-enhanced
contributions in the perturbative series.\footnote{
IR-resummation within TSPT, in principle, can be carried out even without
explicitly assuming a ``wiggly-smooth'' decomposition~\cite{Ivanov:2024xgb}.} Previous works~\cite{Blas:2016sfa,Ivanov:2018gjr} have carried out this program in very general settings that apply to the two-loop galaxy power spectrum as well. 
The formulas derived in~\cite{Blas:2016sfa,Ivanov:2018gjr} are applicable to the two-loop galaxy power 
spectrum provided that the Eulerian bias expansion $\delta_g[\delta]$
does not feature any IR-enhancements, which is the case by construction 
because these enhancements would violate the equivalence principle. 
The only caveat here is that the expansion we are using is hybrid
Eulerian-Lagrangian, so it includes Lagrangian operators as well. 
However, the Lagrangian bias 
is manifestly Galilean-invariant since it is formulated in terms of Lagrangian 
coordinates, and hence it does not violate the assumptions of the general 
derivations of~\cite{Blas:2016sfa,Ivanov:2018gjr}. Alternatively, one can explicitly invert the SPT expansion $\delta=\delta_L+\int F_2 \delta_L^2+...$
in perturbation theory
as $\delta_L=\delta+\int F'_2\delta^2$ and plug them into our Lagrangian
bias operators and confirm that their Eulerian analogs are manifestly 
IR safe. After that, it is straightforward to
repeat the computations of~\cite{Blas:2016sfa,Ivanov:2018gjr} and arrive
at the following answer for the IR-resummed total galaxy power spectrum at 
the two-loop order:
\be 
\label{eq:irres}
\begin{split}
& P^{\mathrm{2-loop,~tot}}_{gg}\Big|_{\rm IR-res}=b_1^2
\left[P_{s}+e^{-\mathcal{S}}P_{w}\left(1+\mathcal{S}+\frac{1}{2}\mathcal{S}^2\right)\right]
+\Delta P^{\mathrm{1-loop}}_{gg}[P_{s}+e^{-\mathcal{S}}P_{w}(1+\mathcal{S})]
+\Delta P^{\mathrm{2-loop}}_{gg}[P_{s}+e^{-\mathcal{S}}P_{w}]\,,
\end{split}
\ee 
where 
\be 
\mathcal{S}\equiv \frac{k^2}{6\pi^2}\int_0^{\Lambda_{\rm IR}} dq~\PL(q)\left[1-j_0(qr_s)+2j_2(qr_s)\right]\,,
\ee 
and $r_s$ is the comoving sound horizon at decoupling 
and $j_\ell(x)$
is the spherical Bessel function of order $\ell$,
and the IR cutoff $\Lambda_{\rm IR}=0.2~\hMpc$.

\subsection{Numerical implementation}

\begin{figure}[t!]
\centering
\begin{tabular}{cc}
\includegraphics[width=0.49\textwidth]{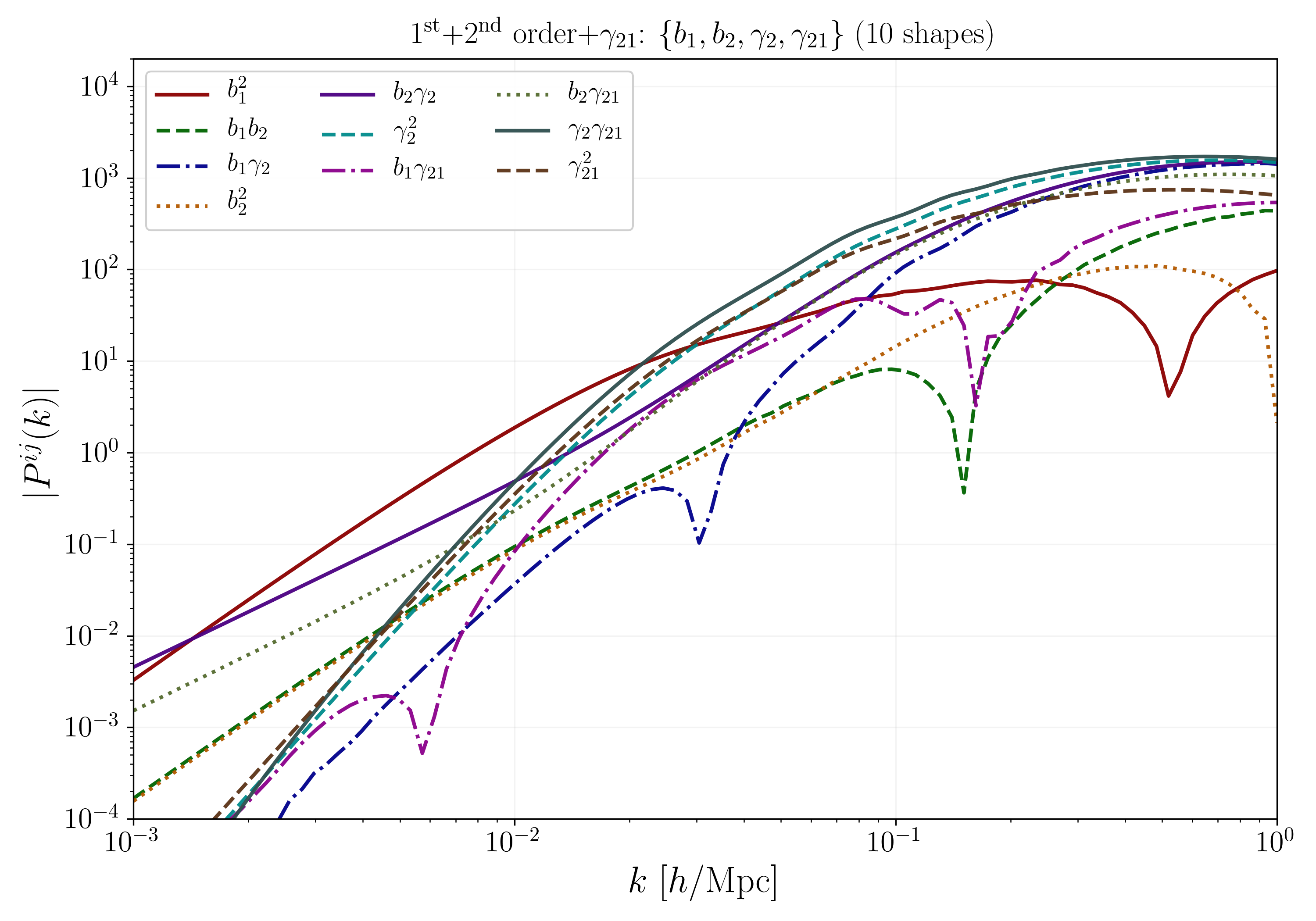} &
\includegraphics[width=0.49\textwidth]{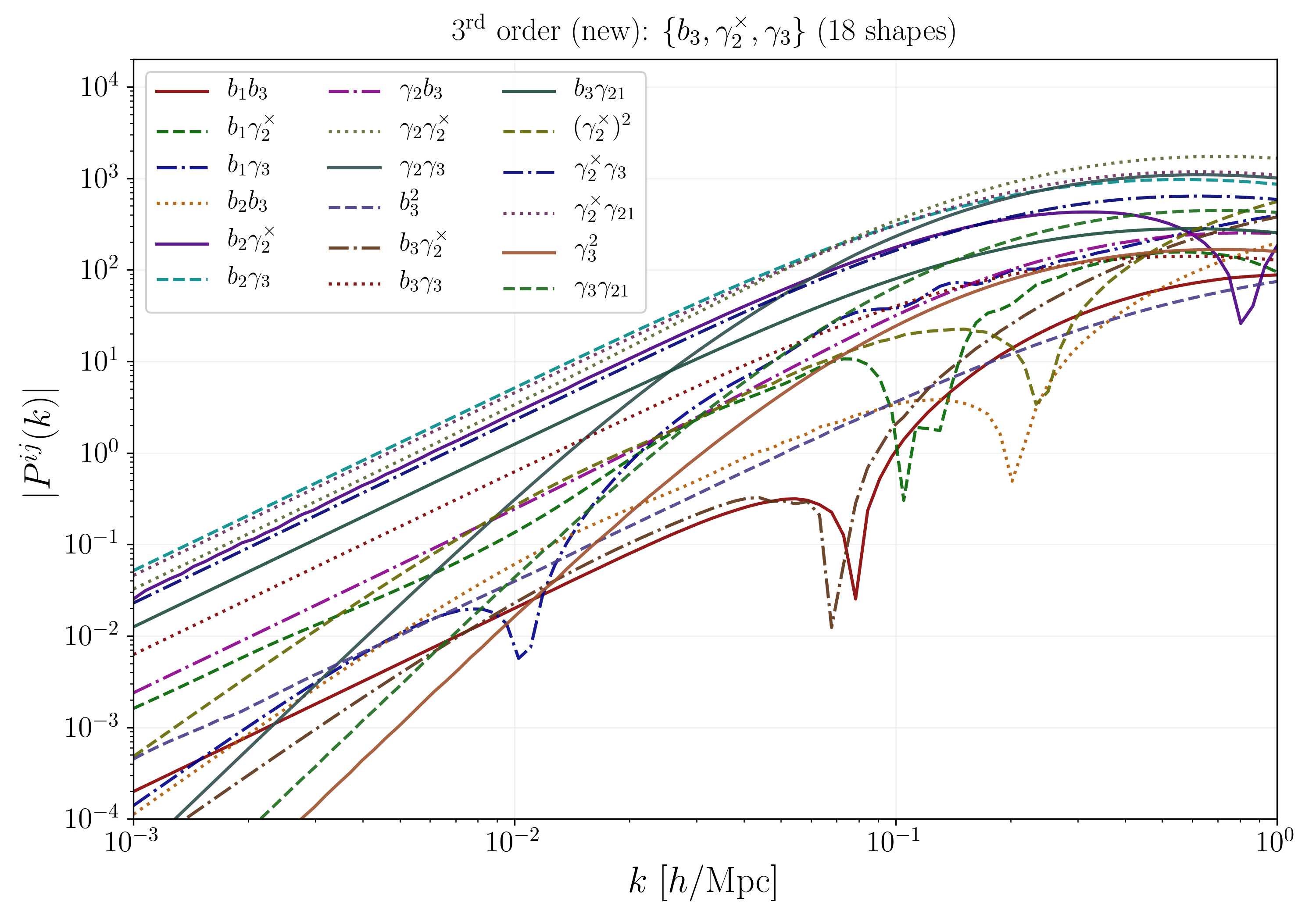} \\
\includegraphics[width=0.49\textwidth]{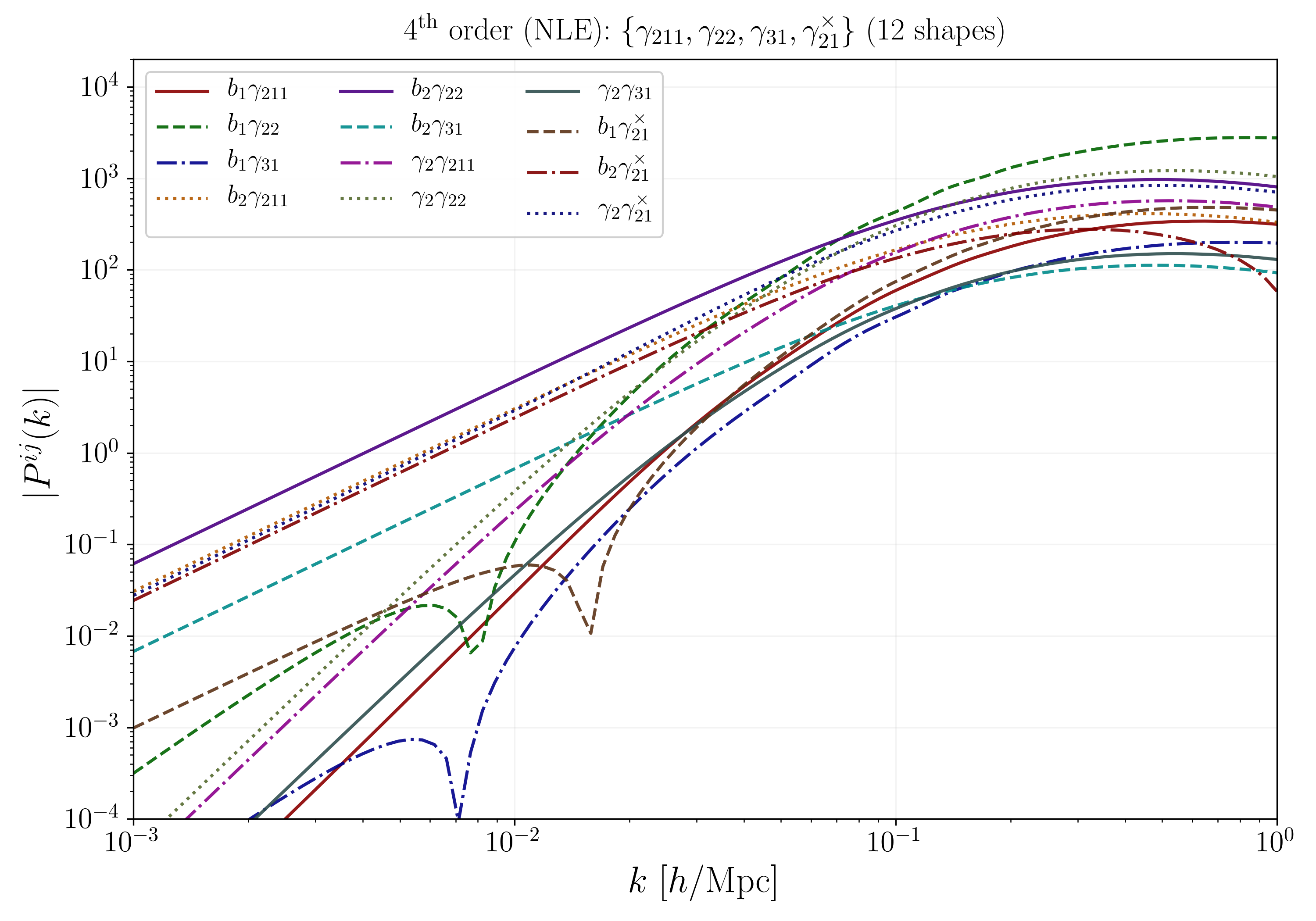} &
\includegraphics[width=0.49\textwidth]{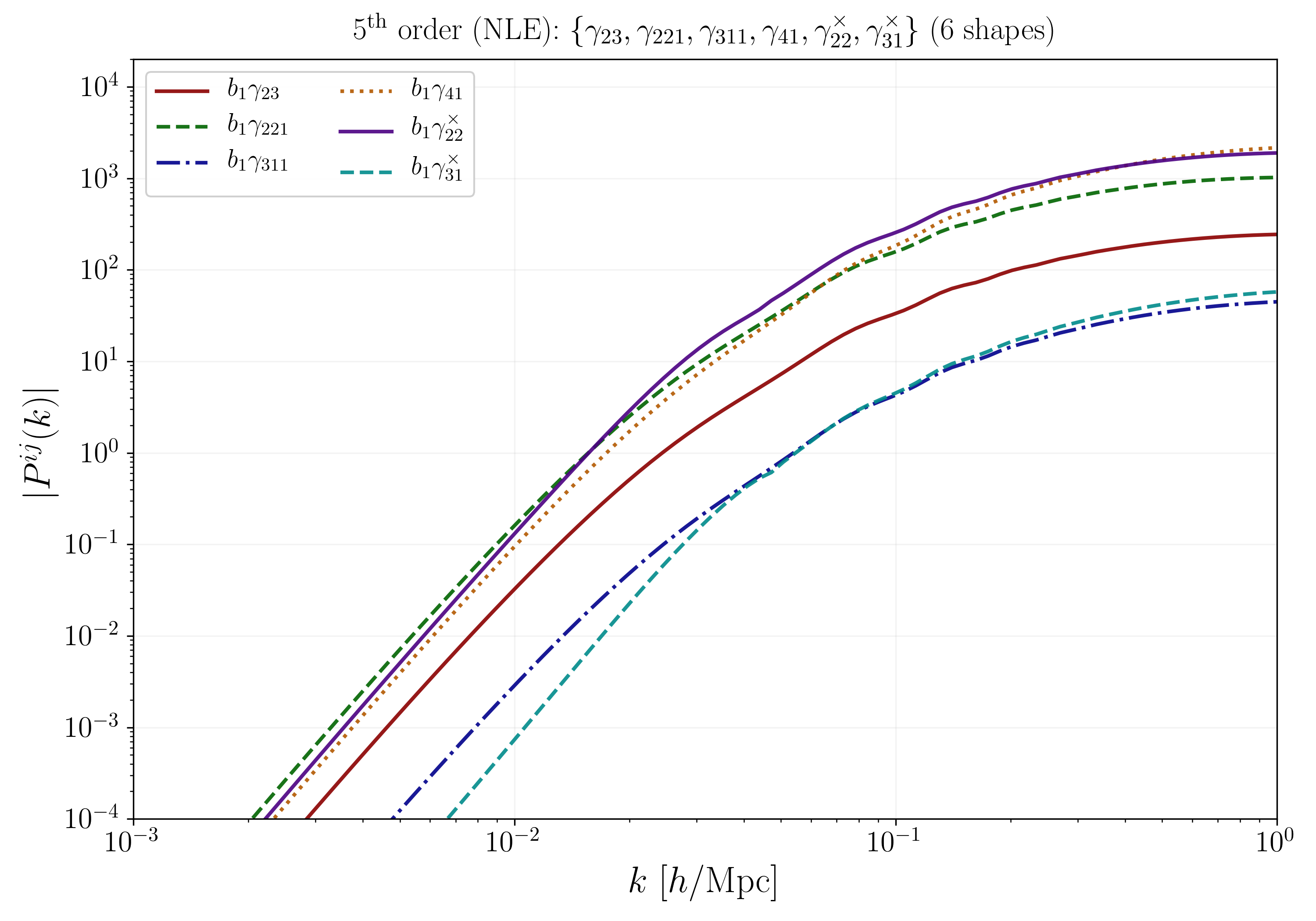}
\end{tabular}
\caption{Shape functions grouped by operator order. \emph{Top left:}
1st+2nd order pairs involving $\{b_1, b_2, \gamma_2, \gamma_{21}\}$
(10 shapes). \emph{Top right:} 3rd order cubic operators $\{b_3,
\gamma_2^\times, \gamma_3\}$ (18 shapes). \emph{Bottom left:} 4th order
NLE operators $\{\gamma_{211}, \gamma_{22}, \gamma_{31},
\gamma_{21}^\times\}$ (12 shapes). \emph{Bottom right:} 5th order NLE
operators $\{\gamma_{23}, \gamma_{221}, \gamma_{311}, \gamma_{41},
\gamma_{22}^\times, \gamma_{31}^\times\}$ (6 shapes).}
\label{fig:shapes_by_order}
\end{figure}

The two-loop galaxy power spectrum decomposes as 
\be 
\Delta P_{gg}^{\text{2-loop}}(k) =\sum_{i\leq j} b_i\, b_j\; (P^{ij}_{15}+P^{ij}_{24}+P^{ij}_{33,\mathrm{I}}+P^{ij}_{33,\mathrm{II}}) =\sum_{i\leq j} b_i\, b_j\; P^{ij}(k)\,,
\ee 
where the sum is over 46 non-trivial two-loop 
shapes $P^{ij}$ that depend on 17 bias parameters.

We discuss now the implementation of numerical 
computations of the relevant two-loop integrals.
At this order we have first encountered 
the bias operators which are non-redundant, 
but produce genuine UV-divergences.
For $P_{24}$ this is the $\gamma_{21}^\times$ operator, 
whose renormalization was previously discussed in
the context of the one-loop bispectrum~\cite{Eggemeier:2018qae}. 
The finite contribution is obtained by a direct subtraction
of the divergent UV limit from a single loop domain. 
The situation is more complicated for $P_{15}$, where
we implement the full BPHZ subtraction procedure
for $\gamma_{22}^\times$, and
SPT-evolved $\gamma_{2}^\times$, 
and $\gamma_{31}^\times$ terms.
 
Upon implementing the appropriate subtractions,
all the remaining integrals are convergent.
However, the $P_{33,\mathrm{I}}$ and $P_{24}$ contributions
have large UV contributions which are constant
in the $k\to 0$ limit. This UV sensitivity
is absorbed by the constant shot noise counterterm.
For this reason in the standard renormalization schemes e.g. in \texttt{CLASS-PT}~\cite{Chudaykin:2020aoj}, these constant 
terms are explicitly subtracted at the integrand level,
which we also do here by subtracting the $k=0$ limit of 
all relevant $P_{24}$ and $P_{33}$ integrands. With this procedure, 
the finite contributions start as $k^2$ or $k^4$ in the $k\to 0$ limit.

Finally, the finite $P_{15}$ contributions are dominated by the $k^2\PL$
UV-limits, which are degenerate with the one-loop higher-derivative bias 
and the UV-part of the one-loop matter-matter power spectrum.
To isolate genuinely new shapes we have subtracted the low-$k$ 
asymptotics $\propto k^2$ at the integrand level
in $\gamma_{23}$, $\gamma_{221}$, $\gamma_{311}$,
$\gamma_{41}$, and for the rest of relevant $P_{15}$ terms in post-processing.
This guarantees the $k\to 0$ asymptotic of all $P_{15}$ terms starts at 
$k^4\PL(k)$. For the matter-matter ($b_1^2$) contribution we implement IR-safe integrands
following~\cite{Carrasco:2013sva}. For the remaining terms the 
spurious IR enhancements are much milder, 
so their computation with high precision is enough
to resolve the physical remainders left after the cancellation
of the IR-enhanced terms.

We implement numerical integration using the Cuba library in Julia, employing
the VEGAS adaptive Monte Carlo algorithm for all two-loop integrals. 
All operator kernels are evaluated using exact
symmetrized SPT recursion relations. The loop momenta are parameterized following~\cite{Blas:2013aba}.
Each
loop momentum in the five-dimensional integrals is sampled logarithmically over the range $q \in [10^{-4}, 100]\,h/\mathrm{Mpc}$.
The computation is parallelized across $k$-points using Julia's distributed
framework. We target a relative tolerance of $10^{-3}$ with up to $5 \times
10^7$ integrand evaluations per $k$-point, with increased sampling at low $k$
where constant subtraction leads to large cancellations.
The final result comprises 46 shape functions $P^{ij}(k)$ associated
with 17 bias parameters, evaluated on a grid of 96 wavenumbers spanning $k \in
[10^{-3}, 1]\,h/\mathrm{Mpc}$. 

Our numerical results are shown in Fig.~\ref{fig:shapes_by_order}.
As one can see, the two-loop
shapes are fairly degenerate on large scales, 
but in general have distinctive scale-dependence 
around $k\sim 0.5~h/\mathrm{Mpc}$
where 
they start to contribute significantly 
to the total power spectrum.

\section{Counterterms and Stochasticity}
\label{sec:HD}

In this Section we work out the new higher-derivative and stochastic contributions.

\subsection{Higher-derivative bias}\label{sec:hdbias}

We start with the higher-derivative
operators. The two-loop order power counting 
suggests that we need to 
consider terms of the order $O(k^4\delta)$,
$O(k^2\delta^2)$ and $O(k^2\delta^3)$.
Let us consider them one-by-one. 
In what follows, 
we use the one-loop integral basis of Ref.~\cite{Chudaykin:2020aoj}.
These include the following $P_{22}$-type integrals:
\begin{subequations}
\label{eq:P22stand}
\begin{align}
\Ids &\equiv 2\!\int_{\vq} F_2(\vq,\k\!-\!\vq)\,
  \PL(|\k\!-\!\vq|)\,\PL(q)\,,
\label{eq:Ids}\\[3pt]
\IG &\equiv 2\!\int_{\vq} \Kcal(\vq,\k\!-\!\vq)\,
  F_2(\vq,\k\!-\!\vq)\,\PL(|\k\!-\!\vq|)\,\PL(q)\,,
\label{eq:IG}\\[3pt]
\Idsds &\equiv 2\!\int_{\vq} \PL(|\k\!-\!\vq|)\,\PL(q)
  \,,
\label{eq:Idsds}\\[3pt]
\IdsG &\equiv 2\!\int_{\vq}\Kcal(\vq,\k\!-\!\vq)\,
  \PL(|\k\!-\!\vq|)\,\PL(q)\,,
\label{eq:IdsG}\\[3pt]
\IGG &\equiv 2\!\int_{\vq}\Kcal^2(\vq,\k\!-\!\vq)\,
  \PL(|\k\!-\!\vq|)\,\PL(q)\,,
\label{eq:IGG}
\end{align}
\end{subequations}
and a single $P_{13}$-type NLE integral:
\be\label{eq:FG}
\FG \equiv 4\,\PL(k)\!\int_{\vq}\Kcal(\vq,\k\!-\!\vq)\,
F_2(\k,-\vq)\,\PL(q)\,.
\ee
First, we have a single new operator at $O(k^4\delta)$ in addition to the usual higher-derivative bias:
\be 
\delta_g\big|_{\nabla^2\delta}+\delta_g\big|_{\nabla^4\delta}=\beta_{1} \nabla^2\delta+ \tilde{\beta}_{1} \nabla^4 \delta\,,
\ee 
where $\beta_1\equiv b_1c_s+b_{\nabla^2\delta}$ ($c_s$ is the dark matter sound speed and $b_{\nabla^2\delta}$ is the pure higher-derivative bias parameter).
At two loop order, the above leading 
higher derivative bias term  
$\delta_g\big|_{\nabla^2\delta}$
has to be expanded to second and third orders 
in SPT, which gives rise to the following contribution:
\be 
P_{gg}\Big|_{\nabla^n\delta-\mathrm{like}}=-2b_1\beta_{1}k^2\PL(k)-2\,b_1\,\beta_{1}\,k^2\,P_{\rm 1\text{-}loop}^{\rm SPT}(k)
+2b_1\tilde{\beta}_{1}k^4\PL(k)-\beta_1\,k^2\Big[b_2\,\Ids + 2\,b_{\Gal_2}\,\IG\Big]\,,
\ee 
where $b_{\Gal_2}\equiv\gamma_2$.
The quadratic higher-derivative operators 
$O(k^2\delta^2)$ have already been
studied in detail in the context of the one-loop 
galaxy bispectrum~\cite{Eggemeier:2018qae,Philcox:2022frc,Bakx:2025pop}. 
The four non-redundant 
operators at this order are:
\begin{equation}\label{eq:k2d2ops}
  \delta_g\big|_{\nabla^2\delta^2}=\beta_{2,1}\,\nabla^2\delta^2 +
  \beta_{2,2}\,\nabla_i\delta\nabla^i\delta +
  \beta_{2,3}\,\nabla^2\Gal_2(\Phi_v) +
  \beta_{2,4}\,\Gal_2(\nabla_i\Phi_v,\nabla_i\Phi_v)\,.
\end{equation}
In Fourier space this produces the kernel:
\begin{equation}\label{eq:k2d2_fourier}
K_{\rm h.d.}^{(2)}=
-\beta_{2,1}\,k^2
-\beta_{2,2}\,(\vq_1\!\cdot\!\vq_2)
-\beta_{2,3}\,k^2\Kcal(\vq_1,\vq_2)
-\beta_{2,4}\,(\vq_1\!\cdot\!\vq_2)\,\Kcal(\vq_1,\vq_2)\,.
\end{equation}
These corrections generate P22-like integrals when contracted with the bias 2nd order
kernel,
\be 
P_{gg}\Big|_{k^2P_{22}-\mathrm{like}}=4\int_{\vq}
K_g^{(2)}(\k-\vq,\vq)
K_{\rm h.d.}^{(2)}(-\k+\vq,-\vq)\PL(q)\PL(|\k-\vq|)~\,.
\ee 
$P_{gg}\Big|_{k^2P_{22}-\mathrm{like}}$ 
consists of 12 terms, but many of the loop
integrals can be reduced to the standard bias
$P_{22}$ integrals~\eqref{eq:P22stand}. 
To see this, it is convenient 
to switch to a new basis of $O(k^2\delta^2)$ counterterms~\cite{Chen:2026usz}:
\begin{align}\label{eq:Endef}
E_1&=k^2\,,&
E_2&=k^2\,\Kcal(\vq_1,\vq_2)\,,\nonumber\\[3pt]
E_3&=\tfrac{(\vq_1\cdot\vq_2)}{2}
\Big(\tfrac{\k\cdot\vq_2}{q_2^2}+\tfrac{\k\cdot\vq_1}{q_1^2}\Big),&
E_4&= \vq_1\cdot\vq_2 \,.
\end{align}
The 
counterterms in the $\{E_1,E_2,E_3,E_4\}$ basis read
\be\label{eq:dg2HD}
\delta_g^{(2)}\big|_{\rm HD} =
  -c_1\,E_1 - c_2\,E_2 + c_4\,E_3 - c_3\,E_4\,,
\ee
with
\be\label{eq:ci_def}
c_1 = \beta_{2,1}+\tfrac{1}{2}\beta_{2,4}\,,\quad
c_2 = \beta_{2,3}+\tfrac{1}{2}\beta_{2,4}\,,\quad
c_3 = \beta_{2,2}\,,\quad
c_4 = \beta_{2,4}\,.
\ee
The kernels $E_1$ and $E_2$ map
onto precomputed $\mathcal{I}$ integrals~\eqref{eq:P22stand} multiplied by 
$k^2$. However,
$E_3$ and $E_4$ have
non-trivial loop-momentum dependence that cannot be reduced to that of the one-loop bias basis. 
These produce new
integrals defined as:
\begin{subequations}\label{eq:JE3}
\begin{align}
\mathcal{J}_{F_2}^{E_3}&\equiv
  2\!\int_{\vq}\!E_3(\k\!-\!\vq,\vq)\,
  F_2(\k\!-\!\vq,\vq)\,\PL(|\k\!-\!\vq|)\,\PL(q)\,,
\label{eq:JF2E3}\\[3pt]
\mathcal{J}_{\delta^2}^{E_3}&\equiv
  2\!\int_{\vq}\! E_3(\k\!-\!\vq,\vq)\,
  \PL(|\k\!-\!\vq|)\,\PL(q)\,,
\label{eq:JdsE3}\\[3pt]
\mathcal{J}_{\Gal_2}^{E_3}&\equiv
  2\!\int_{\vq}\! E_3(\k\!-\!\vq,\vq)\,
  \Kcal(\k\!-\!\vq,\vq)\,
  \PL(|\k\!-\!\vq|)\,\PL(q)\,,
\label{eq:JGE3}\\[3pt]
\mathcal{J}^{E_4}_{F_2}&\equiv
  2\!\int_{\vq}\!
  (\k\cdot\vq-q^2)\,
  F_2(\k\!-\!\vq,\vq)\,\PL(|\k\!-\!\vq|)\,\PL(q)\,,
\label{eq:JF2E4}\\[3pt]
\mathcal{J}^{E_4}_{\delta^2}&\equiv
  2\!\int_{\vq}\!
  (\k\cdot\vq-q^2)\,
  \PL(|\k\!-\!\vq|)\,\PL(q)\,,
\label{eq:JdsE4}\\[3pt]
\mathcal{J}^{E_4}_{\Gal_2}&\equiv
  2\!\int_{\vq}\!
  (\k\cdot\vq-q^2)\,
  \Kcal(\k\!-\!\vq,\vq)\,
  \PL(|\k\!-\!\vq|)\,\PL(q)\,.
\label{eq:JGE4}
\end{align}
\end{subequations}
With some abuse of language, we will call these 
terms 
``two-loop counterterms,''
even though their complexity is actually that
of the one-loop integrals. In Appendix~\ref{sec:2lmatter}
we additionally show that the P22-like dark matter 
counterterm can be rewritten in terms of our 
template integrals.
{Note three redundancies among the template integrals:}
\be 
\begin{split}
&\mathcal{J}^{E_3}_{\delta^2}
= \frac{1}{2}\,k^2\left(\Idsds + \IdsG\right)
- \mathcal{J}^{E_4}_{\Gal_2}\,,\\
&\mathcal{J}^{E_4}_{F_2}
= \frac{1}{7}\,k^2\left(\Idsds + \IdsG\right)
+ \frac{5}{7}\,\mathcal{J}^{E_3}_{\delta^2}\,,\\
&\mathcal{J}^{E_3}_{\Gal_2}
= \frac{7}{10}\,k^2\left(\Ids + \IG\right)
- \frac{1}{5}\,k^2\left(\IdsG + \IGG\right)
- \frac{7}{5}\,\mathcal{J}^{E_3}_{F_2}\,.
\end{split}
\ee

At the cubic order we find the following expression with thirteen  
new coefficients:
\begin{equation}\label{eq:k2d3ops}
\begin{split}
\delta_g \big|_{\nabla^2\delta^3}= \;&
\beta_{3,1}\;\delta^2\nabla^2\delta
+\beta_{3,2}\;\delta\,(\boldsymbol{\nabla}\delta)^2
\\[4pt]
+&\beta_{3,3}\;(\nabla^2\delta)\,\mathcal{G}_2
+\beta_{3,4}\;\delta\,\nabla^2\mathcal{G}_2
+\beta_{3,5}\;\delta\,\mathcal{G}_2(\nabla_i\Phi_v,\nabla_i\Phi_v)
+\beta_{3,6}\;(\nabla_i\delta)(\nabla_i\mathcal{G}_2)
\\[4pt]
+&\beta_{3,7}\;\nabla^2\mathcal{G}_3(\Phi_v)
+\beta_{3,8}\;\mathcal{G}_3(\nabla_i\Phi_v,\nabla_i\Phi_v,\Phi_v)
+\beta_{3,9}\;\nabla^2\!\big[\mathcal{G}_2(\varphi_2,\varphi_1)\big] \\[4pt]
+&\beta_{3,10}\;\mathcal{G}_2(\nabla_i\varphi_2,\nabla_i\varphi_1)
+\beta_{3,11}\;\mathcal{G}_2(\nabla^2\varphi_2,\varphi_1)
+\beta_{3,12}\;\Pi_{ij}\,\nabla_i\delta\,\nabla_j\delta \\
 + & 2\gamma_2\;\mathcal{G}_2\bigl(\Phi_{v,\rm ct}^{(2)},\Phi_v^{(1)}\bigr)
+\check{\beta}_{3,13}\; \Gal_2(\vec{\Psi}^{(2)}_{\mathrm{ct}},\vec{\Psi}^{(1)})
\\[4pt]
+&2\beta_{2,1}\;\nabla^2(\delta^{(1)}\delta^{(2)})
+2\beta_{2,2}\;\nabla_i \delta^{(1)}\nabla^i\delta^{(2)}\\[4pt]
+&2\beta_{2,3}\;\nabla^2\mathcal{G}_2(\Phi_v^{(2)},\Phi_v^{(1)})
+2\beta_{2,4}\;\mathcal{G}_2(\nabla_i\Phi_v^{(2)},\nabla_i\Phi_v^{(1)})\,,
\end{split}
\end{equation}
{where $\Pi_{ij}\equiv(\nabla_i\nabla_j/\nabla^2)\delta$, and the last two lines are
SPT-evolved quadratic operators.}
{In addition to the higher-derivative terms due to galaxy biasing, there are 
higher-derivative operators that originate from the dark matter 
stress-tensor~\cite{Baldauf:2014qfa,
Anastasiou:2025jsy,Chen:2026usz}.
These enter the bias expansion through 
the counterterm dressing of non-linear matter and velocity 
fields and their potentials
in the Eulerian bias expansion, 
\be 
  b_1\delta\to b_1(\delta+\delta_{\rm ct}) +\cdots \,,\quad 
  \Phi_v\to\Phi_v+\Phi_{v,\mathrm{ct}}\,,\quad \text{etc.}
\ee
The $2\gamma_2\;\mathcal{G}_2\bigl(\Phi_{v,\rm ct}^{(2)},\Phi_v^{(1)}\bigr)$ above 
precisely corresponds to a cubic operator generated this way. 
The other contributions generated from the renormalized velocity are degenerate with the higher-derivative bias expansion. We have singled out 
this term because it generates a new non-local 
shape not spanned by the higher-derivative bias operators.
The extra counterterms multiplying $b_1$ will be discussed momentarily.}

{The above operator $\Gal_2(\vec{\Psi}^{(2)}_{\mathrm{ct}},\vec{\Psi}^{(1)})$ deserves a special mention. Normally, in the Lagrangian bias
expansion one builds operators from 
the dark matter displacements at various orders in LPT, e.g. $\Gal_2(\vec{\Psi}^{(2)},\vec{\Psi}^{(1)})$. However, up to the third order it produces results indistinguishable from the simple Eulerian bias expansion built with operators like $\mathcal{G}_2(\Phi_v)$. Therefore one can choose to build the lowest order operators in the Galileon Eulerian scheme. 
The same logic is applicable to the higher-derivative terms as well, and it produces most of the operators in Eq.~\eqref{eq:k2d3ops} above. However, the LPT expansion is consistent only when the counterterms 
are included, in which case the Lagrangian basis of bias operators must include separate terms built from the
displacement counterterms. 
Indeed, $\vec{\Psi}^{(n)}_{\rm ct}$
are as fundamental as $\vec{\Psi}^{(n)}$ in the EFT framework.
At the first order this produces an operator $\Gal_2(\vec{\Psi}^{(1)}_{\mathrm{ct}},\vec{\Psi}^{(1)})$ that is degenerate 
with the Eulerian operators parameterized by $\beta_{2,3},\beta_{2,4}$.} 
{However, already at second order one can write
$\Gal_2(\vec{\Psi}^{(2)}_{\mathrm{ct}},\vec{\Psi}^{(1)})$,
which cannot be reduced to the simple local gradients 
of the Eulerian operators. Therefore, this operator must be 
included as a separate term in the higher-derivative bias 
expansion.}

{Since   
$\vec{\Psi}^{(2)}_{\mathrm{ct}}$ carries a set of dimensionful
dark-matter counterterms, the coefficient 
$\check{\beta}_{3,13}$ is dimensionless, in contrast to the other coefficients in the $\beta_{3,a}$ family. 
$\vec{\Psi}^{(2)}_{\mathrm{ct}}$ also depends on a number
of free dark matter counterterms.  
To homogenize the notation and facilitate the study 
of the renormalization conditions
it is convenient to fix the ratios of the dark matter 
counterterm coefficients by the UV divergences and pull out 
a single free parameter in front of $\vec{\Psi}^{(2)}_{\mathrm{ct}}$. 
This makes it natural to use the dark matter UV-divergence
normalization and 
absorb the loop factors into the counterterms themselves,
\be\label{eq:hatunits}
\beta_{3,13}\equiv 2\sigma^2_{-2}\,\check\beta_{3,13}\,,
\qquad
\hat\gamma_2\equiv 2\sigma^2_{-2}\,\gamma_2\,,
\qquad
\sigma^2_{-2}=\int_{\vq}\frac{\PL(q)}{q^2}\,,
\ee
so that $\beta_{3,13}$ and $\hat\gamma_2$ carry momentum dimension
$k^{-2}$ like every other $\beta_a$ and multiply pure dimensionless
coefficients. Note that $\hat\gamma_2$ is not a free parameter, it is completely fixed by dark matter. The explicit expression
for $\vec{\Psi}^{(2)}_{\mathrm{ct}}$ and dark-matter trispectrum-level renormalization conditions are presented in Appendices~\ref{app:Psi2ctr} and~\ref{app:trispec}.
}

When contracted with linear bias term
$b_1\delta$, the above terms will generate the 
P13-type integrals,
\be 
P_{gg}\Big|_{k^2P_{13}-\mathrm{like}}=6b_1\PL(k)\int_{\vq}
K_{\rm h.d.}^{(3)}(\k,\vq,-\vq)\PL(q)+6K^{(1)}_{\rm h.d.}\PL(k)\int_{\vq}
K_g^{(3)}(\k,\vq,-\vq)\PL(q)~\,,
\ee 
where $K^{(1)}_{\rm h.d.}\equiv -\beta_{1}k^2$.
Due to the specific momentum configuration
in this P13-like integral
many of the above terms turn out to be redundant.
{Indeed, twelve
of the eighteen kernels are purely trivial: their entire
contribution is absorbed into $b_1$ and
$b_{\nabla^2\delta}$:
\be\label{eq:k2d3_conditions}
\begin{split}
\delta b_1\big|_{\nabla^2\delta^3} &= \tfrac12\Bigl[
-4\beta_{3,1}+2\beta_{3,2}+\tfrac83\beta_{3,3}+\tfrac83\beta_{3,4}
-\tfrac83\beta_{3,6}+\tfrac23\beta_{3,12}
+\tfrac{164}{21}\beta_{2,2}\Bigr]\int_{\vq}q^2\PL(q)\,,\\
\delta b_{\nabla^2\delta}\big|_{\nabla^2\delta^3} &= -\tfrac12\Bigl[
-2\beta_{3,1}+\tfrac83\beta_{3,4}-\tfrac{32}{15}\beta_{3,11}
-\tfrac{136}{21}\beta_{2,1}+\tfrac43\beta_{2,2}
-\tfrac{8}{15}\beta_{2,4}
+\tfrac{172519}{2546775}\beta_{3,13}
-\tfrac{703784}{2546775}\hat\gamma_{2}
\Bigr]\int_{\vq}\PL(q)\,.
\end{split}
\ee
The non-locality in $\beta_{3,11}\mathcal{G}_2(\nabla^2\varphi_2,
\varphi_1)$ is canceled by the Laplacian in
$\nabla^2\varphi_2$.}  
{What
survives as genuine shape content from the whole block is}
\be\label{eq:k2d3_shapes}
\begin{split}
6\PL(k)\int_{\vq}
K_{\rm h.d.}^{(3)}(\k,\vq,-\vq)\PL(q)\to 
&\Bigl[\tfrac{7}{5}\beta_{3,9}-\tfrac{6}{5}\beta_{2,3}
+\tfrac{167}{40425}\beta_{3,13}+\tfrac{11419}{80850}\hat\gamma_{2}
\Bigr] k^2\FG \\
& 
+\Bigl[\beta_{3,10}-\tfrac{6}{7}\beta_{2,4}
-\tfrac{167}{56595}\beta_{3,13}-\tfrac{2551}{11319}\hat\gamma_{2}
\Bigr]
\mathcal{F}^{E_4}_{\Gal_2}\,,
\end{split}
\ee
where we have introduced a new basis
integral:
\be\label{eq:FGE4}
\mathcal{F}^{E_4}_{\Gal_2}\equiv
4\,\PL(k)\!\int_{\vq}\!(q^2-\k\cdot\vq)\,
\Kcal(\vq,\k\!-\!\vq)\,
\Kcal(\k,\vq)\,\PL(q)\,.
\ee
Crucially, this leaves us with only two 
non-trivial $P_{13}$-like integrals, one of which  
is simply $k^2\FG$, and the other one is~\eqref{eq:FGE4}.
{Using them the non-redundant P13-counterterm can be rewritten as 
\be 
P_{gg}\Big|_{k^2P_{13}-\mathrm{like}}= b_1 c_5k^2\mathcal{F}_{\Gal_2}+ b_1c_6\mathcal{F}^{E_4}_{\Gal_2} -
b_1\beta_1 k^2\left[2P^{\rm SPT}_{13}+2\left(\tfrac{3}{5}\gamma_2-\tfrac{7}{10}\gamma_{21}\right)\F_{\Gal_2}\right]~\,,
\ee 
where $c_5$ and $c_6$ are given by the brackets in Eq.~\eqref{eq:k2d3_shapes},
and $P^{\rm SPT}_{13}$
is the standard matter P13 contribution. }

Let us discuss now additional structures from the renormalization-enforced dark matter 
counterterms $b_1\delta_{\rm ct}$.
We show in Appendix~\ref{sec:2lmatter} that the complete 
two-loop dark matter power spectrum counterterm contribution can be 
written at the field level as
\be\label{eq:matter_ctr2}
\begin{split}
\delta_m^{\rm ct} 
=&-c_s k^2\delta  + \tilde{c}_s k^4\delta+ \sum_{n=1,2,3}e_n\int_{\vq_1,\vq_2}(2\pi)^3\delta_D^{(3)}(\k-\vq_{12})\,E_n(\vq_1,\vq_2)\,\delta_L(\vq_1)\delta_L(\vq_2)\,,\\
&+\sum_{n=3,4,5}e_n \int_{\vq_1,\vq_2,\vq_3}(2\pi)^3\delta_D^{(3)}(\k-\vq_{123})\,E^{(3)}_n(\vq_1,\vq_2,\vq_3)\,\delta_L(\vq_1)\delta_L(\vq_2)\delta_L(\vq_3)\,,
\end{split}
\ee
{where $c_s,\tilde{c}_s,e_{1,...,5}$ 
are dark matter counterterms,
$\delta$ in the ``sound speed'' operator 
has to be expanded to higher order in SPT, {the coefficient of the cubic kernel $E^{(3)}_3$ is locked to that of the quadratic kernel $E_3$, see Appendix~\ref{sec:2lmatter}}, and the 
non-redundant cubic operators evaluated on the 
relevant momentum configurations are:
\be 
\begin{split}
& E^{(3)}_3 
 =-\frac{7}{18}\left( k^2F_3(\k,\q,-\q) + \frac{k^4}{18\,q^2}\right)~\,,\quad E^{(3)}_4 
 = \frac{1}{3}\Kcal(\vq,\k\!-\!\vq)\,
F_2(\k,-\vq)~\,,\quad  
E^{(3)}_5 =\frac{1}{3} (q^2-\k\cdot\vq)\,
\Kcal(\vq,\k\!-\!\vq)\,
\Kcal(\k,\vq) ~\,.
\end{split}
\ee 
Clearly, the contributions generated in this way are highly degenerate with those 
coming from the higher-derivative bias, and hence almost all of them  can be 
absorbed by a redefinition of the bias parameters $c_a$. This logic extends to other Eulerian operators as well, except $\gamma_2$.
For instance, $b_1\delta_{\rm ct}$ gives:
\be 
c_{1,2}\to c_{1,2}-b_1e_{1,2}\,,\quad 
c_3\to c_3+b_1e_3\,, \quad c_5\to c_5-\frac{5}{7}b_1^2 e_4\,,
\quad c_6\to c_6 +b_1^2e_5\,.
\ee 
The exception here is the advected $e_3$ term that is not explicitly present in the bias expansion, but whose 
kernel for the power spectrum momentum configuration is reduced to 
the IR-safe version of $k^2 P^{\rm SPT}_{13}$, 
\be 
\bar P^{\rm SPT}_{13}=P^{\rm SPT}_{13}-\tfrac{k^2}{3}\PL(k)\sigma_v^2\,,
\ee 
where $\sigma_v^2\equiv \int_{\q}\tfrac{\PL(q)}{q^2}$ is the IR displacement
variance. This term, however, does not 
introduce any new free parameters: its Wilson coefficient $e_3$ is fully determined
by matter and therefore, it can be matched e.g. from N-body data. }

\begin{figure}[t]
\centering
\includegraphics[width=0.49\textwidth]{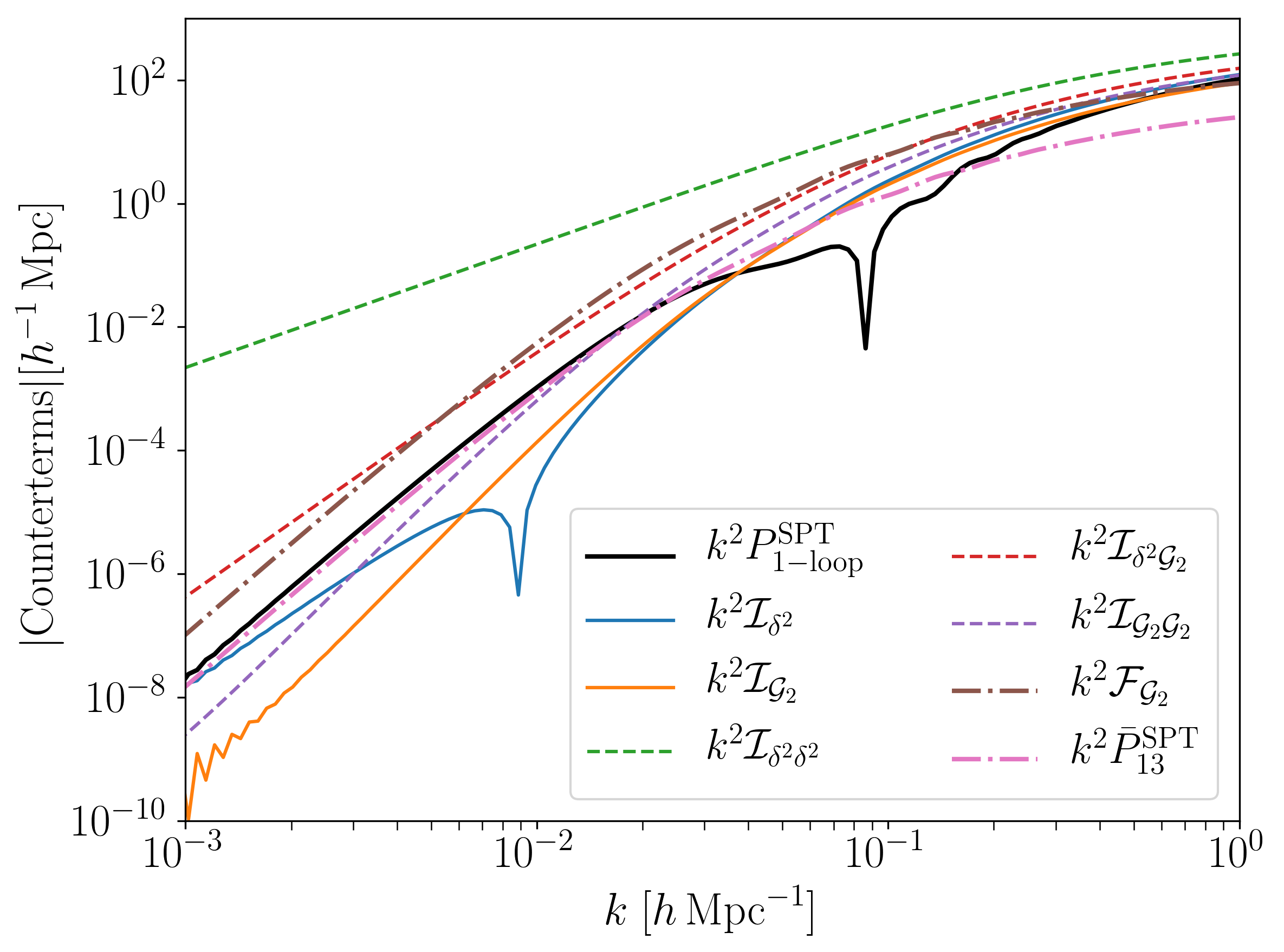}%
\hfill
\includegraphics[width=0.49\textwidth]{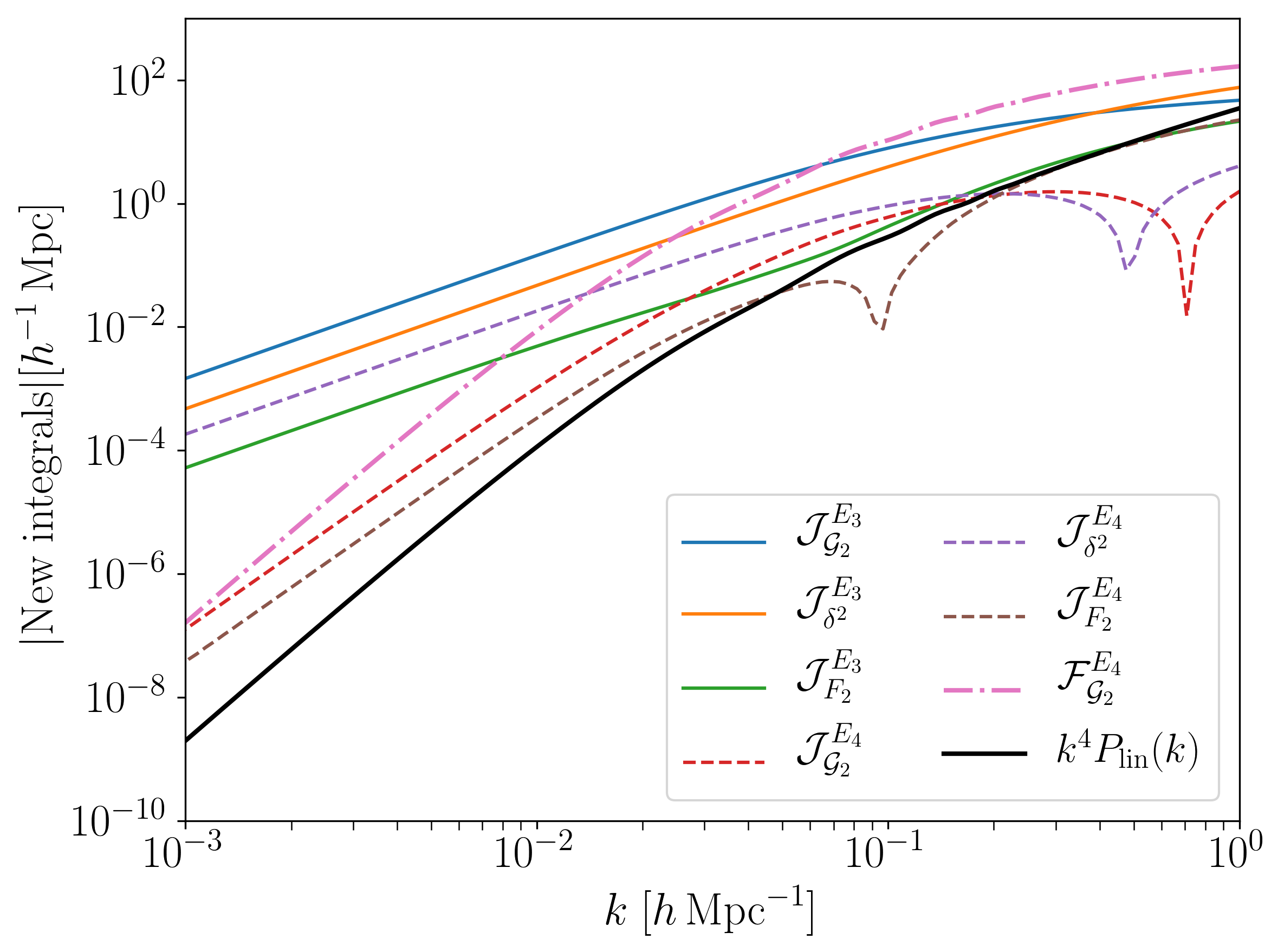}
\caption{\emph{Left:} 
Power spectrum counterterm shapes based on the standard 
one-loop
bias templates, entering the two-loop counterterms 
$P_0^{\rm HD}$--$P_2^{\rm HD}$ plus $k^2\bar P^{\rm SPT}_{13}$.
\emph{Right:} New integrals entering $P_3^{\rm HD}$--$P_7^{\rm HD}$ (including the $k^4\PL$ counterterm). }
\label{fig:templates}
\end{figure}

Using the known one-loop bias templates and the seven genuinely new two-loop counterterm shapes, the 
total higher-derivative contribution can be written as 
\be \label{eq:master}
\Delta P_{gg}^{\rm HD}(k)
= \sum_{i=0}^{7} c_i\;P^{\rm HD}_i(k)- \tfrac{7}{18}\,e_3\,b_1^2\,k^2\bar P^{\rm SPT}_{13}(k)\,,
\ee
where 
$c_0=\beta_1=b_1c_s+b_{\nabla^2\delta}$, while $c_{1,\dots,6}$ and $c_{7}\equiv \beta_1^2 + 2b_1\tilde\beta_1$ are seven new two-loop counterterm parameters, together multiplying the basis shapes
\begin{subequations}\label{eq:shapes}
\begin{align}
P^{\rm HD}_0(k) &=
  -2b_1\,k^2\,P_{\rm 1\text{-}loop}^{\rm SPT}(k)
  -k^2\Big[b_2\,\Ids + 2b_{\Gal_2}\,\IG
  +\tfrac{1}{5}\big(6b_{\Gal_2}-7\gamma_{21}\big)\FG\Big]\,,
\label{eq:PHD0}\\[6pt]
P^{\rm HD}_1(k) &=
  -k^2\Big[2b_1\,\Ids
  +b_2\,\Idsds
  +2b_{\Gal_2}\,\IdsG\Big]\,,
\label{eq:PHD1}\\[6pt]
P^{\rm HD}_2(k) &=
  -k^2\Big[2b_1\,\IG
  +b_2\,\IdsG
  +2b_{\Gal_2}\,\IGG
  +{\tfrac{6}{5}}\,b_1\,\FG\Big]\,,
\label{eq:PHD2}\\[6pt]
P^{\rm HD}_3(k) &=
  -2b_1\,\mathcal{J}^{E_4}_{F_2}
  -b_2\,\mathcal{J}^{E_4}_{\delta^2}
  -2b_{\Gal_2}\,\mathcal{J}^{E_4}_{\Gal_2}\,,
\label{eq:PHD3}\\[6pt]
P^{\rm HD}_4(k) &=
  2b_1\,\mathcal{J}_{F_2}^{E_3}
  +b_2\,\mathcal{J}_{\delta^2}^{E_3}
  +2b_{\Gal_2}\,\mathcal{J}_{\Gal_2}^{E_3}
  +{\tfrac{3}{5}}\,b_1\,k^2\,\FG
  {-\tfrac{6}{7}}\,b_1\,\mathcal{F}^{E_4}_{\Gal_2}\,,
\label{eq:PHD4}\\[6pt]
P^{\rm HD}_5(k) &=
  -\tfrac{7}{5}\,b_1\,k^2\,\FG\,,
\label{eq:PHD5}\\[6pt]
P^{\rm HD}_6(k) &=
 b_1\,\mathcal{F}^{E_4}_{\Gal_2}\,,
\label{eq:PHD6}\\[6pt]
 P^{\rm HD}_7(k) &=
  2b_1\,k^4\PL(k)\,.
\label{eq:PHD7}
\end{align}
\end{subequations}
Note that all contributions above, including $P_{0}^{\rm HD}$,
are IR-safe. Since the complexity of the 
new seven two-loop counterterm 
integrals is the same as that of the one-loop
EFT bias contributions,
they can be easily computed with 
the 
FFTLog technique~\cite{Simonovic:2017mhp}.
The counterterms above 
computed for our fiducial PT Challenge cosmology
are shown in Fig.~\ref{fig:templates}.
We implement  the damping of BAO wiggles (IR resummation)
in these contributions as in~\eqref{eq:irres}
following~\cite{Blas:2016sfa}.

{Before closing this section, let us note that we have 
explicitly checked that the four quadratic operators
\eqref{eq:k2d2ops} and thirteen 
cubic operators \eqref{eq:k2d3ops}
are sufficient to renormalize the 
two-loop galaxy power spectrum, one-loop bispectrum, and one-loop trispectrum. 
In Appendix~\ref{app:trispec}
we show that 
to derive the trispectrum renormalization conditions
it is essential to first renormalize the pure dark matter
one-loop trispectrum. The bias expansion itself only requires the 
thirteen operators in \eqref{eq:k2d3ops}, of which only one new non-local
structure $\Gal_2(\vPsi^{(2)}_{\rm ct},\vPsi^{(1)})$ is needed. 
This structure, however, produces a purely redundant contribution 
to the power spectrum, as we have shown in Eqs.~(\ref{eq:k2d3_conditions},\ref{eq:k2d3_shapes}).
We conclude that the operators we consider here are sufficient
for the two-loop galaxy power spectrum
renormalization, and the power spectrum templates introduced above fully exhaust 
the non-trivial two-loop counterterm shape space.}

\subsection{Galaxy-matter higher-derivative contributions}

To obtain the galaxy-matter cross-spectrum
counterterms we use the full 
set 
of two-loop dark matter counterterms from Eq.~\eqref{eq:matter_ctr2}.
Contracting $\delta_m$ with the galaxy density we get:
\begin{align}
\Delta P^{\rm ctr}_{gm} ={}& -(b_{\nabla^2\delta}+2b_1c_s)\,
 k^2\bigl[\PL + P^{\rm SPT}_{\rm 1\text{-}loop}\bigr]
 + \tilde{c}_{gm}\,k^4\PL
 - c_s k^2\Bigl[\tfrac{b_2}{2}\,\Ids + b_{\Gal_2}\,\IG\Bigr]
\nonumber\\
&-(c_1-e_1b_1)\,k^2\Ids - (c_2-e_2b_1)\,k^2\IG
\nonumber\\
&+ e_1k^2\Bigl[\tfrac{b_2}{2}\,\Idsds + b_{\Gal_2}\,\IdsG\Bigr]
 + e_2k^2\Bigl[\tfrac{b_2}{2}\,\IdsG + b_{\Gal_2}\,\IGG\Bigr]
\nonumber\\
&+(c_4+e_3b_1)\,\mathcal{J}^{E_3}_{F_2}
 + e_3\Bigl[\tfrac{b_2}{2}\,\mathcal{J}^{E_3}_{\delta^2}
 + b_{\Gal_2}\,\mathcal{J}^{E_3}_{\Gal_2}
 - \tfrac{7}{18}\,b_1\,k^2\bar P_{13}\Bigr]
 - c_3\,\mathcal{J}^{E_4}_{F_2}
\nonumber\\
&+\Bigl(-\tfrac{3}{5}c_2+\tfrac{3}{10}c_4-\tfrac{7}{10}c_5
 -\tfrac{c_s}{10}\bigl(6b_{\Gal_2}-7\gamma_{21}\bigr)
 + \tfrac{b_1e_4}{2}\Bigr)k^2\FG
 +\Bigl(-\tfrac{3}{7}c_4+\tfrac{c_6}{2}
 + \tfrac{b_1e_5}{2}\Bigr)\mathcal{F}^{E_4}_{\Gal_2}\,,
\label{eq:Pgm_ctr_final}
\end{align}
where we defined $\tilde c_{gm}\equiv\bigl(\beta_1 c_s + b_1\tilde c_s + \tilde\beta_1\bigr)$. 
Importantly, the cross-spectrum depends
only on the galaxy and matter 
counterterms that we have encountered above, 
i.e. it does not require 
any free parameters.

\subsection{Stochastic terms}

The stochastic terms come from the auto-spectrum
of the stochastic density component $\epsilon_g$. 
In EFT the spectrum is a power-law series:
\be \label{eq:gg-stoch}
\Delta P_{gg}^{\rm stoch}=\langle \epsilon_g(\k)\epsilon_g(-\k)\rangle'=\frac{1}{\bar n_g}\left(1+P_{\rm shot}+a_0\left(\frac{k}{k_{\rm stoch}}\right)^2+a_4\left(\frac{k}{k_{\rm stoch}}\right)^4+\cdots\right)\,,
\ee 
where we inserted the galaxy number density
$\bar n_g$ and non-local stochasticity scale 
$k_{\rm stoch}=1~\hMpc$ to make stochastic counterterms dimensionless,
and retained only the contributions
relevant at the two-loop order. Likewise, 
for the galaxy-matter cross-spectrum the stochastic term 
reads
\be \label{eq:gm-stoch}
\Delta P_{gm}^{\rm stoch}=\langle \epsilon_g(\k)\epsilon_m(-\k)\rangle'=\frac{1}{\bar n_g}\left(a'_0\left(\frac{k}{k_{\rm stoch}}\right)^2+a'_4\left(\frac{k}{k_{\rm stoch}}\right)^4+\cdots\right)\,,
\ee 
which does not have the constant term due to the mass
and momentum conservation of  the matter
density. 

Finally, we find that the mixed stochastic-deterministic 
operators $O(\epsilon_g \delta)$, studied systematically
in the context of the bispectrum in~\cite{Ivanov:2021kcd,Bakx:2025pop},
produce only purely redundant 
contributions at the level of the galaxy
power spectrum. This result is valid to all orders in EFT, see
Appendix~\ref{sec:mixed} for more detail.
Note that a similar conclusion was obtained
in Ref.~\cite{Bakx:2025cvu}.

\subsection{Summary}

All in all, our complete two-loop calculation yields
the following prediction:
\be \label{eq:sum}
\begin{split}
& P_{gg}(k) =  P_{gg}^{\rm 1-loop,~tot} + \Delta P_{gg}^{\rm 2-loop}+\Delta P^{\rm HD}_{gg} + \Delta P_{gg}^{\rm stoch}\,,\\
& P_{gm}(k) =  P_{gm}^{\rm 1-loop,~tot} + \Delta P_{gm}^{\rm 2-loop}+\Delta P^{\rm HD}_{gm} + \Delta P_{gm}^{\rm stoch}\,,
\end{split}
\ee 
where $P_{gg}^{\rm 1-loop,~tot}$, $P_{gm}^{\rm 1-loop,~tot}$
are the total one-loop 
galaxy-galaxy and galaxy-matter power-spectra~\cite{Chudaykin:2020aoj,Ivanov:2026dvl}, $\Delta P^{\rm stoch}_{gg,gm}$ are the stochastic 
power spectra, which depend on $(P_{\rm shot},a_0,a_4)$ and 
$(a'_0,a'_4)$ parameters, respectively.
The two-loop bias contributions read:
\be 
\begin{split}
& \Delta P_{gg}^{\rm 2-loop}=\sum_{i\leq j} b_i\, b_j\; P^{ij}(k)\,, \\
& \Delta P_{gm}^{\rm 2-loop}=\sum_{i\leq j} \frac{1}{2}(b^m_i\, b_j+b_i\, b^m_j)\; P^{ij}(k)\,,
\end{split}
\ee
where $P^{ij}(k)$ are two-loop bias shapes and $b^i$ is the vector of 17 galaxy bias parameters,
\be 
b^i =
\bigl(
b_1,\,
b_2,\,
\gamma_2,\,
\gamma_{21},\,
b_3,\,
\gamma_2^\times,\,
\gamma_3,\,
\gamma_{211},\,
\gamma_{22},\,
\gamma_{31},\,
\gamma_{21}^\times,\,
\gamma_{23},\,
\gamma_{221},\,
\gamma_{311},\,
\gamma_{41},\,
\gamma_{22}^\times,\,
\gamma_{31}^\times
\bigr),
\ee
and $b_i^m=(1,\Vec{0})$ is the formal vector of bias 
parameters for dark matter. 
$\Delta P^{\rm HD}_{gg,gm}$
are the two-loop higher derivative contributions 
from eqs.~(\ref{eq:Pgm_ctr_final},~\ref{eq:master}),
which depend on the higher derivative bias $b_{\nabla^2\delta}$,
the 
dark matter sound-speed counterterm
$c_s$, {7} two-loop galaxy higher-derivative bias parameters $c_{1,...,7}$,
and {six additional two-loop dark matter counterterms
$\tilde{c}_s,e_{1,...,5}$ (the seventh dark matter counterterm, $c_s$, is already listed above; note that $P_{gg}$
constrains only the combination $\beta_1 = b_1c_s+b_{\nabla^2\delta}$,
so it is the cross-spectrum that breaks the degeneracy between $c_s$ and $b_{\nabla^2\delta}$).}
Eqs.~\eqref{eq:sum} are subject to two-loop 
IR resummation from eq.~\eqref{eq:irres}.

The full two-loop galaxy power spectrum computation 
depends on 17 bias parameters, 8 higher-derivative parameters,
and 3 stochastic parameters. The galaxy-matter cross-spectrum introduces
additional 7 higher-derivative counterterms and 2 stochastic parameters,
for a total of {37} free parameters. {Relative to the one-loop $P_{gg}$ model, which depends on 7 parameters ($b_1,b_2,\gamma_2,\gamma_{21},b_{\nabla^2\delta},P_{\rm shot},a_0$), the two-loop $P_{gg}$ model thus introduces $28-7=21$ new parameters, which is the number quoted in the abstract.}

\begin{figure}[htb]
\includegraphics[width=0.49\textwidth]{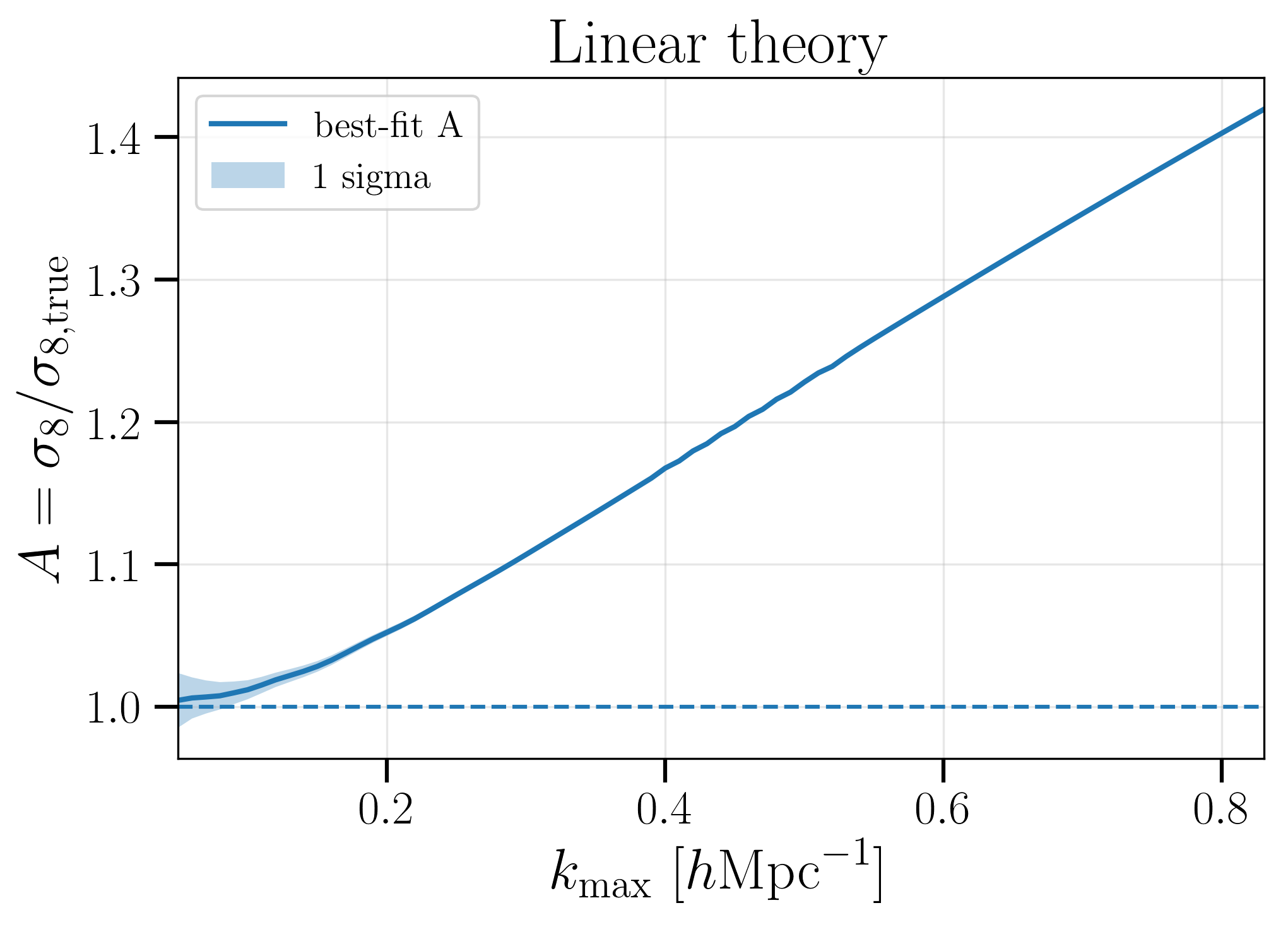}
\includegraphics[width=0.49\textwidth]{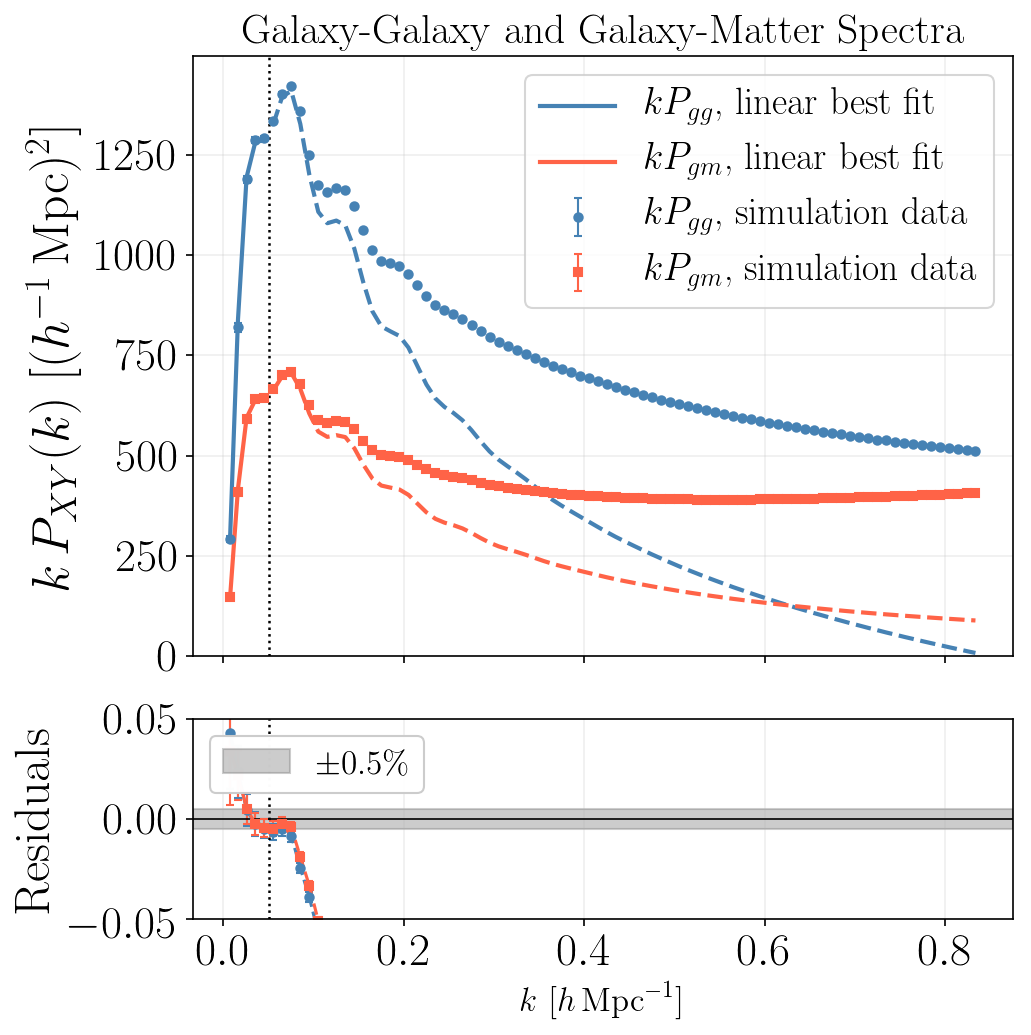}
\caption{\emph{Left:} 
The inference of the mass fluctuation amplitude $\sigma_8$
from the analysis of  the PT Challenge 
galaxy-galaxy and galaxy-matter power spectrum data ($P_{gg}-P_{gm}$) as a function of the scale cut $k_{\rm max}$.
We use a fiducial survey volume $V=1$ [$h^{-1}$Gpc]$^3$.
\emph{Right:} PT Challenge $P_{gg}-P_{gm}$ data and the linear theory best-fit at 
the largest scale cut 
$k_{\rm max}=0.05~\hMpc$ where the linear theory model is still unbiased. 
The fiducial data errors 
are reduced by a factor of 10 for better visibility.}
\label{fig:lin_fit}
\end{figure}

\begin{figure}[htb]
\centering
\includegraphics[width=0.49\textwidth]{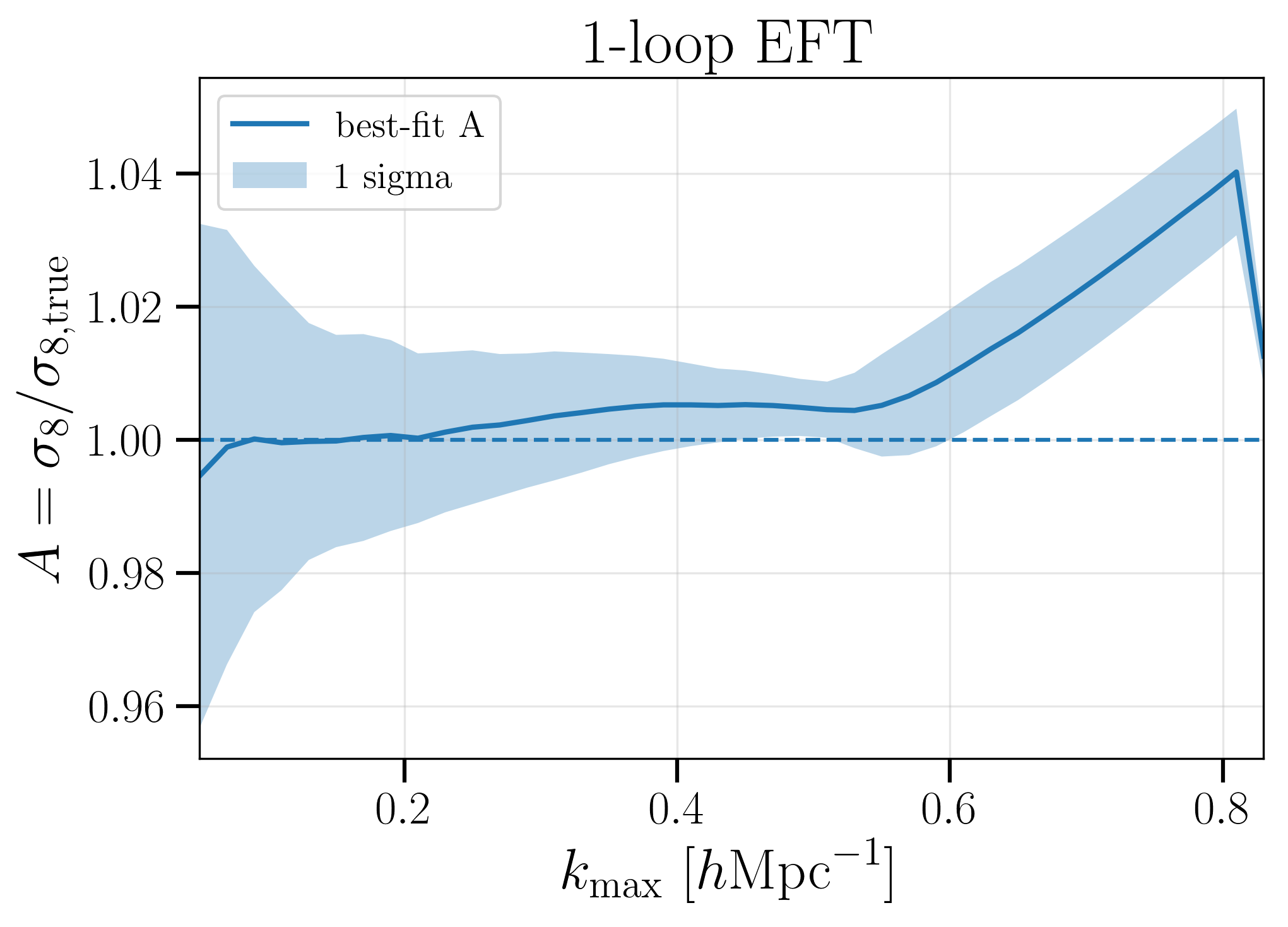}
\includegraphics[width=0.49\textwidth]{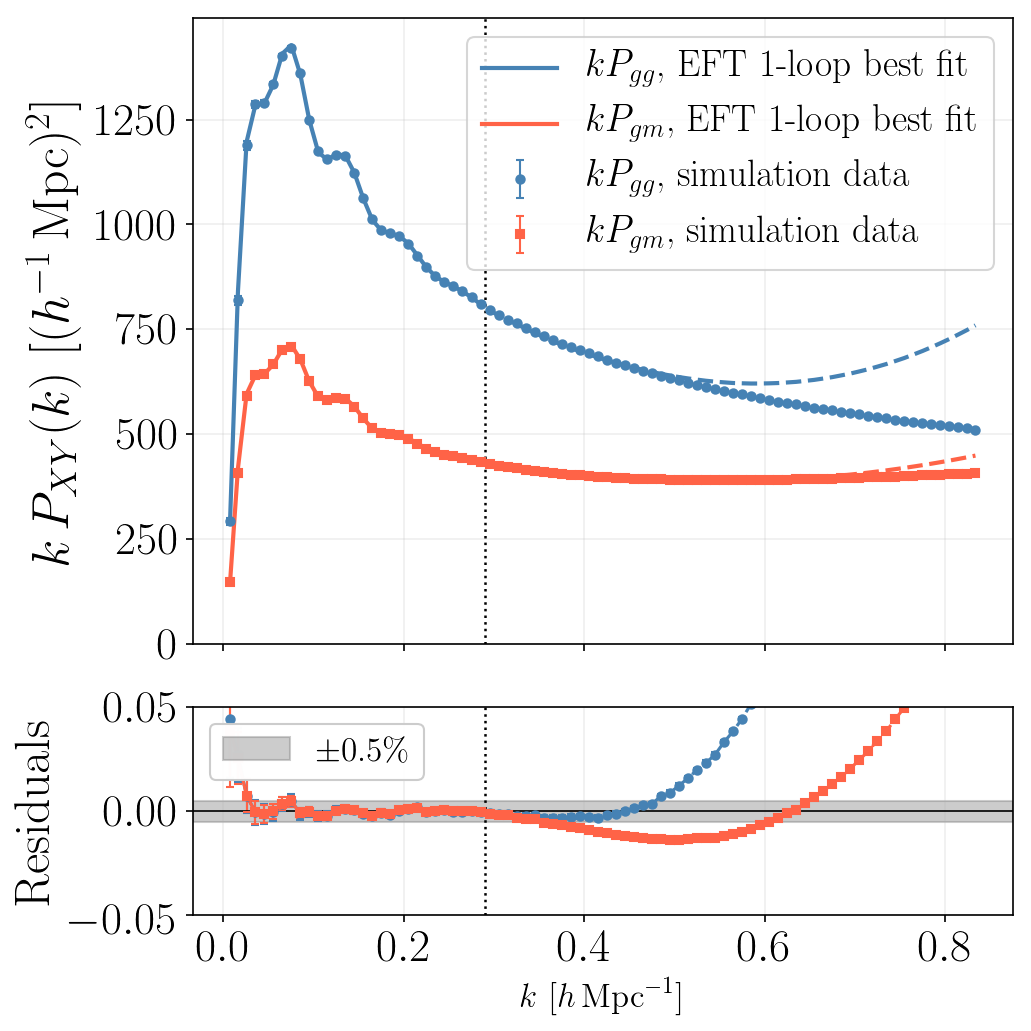}
\caption{Same as Fig.~\ref{fig:lin_fit}, but for the 1-loop EFT model. The best-fit curves in the right plot are shown for $k_{\rm max}=0.29~\hMpc$. }
\label{fig:1loop_fit}
\end{figure}

\begin{figure}[htb]
\centering
\includegraphics[width=0.49\textwidth]{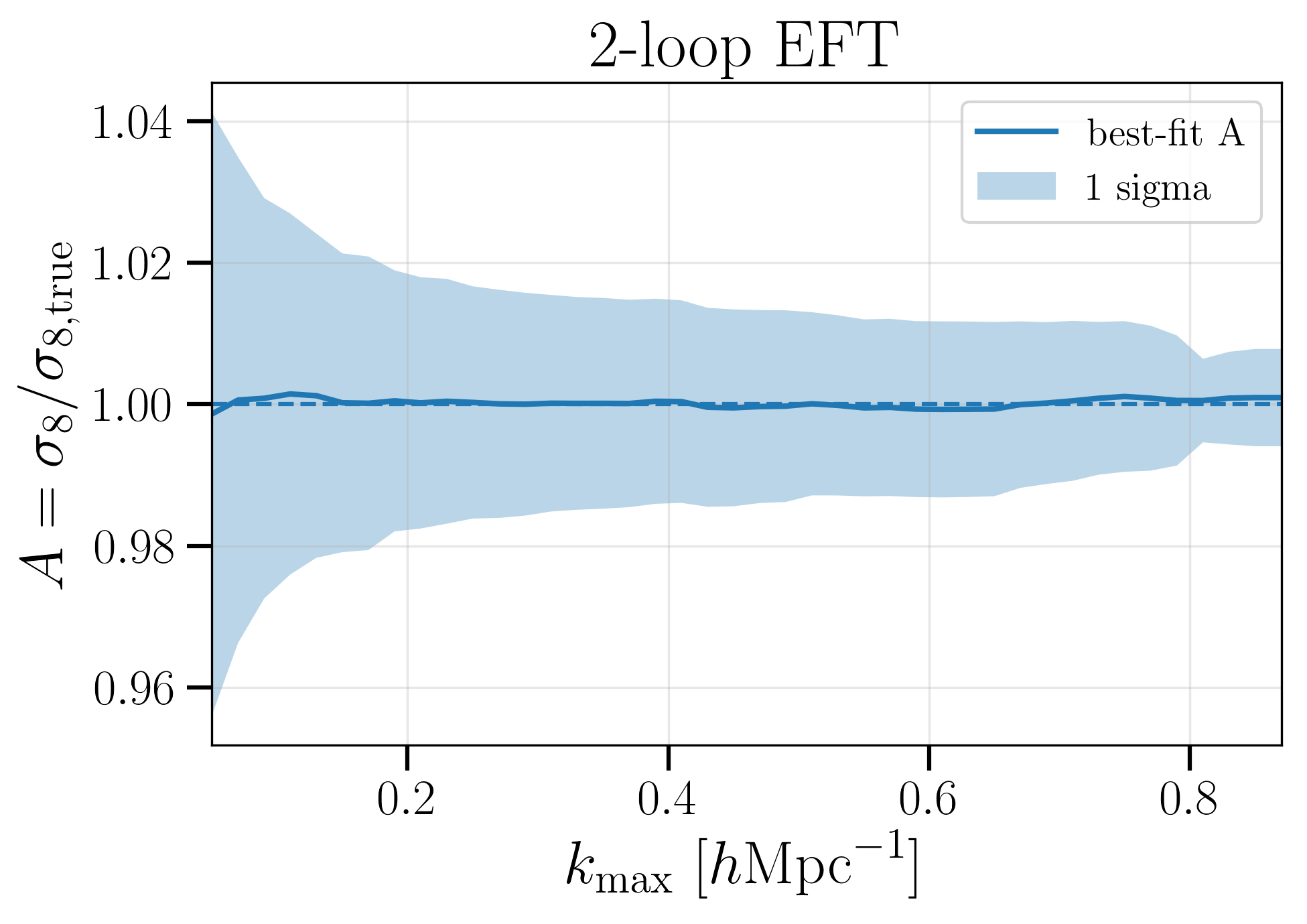}
\includegraphics[width=0.49\textwidth]{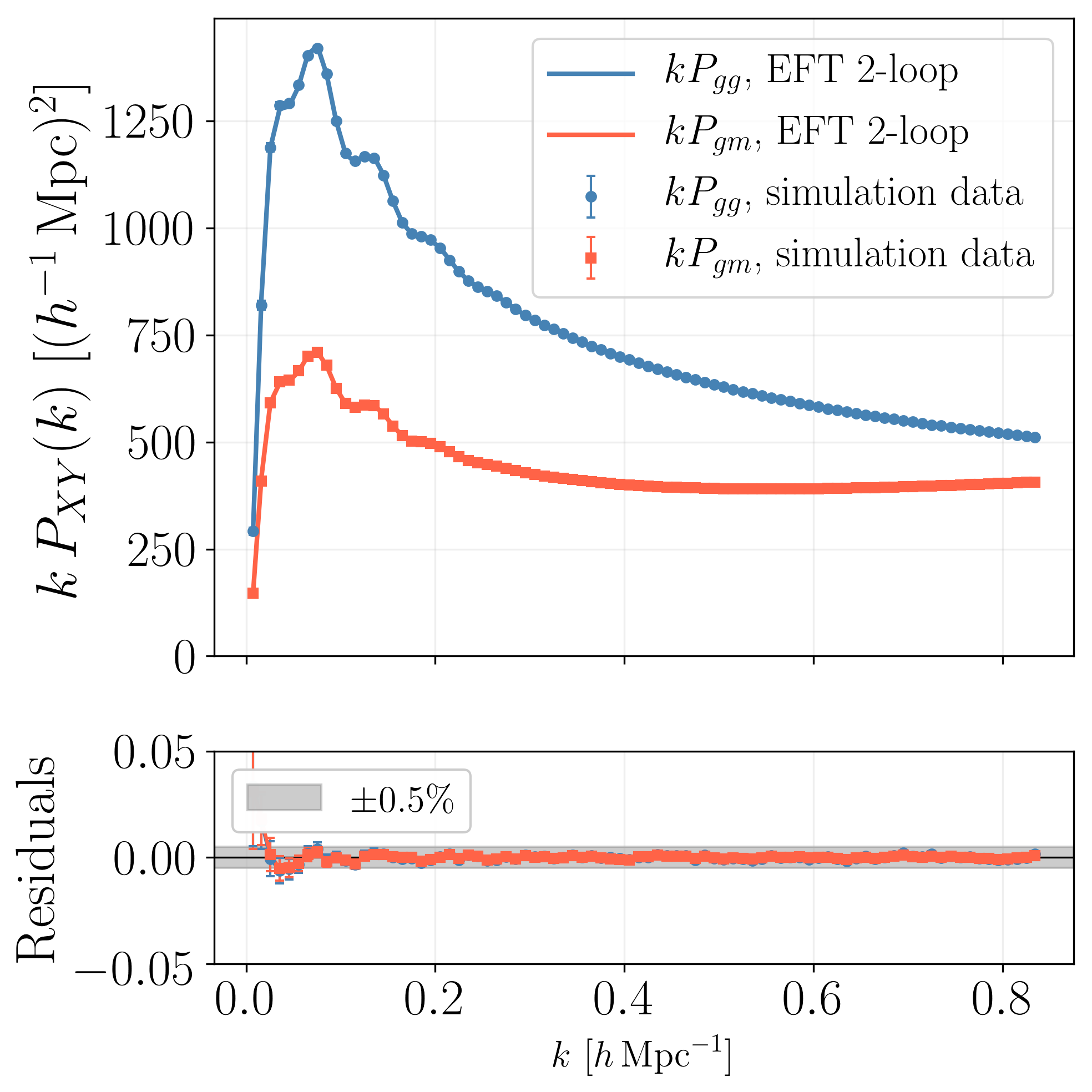}
\caption{Same as Fig.~\ref{fig:lin_fit}, but for the 2-loop EFT model. The best-fit curves in the right plot are shown for $k_{\rm max}=0.85~\hMpc$. }
\label{fig:2loop_fit}
\end{figure}

\section{Comparison with Simulation Data}
\label{sec:data}

In this section we compare our two-loop computation
with the galaxy-galaxy power spectrum and galaxy-matter cross-spectrum
of the PT Challenge simulation data~\cite{Nishimichi:2020tvu} at a fixed redshift $z=0.61$. 
We focus on the $P_{gg}-P_{gm}$
combination in order to understand the implications of the two-loop
galaxy power spectrum model for the $2\times 2$pt analysis of the 
galaxy clustering and galaxy-lensing data from imaging surveys or CMB lensing maps.
We do not include $P_{mm}$ in this combination for two reasons:
first, if we add $P_{mm}$ to $P_{gg}-P_{gm}$ data-vector, 
it will dominate the constraints on the matter clustering amplitude on all scales,
making it difficult to access the validity of the 
galaxy power spectrum model. 
Second, the $P_{gg}-P_{gm}-P_{mm}$
data-vector does not accurately 
approximate the full $3\times 2$pt
data because it does not 
capture the finite extent of the 
lensing kernel and the lensing noise, making the outcome of such
analysis hard to interpret in the context of the realistic gravitational weak
lensing data. In contrast, $P_{gg}-P_{gm}$ data at a fixed redshift is a good proxy for 
the galaxy clustering and galaxy-galaxy lensing spectra because 
these observables 
are mainly sourced by $P_{gg}-P_{gm}$
at the effective redshift of the sample if the redshift 
distribution is narrow enough, which is the case, e.g. for the 
DESI photometric sample~\cite{Maus:2025rvz,Ivanov:2026dvl}.

The PT Challenge simulation is a large-volume N-body simulation,
which consists of 10 boxes with the cumulative volume $566$ 
[$h^{-1}$Gpc]$^3$. The galaxy distribution at $z=0.61$ is 
generated using a Halo Occupation Distribution model 
for the BOSS-like luminous red galaxies (LRG)
described
in Ref.~\cite{Nishimichi:2020tvu}.
We use the mean real space galaxy auto spectrum 
and galaxy-matter cross spectrum from the entire suite.
The power spectra are binned linearly with a bin width $\Delta k=0.01~\hMpc$. 
We use the data up to $k=0.85~\hMpc$ in this analysis. We subtract the Poisson 
shot noise contribution $1/\bar{n}_g$ from the $P_{gg}$ data.
We use a Gaussian likelihood with the Gaussian covariance matrix for the $P_{gg}-P_{gm}$
data vector built from the measurements themselves,
\be
\mathrm{Cov}\!\left[
\begin{pmatrix}
P_{gm}(k) \\
P_{gg}(k)
\end{pmatrix},
\begin{pmatrix}
P_{gm}(k) \\
P_{gg}(k)
\end{pmatrix}
\right]
=
\frac{1}{N_k}
\begin{pmatrix}
P_{gg}^{\rm tot}(k)\,P_{mm}(k) + P_{gm}^2(k)
&
2\,P_{gg}^{\rm tot}(k)\,P_{gm}(k)
\\[6pt]
2\,P_{gg}^{\rm tot}(k)\,P_{gm}(k)
&
2\left(P_{gg}^{\rm tot}(k)\right)^2
\end{pmatrix},
\ee 
where $N_k=4\pi k^2\Delta k V/(2\pi)^3$  is the 
number of modes in a $k$-bin, 
$P_{gg}^{\rm tot}(k) = P_{gg}(k) + \frac{1}{\bar n_g}$
and $P_{mm}$ is the matter power spectrum of the simulation.
Note that we use $P_{mm}$ measurements 
only for the covariance
matrix estimation.
To make the analysis more relevant to the actual imaging surveys,
we use the fiducial volume $V=1$~[$h^{-1}$Gpc]$^3$, which roughly reproduces
typical error bars from surveys like DES~\cite{DES:2026mkc}.
While the Gaussian approximation
to the covariance is fairly accurate on large-scales, one might be
worried about its validity on much smaller scales relevant for the 
two-loop fits. In this regime, however, the parameter constraints
are driven by the marginalization over the EFT nuisance parameters,
producing a large ``theoretical error'' covariance~\cite{Chudaykin:2020hbf,Baldauf:2016sjb}, which is more 
significant than any connected contributions. 
This was explicitly tested on the BOSS data in Ref.~\cite{Wadekar:2020hax}.

In this work, we carry out a simplified analysis 
that (a) varies only the mass fluctuation amplitude 
$\sigma_8$ and the EFT parameters, 
and (b) uses a Fisher matrix 
approximation to the posterior distribution.
Our fitting parameters are the scaling parameter $A\equiv \sigma_{8}/\sigma_{8,\mathrm{true}}$ plus the relevant EFT parameters for each model. 
We assume Gaussian priors on all nuisance parameters $\mathcal{N}(\bar{p}_{\rm EFT},\sigma^2_{p_{\rm EFT}})$ with $\bar{p}_{\rm EFT}=2$ for $b_1$
and $\bar{p}_{\rm EFT}=0$  for all other parameters. We use
$\sigma_{p_{\rm EFT}}=3$  in appropriate units of $[\Mpch]^n$ with $n=0,2,4$ for bias operators, $k^2$- and $k^4$-counterterms, respectively.
Our prior is
consistent with requirements of perturbativity
of EFT and the notion of naturalness.  
For each model and scale
cut $\kmax$ choice,
we minimize the joint Gaussian likelihood 
of the data and the priors, and compute the covariance matrix 
for the relevant parameters around the best-fit. 
The $C^{-1}_{AA}$ element
of that covariance is the 1D marginalized 
variance of $A$.
While this Gaussian approach
is approximate, it has the great advantage of being extremely computationally efficient, 
allowing us to scan over the vast range of $k_{\rm max}$ in order
to determine the optimal scale cuts. We leave a more careful
MCMC analysis for future work.

Let us start off with a linear theory analysis. 
This analysis is similar to the fiducial analysis of the DES-Y6
collaboration based on 
the linear bias model~\cite{DES:2026lhi}. 
In this case the models are simply:
\be 
P_{gg}(k) = A^2\,b_1^2\,\PL(k) + P_{\rm shot}\,\bar n^{-1}_g,\quad 
P_{gm}(k) = A^2\,b_1\,\PL(k)\,,
\ee 
where we use the pre-computed $\PL$ for true PTC cosmology.
Note that we vary the leading order stochasticity parameter 
$P_{\rm shot}$ in order to be consistent with the EFT power counting. Indeed, the magnitude of the shot noise of the PT Challenge galaxies is of the same order of magnitude as the linear theory model at $k\sim 0.1~\hMpc$, consistent with our power counting. 

The results of the linear theory data analysis are shown in Fig.~\ref{fig:lin_fit}, which demonstrates the drift plot $A(\kmax)$
and shows the best-fit theory model for the maximal value of 
$\kmax$ where the fit is unbiased, which is $0.05~\hMpc$.
We choose this $k_{\rm max}$ by requiring the bias on $A$ not to exceed
$0.33\sigma_A$. 
At $\kmax=0.05~\hMpc$ our fitting procedure 
gives the 1D marginalized limit 
$A=1.0045 \pm  0.0192$.

We see that the linear theory model becomes biased for $k>0.05~\hMpc$ very quickly,
and the bias grows with $\kmax$ quite steeply. For instance, at $\kmax=0.3~\hMpc$ the recovery of $\sigma_8$ is biased already by $10\%$, which 
significantly exceeds the precision of the current $2\times 2$pt analyses~\cite{DES:2026lhi}. 

We move on now to the one-loop model, 
\be 
\begin{split}
& P_{gg}(k) = A^2\,b_1^2\,\PL(k) +A^4 P^{\rm 1-loop}_{gg}(k) -2(b_1^2c_s+b_1b_{\nabla^2\delta})A^2k^2\PL(k) + \bar n^{-1}_g\left[
P_{\rm shot}+a_0\left(\frac{k}{k_{\rm stoch}}\right)^2
\right]\,,\\ 
& P_{gm}(k) = A^2\,b_1\,\PL(k) +A^4 P^{\rm 1-loop}_{gm}(k) -
(2b_1c_s+b_{\nabla^2\delta})A^2k^2\PL(k) + \bar n^{-1}_g
a'_{0}\left(\frac{k}{k_{\rm stoch}}\right)^2
\,,
\end{split}
\ee 
where $P^{\rm 1-loop}_{gg,gm}$ are the SPT contributions
computed for the PTC cosmology 
with \texttt{CLASS-PT}. 
We also implement the appropriate 
one-loop
IR resummation as in~\cite{Chudaykin:2020aoj}. 
The one-loop results are shown in Fig.~\ref{fig:1loop_fit}. 
In the left panel we see that the one-loop
model becomes biased around $\kmax\simeq 0.3~\hMpc$.
A more precise answer is $\kmax= 0.29~\hMpc$,
where we get $A=1.0028\pm  0.0101$. 
We observe an excellent agreement
between the best-fit 
theory and the data up to $\kmax\simeq 0.29~\hMpc$
in the right panel. 
We also observe now that the biasing 
of the $A$ inference is much more smooth than
in the case of the linear bias model, implying 
that the errors due to the omitted two-loop
contributions are now effectively absorbed by
the one-loop bias parameters. 
The $A(\kmax)$
drift curve shows a 
non-monotonic behavior for $\kmax= 0.6~\hMpc$, 
which is 
a consequence of the significant biasing of the one-loop theory model
on small scales. In particular, this leads to a jump 
around $\kmax=0.8~\hMpc$. 
This behavior is simply an artifact
of the one-loop model used much beyond its validity regime. 

Finally, 
Fig.~\ref{fig:2loop_fit} displays the results for the two-loop model summarized in Eq.~\eqref{eq:sum},
where we  re-scale the new corrections
with appropriate powers of $A$. 
We see now that the recovery of $A$ is unbiased
on all scales used in the analysis. We stop at $\kmax=0.85~\hMpc$ where 
we get $A = 1.0011\pm  0.0067$. 
This constraint is a factor of three better than the linear theory
measurement, and $\approx 40\%$ better than the 1-loop result.

Compared to the linear and one-loop results, 
we see a much more gradual narrowing 
of the error-bars as a function of $\kmax$,
which is a result of having many nuisance parameters in the fit. 
A somewhat fast 
narrowing of the $A$ error at $\kmax=0.85~\hMpc$
happens due to breaking of degeneracies between the fitting parameters
around this scale. 
The right panel of Fig.~\ref{fig:2loop_fit}
shows an excellent $\lesssim 0.5\%$ fit to the data. 
While in principle the two-loop model can fit the data even on smaller
scales, we stop the analysis at $\kmax=0.85~\hMpc$ because smaller scale data
is not available to us, and because the resolution
effects of the simulation might be important at our
precision level.  
It will be interesting to re-do our analysis
for other simulation data, especially galaxy
samples matching the ones used in actual imaging
surveys. 

\begin{figure}[htb!]
\centering
\includegraphics[width=0.99\textwidth]{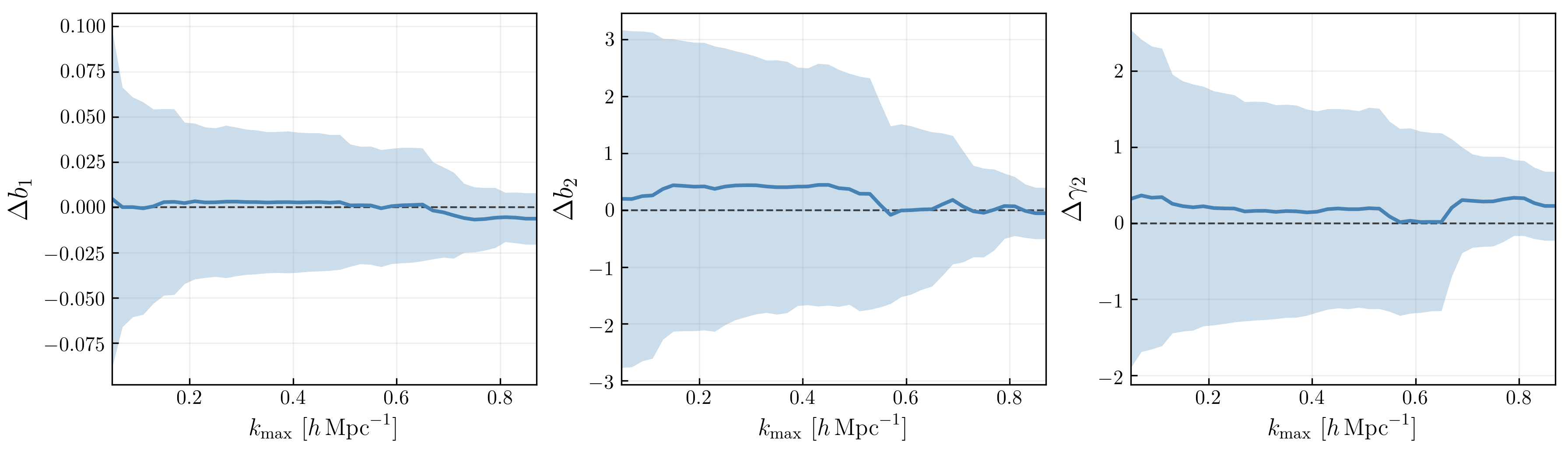}
\caption{The drift plot for the linear and quadratic bias parameters from the joint 
$P_{gm}-P_{gg}$ fit of PT Challenge data. The results are presented as $\Delta p = p-p^{P+B}$ shifts from the 
best-fit values $p^{P+B}$
from a joint one-loop power spectrum and bispectrum analysis of the  
PT Challenge data from~\cite{Bakx:2025pop}. }
\label{fig:bias_pull}
\end{figure}

\begin{figure}[htb!]
\centering
\includegraphics[width=0.7\textwidth]{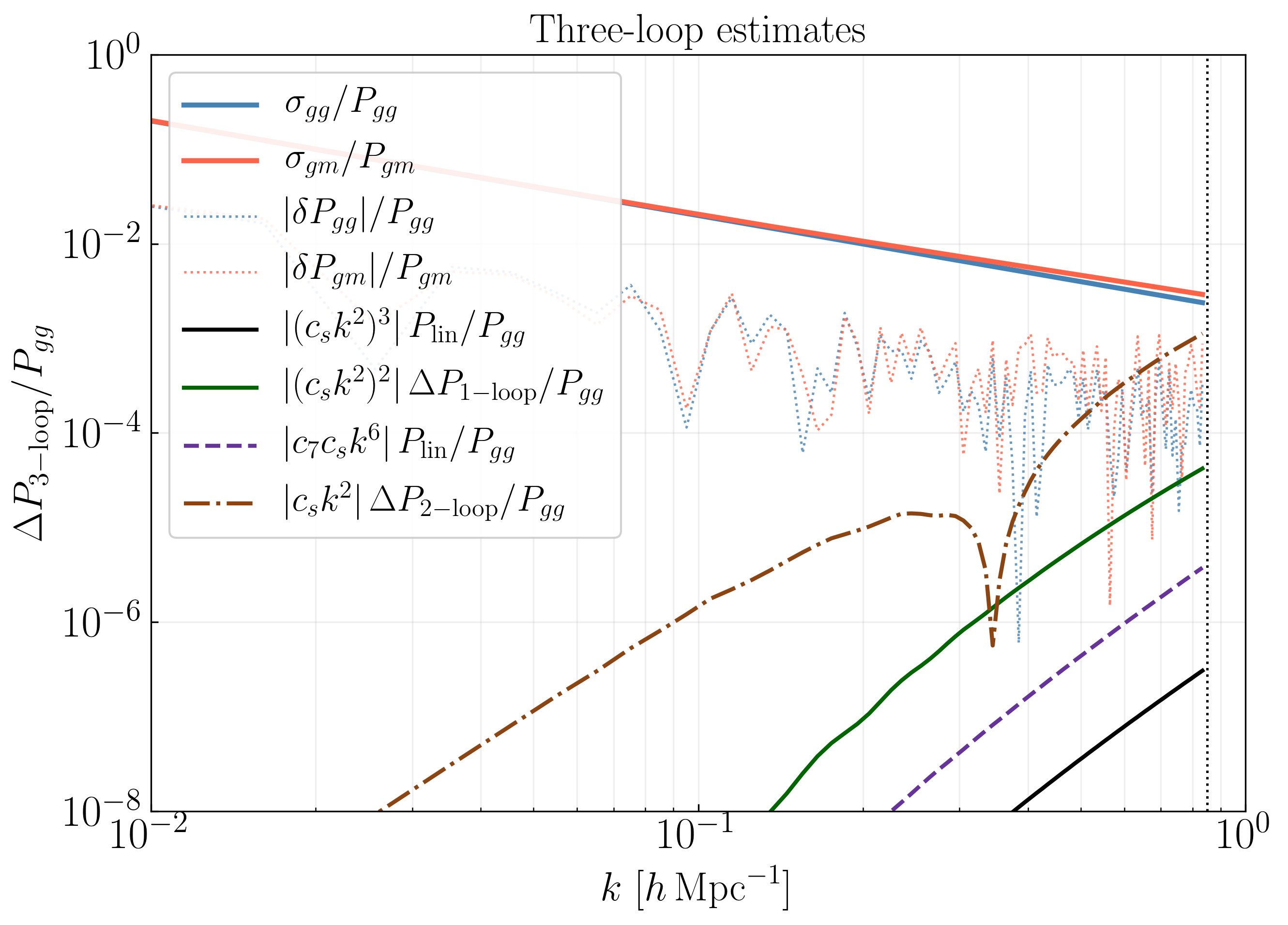}
\caption{Estimates of the three-loop effects for our two-loop 
best-fit at $k_{\rm max}=0.85~\hMpc$. We display the diagonal covariance errors $\sigma_{gg},\sigma_{gm}$ for galaxy-galaxy and galaxy-matter power spectra normalized to these spectra, for the fiducial survey volume $V=1~[h^{-1}\text{Gpc}]^3$, along with the actual residuals $|\delta P_{gg,gm}|/P_{gg,gm}$
which correspond to the effective volume of the 
PT Challenge simulation
$V=566~[h^{-1}\text{Gpc}]^3$, which illustrate that the three-loop
effects are negligible for the precision of both 
fiducial and the full simulation volume. We also show several estimates for the three-loop order diagrams, which can be 
extracted from our best-fit. These corrections are 
significantly smaller
than the fiducial errors at $k_{\rm max}=0.85~\hMpc$.
}
\label{fig:tree-loop}
\end{figure}

{Before closing this section, let us present some 
additional tests of our two-loop matching.
We have carried out several robustness checks of our $k_{\rm max}$ choice.
First, we have studied the drift of the bias parameters as a function of $k_{\rm max}$. 
Fig.~\ref{fig:bias_pull} demonstrates 
the results for the best-measured linear and quadratic bias parameters $b_1,b_2,\gamma_2$,
which we present as pulls from the redshift-space 
power spectrum and bispectrum results $p^{P+B}$ 
for the same simulation from Ref.~\cite{Bakx:2025pop} obtained at $k_{\rm max}^{P}=0.16~\hMpc$,
$k_{\rm max}^{B}=0.15~\hMpc$. 
We observe both 
the stability  of the inferred values w.r.t. $k_{\rm max}$, and an excellent consistency
between the independent measurements of these parameters from the small-scale real-space
galaxy power spectrum and the large-scale redshift-space galaxy 
power spectrum and bispectrum.}

{As a second test, we estimate the impact of the three-loop
effects based on the amplitudes of some of the diagrams 
which can be readily evaluated 
given our best-fit EFT parameters:\footnote{We use $P^{gm}_{\rm 1,2-loop}$ for galaxy-matter in these examples since their counting of bias parameters matches that of the 
corresponding three-loop diagrams, though the conclusions are
the same if we formally use $P^{gg}_{\rm 1,2-loop}$ instead.} 
$(c_sk^2)^3\PL$,  $(c_sk^2)^2 P_{\rm 1-loop}$,
$c_s c_7 k^6 \PL$ and $c_s k^2 P_{\rm 2-loop}$.
While these terms are all smaller than the two-loop power spectrum itself, 
the relevant parameter here is the magnitude
of these terms w.r.t. the data 
errorbars because 
if the experiment is very sensitive, the three-loop
effects can be important even at low $k$. 
We focus on two particular survey configurations
with the fiducial effective volume
$V=1~[h^{-1}\text{Gpc}]^3$
and the actual simulation volume 
$V=566~[h^{-1}\text{Gpc}]^3$. These two setups 
illustrate whether the three-loop effects are important 
for our fiducial experiment or the full simulation. 
Our results are demonstrated in Fig.~\ref{fig:tree-loop}.
We see that the three-loop effects are negligibly small
for the entire range of $k$ of interest both in the  case of the 
fiducial experiment and the full simulation volume.
While some of these corrections become comparable 
to $1\sigma$ errors of the full simulation 
volume around $\kmax=0.85~\hMpc$, 
the small range of these terms and the stability of our 
fits 
suggest that their effect can be absorbed by the nuisance parameters.
This
suggests that our key findings,
formulated for the realistic fiducial experiment,
are robust with respect to the three-loop effects.}

{Let us also comment on the relation between $\kmax=0.85~\hMpc$ at $z=0.61$ and the non-linear scale
$\knl\simeq 0.45\,[D(z)]^{-2/(n+3)}~\hMpc$ of Section~\ref{sec:power}. The latter is defined by the condition
$\Delta^2_{\rm lin}(\knl)=1$ for a power-law approximation to the linear spectrum and only fixes the parametric
scaling of the loop corrections, $(k/\knl)^{(1+L)(3+n)}$. 
However, we stress that 
$\knl\simeq 0.45\,[D(z)]^{-2/(n+3)}~\hMpc$
is only an order-of-magnitude estimate. 
The actual size of each contribution is 
set by its numerical coefficient,
and in $\Lambda$CDM the loop corrections are known to be numerically smaller than this naive estimate, see e.g.~\cite{Baldauf:2015zga,Bakx:2025jwa,Chen:2026usz},
which suggests that the actual EFT cutoff $\Lambda_{\rm UV}$
is somewhat larger than $\knl$.
The relevant question for the data analysis is therefore not whether $k/\knl<1$, but 
whether the omitted three-loop terms are below
the error bars, which is what the tests above verify.}

\section{Implications for Weak Lensing Data and New Physics}
\label{sec:implications}

\begin{figure}[htb]
\centering
\includegraphics[width=0.6\textwidth]{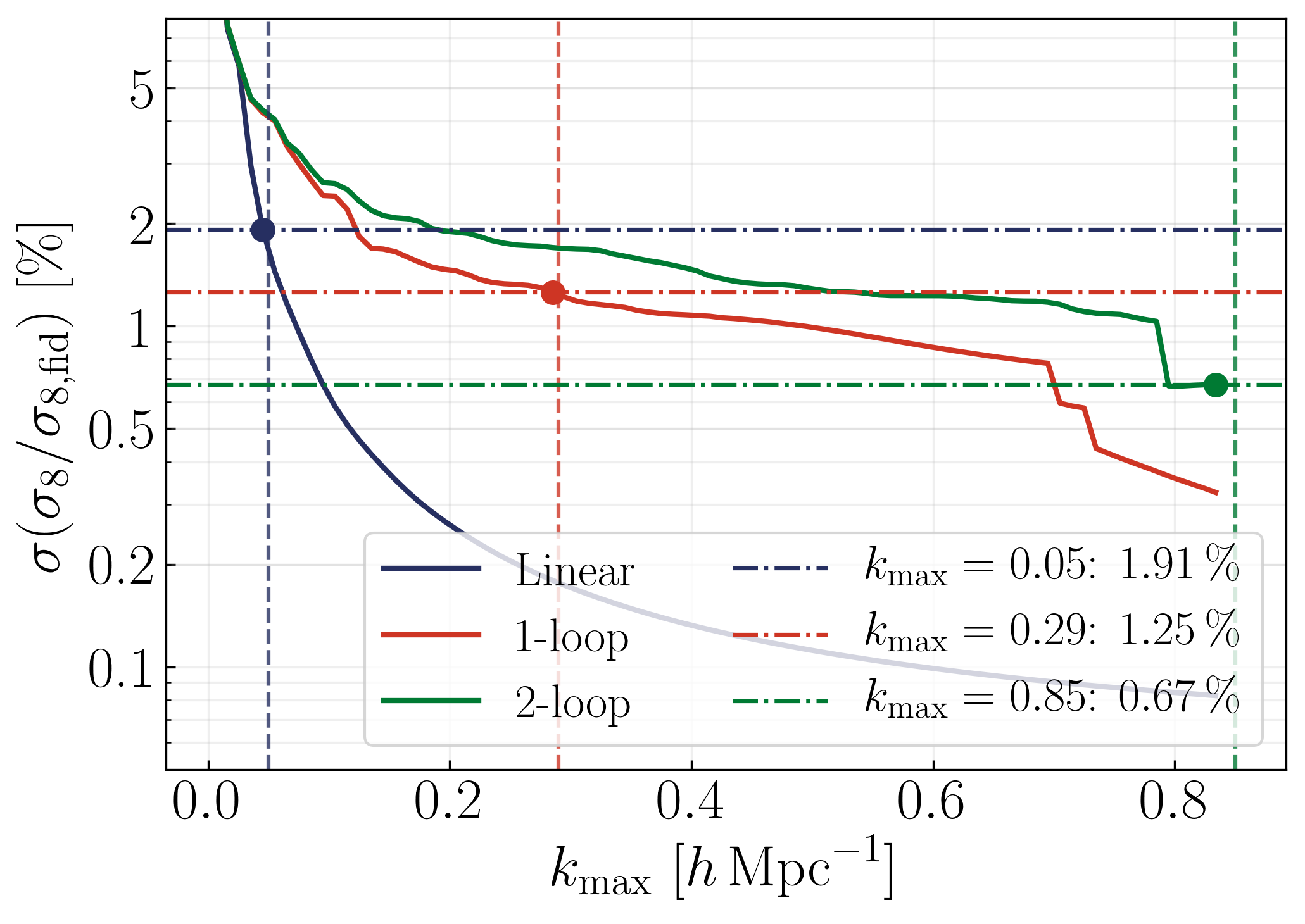}
\caption{Comparison of the Fisher forecast constraints on the 
mass fluctuation amplitude $\sigma_8$ as a function of scale cut for 
the linear, one-loop EFT, and two-loop EFT models. Vertical dashed lines 
show the limiting $\kmax$ values where each model is reliable. }
\label{fig:forecast}
\end{figure}

Having determined the reach of the galaxy power spectrum 
models in real space (at $z=0.61$ for the simulated LRG),
let us now discuss the implications for the weak lensing data. 
To keep our discussion more robust w.r.t. the numerical
artifacts due to the mis-modeling of the data,  
we will use a Fisher matrix forecast 
for $\sigma_8$ measurements 
for the $\{P_{gm},P_{gg}\}$
model vector, whose results are shown in Fig.~\ref{fig:forecast}.
This forecast is produced for the best-fit two-loop
galaxy power spectrum model at $\kmax=0.85~\hMpc$.
Our Fisher forecast results are in perfect agreement
with the analysis of the actual data. Dots in~Fig.~\ref{fig:forecast}
correspond to values at optimal $\kmax$ where the 
corresponding models are unbiased. 

Many current $2\times 2$pt imaging data 
analyses are
based on the 
linear bias model, see e.g.~\cite{DES:2026lhi}. 
This model has an advantage 
of having very few fitting 
parameters ($b_1$ and optionally $P_{\rm shot}$). 
In addition, the error on $\sigma_8$ shrinks very quickly with
$\kmax$, which increases the constraining power quite significantly if one 
were to naively push the analysis to small scales. 
However, we have seen that the model becomes 
biased very quickly as one increases $\kmax$. Requiring the 
unbiased recovery of $\sigma_8$ selects the scale cut $\kmax=0.05~\hMpc$, which leads to a relatively poor $\approx 2\%$ determination of $\sigma_8$. 

In contrast, the one-loop EFT model leads to a much slower
improvement of $\sigma_8$ constraints as a function of $\kmax$
due to a large number of EFT parameters. Nevertheless, we get 
a $1.25\%$ measurement at $\kmax=0.29~\hMpc$, which is about a factor of 2 better than the linear theory result. This illustrates that adding more fitting parameters (the one-loop model has 9) and
pushing the analysis to small scales is actually more beneficial
for the cosmological constraints than 
using a more minimal
model restricted to linear scales only. 

Finally, the two-loop model produces 
a relatively slow
$\sigma(\sigma_8/\sigma_{8,\mathrm{fid}})(\kmax)$ curve. 
This model has 37 nuisance parameters,
but it leads to more significant 
eventual gains than the linear theory
or the one-loop EFT models. 
Note that our Fisher forecast reproduces a jump of the 
$\sigma(\sigma_8/\sigma_{8,\mathrm{fid}})(\kmax)$ curve at $\kmax=0.85~\hMpc$
identical to the jump we have seen before in the actual data. This confirms that 
it is not an artifact of the mis-modeling, 
but a genuine effect of
degeneracy breaking. 

Finally, we get a $0.67\%$ error on $\sigma_8$
at $\kmax=0.85~\hMpc$.
This is a factor 
of three and $\approx 40\%$ more precise than the linear theory and one-loop results, respectively.
Note that our simulated galaxies are more biased
than the actual imaging galaxy lens samples.
Less biased tracers are typically better modeled in perturbation
theory (see e.g.~\cite{Ivanov:2021zmi,Ivanov:2024dgv,Sullivan:2025eei,deBelsunce:2026tks}), which suggests an even larger gain for realistic samples. 

{A comment on the treatment of free parameters is in order. Realistic weak lensing analyses are based on multiple galaxy samples, which would inevitably increase the number of 
free parameters. For instance, the DES-Y6 analysis~\cite{DES:2026fyc} is based on four tomographic bins, which naively increases the number of free parameters to $37\times 4=148$. Since the majority of these parameters enter the theory model linearly (27 per sample), they can be analytically marginalized over. There are additional assumptions that can be used to deal with the remaining 10 parameters per sample. For instance, they can be constrained by
priors from halo-occupation distribution or hydrodynamical simulations~\cite{Ivanov:2024hgq,Ivanov:2024xgb,Ivanov:2024dgv,Shiferaw:2024ehr,Chudaykin:2026nls}, or the semi-analytic halo-based models~\cite{Desjacques:2016bnm,Ivanov:2025qie}. In practice, one can linearize the theory model around the simulation-based priors and then analytically marginalize the likelihood over them. This approach will allow 
one to reduce the number of explicitly 
sampled free parameters to only a few, e.g. $\{b_1,b_2,\gamma_2\}$
per sample. Alternatively, one can use model compression techniques 
along the lines of~\cite{Philcox:2020zyp,Karacayli:2026wtw}.   }

{As far as the dark matter parameters are concerned, 
one can implement their joint fit to the $3\times 2$pt data
that includes the lensing power spectrum probing dark matter. 
Since the lensing power spectrum has a broad redshift kernel, 
a joint fit will require a model for the time-evolution
of the dark matter counterterms. In practice, it can be captured
by a linear spline with a few nodes, see e.g.~\cite{Chen:2026usz}.
The same spline approach can be used to model the time-evolution
of the galaxy bias parameters, in which case the model input 
can be the values of the bias parameters 
at the reference nodes.
}

\begin{figure}[htb!]
\centering
\includegraphics[width=0.7\textwidth]{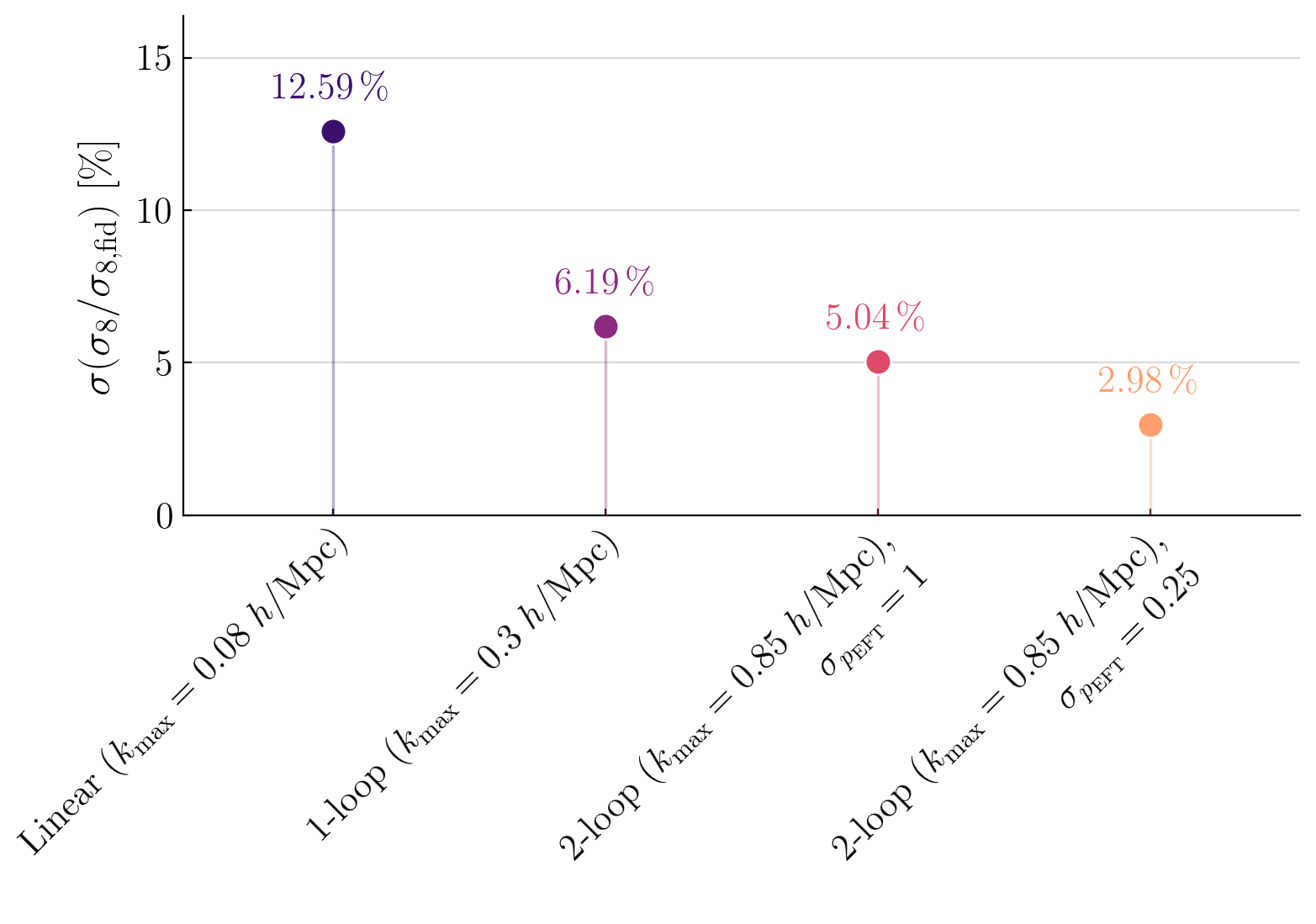}
\caption{Forecast error on the clustering amplitude
$\sigma_8/\sigma_{8,\rm fid}$ from the $2\times 2$pt analysis 
of $C_\ell^{gg}+C_\ell^{\kappa g}$ data from DESI spectroscopic galaxies and CMB lensing from \textit{Planck} PR4 and ACT DR6 
for four analysis configurations. 
The scale
cut $k_{\rm max}$ is applied through $\ell_{\rm max}=k_{\rm max}\chi(z_{\rm eff})-1/2$
per redshift bin, and each case is marginalized over the full set of bias, counterterm, and
stochastic parameters of the corresponding model. $\sigma_{p_{\rm EFT}}$ is the width of
the Gaussian prior on the two-loop EFT nuisance parameters.}
\label{fig:desi}
\end{figure}

{To estimate improvements for projected clustering 
experiments in a more realistic setup, we run a Fisher 
matrix forecast for the measurements of the clustering
amplitude $\sigma_8$ from the projected 
clustering of DESI DR1 spectroscopic galaxies
used in DESI full-shape analyses~\citep{DESI:2024aax,DESI:2024jis}
and their cross-correlations with \textit{Planck} PR4~\cite{Carron22}+ACT DR6 CMB lensing maps~\cite{Qu24,MacCrann24,Madhavacheril24}.
Our Fisher code is based on the pipeline of Ref.~\cite{Ivanov:2026dvl} which extracted cosmological information
from the $2\times 2$pt data 
of BGS, LRG1, LRG2, and LRG3 spectroscopic galaxy catalogs 
of DESI DR1
and their cross-correlations with CMB lensing. 
We use the actual galaxy distributions, window functions,
and covariance matrices from the data analysis of Refs.~\cite{Sailer:2024jrx,Maus:2024dzi,Ivanov:2026dvl}. We consider the following configurations: the linear theory analysis with flat $\kmax=0.08~\hMpc$, the one-loop EFT analysis with $\kmax=0.3~\hMpc$,
using the actual data analysis pipeline from Ref.~\cite{Ivanov:2026dvl}, and the two-loop EFT model with $\kmax=0.85~\hMpc$
and two versions of the Gaussian prior on EFT parameters: conservative, with the flat standard deviations $\sigma_{p_{\rm EFT}}=1$,
and optimistic, with $\sigma_{p_{\rm EFT}}=0.25$, which 
is designed to model the addition of simulation-based 
priors. In all analyses we use flat redshift-independent $\kmax$ for all four 
galaxy samples, and compute the angular multipole scale cut as
$\ell_{\rm max}=\chi(z_{\rm eff})k_{\rm max}-1/2$,
where $\chi$ is the comoving distance and $z_{\rm eff}$
is the effective redshift of a galaxy sample. 
The two-loop analyses include a full marginalization
over 148 nuisance parameters.
The technical details of the forecast are presented 
in Appendix~\ref{app:desi_forecast}; the results are displayed
in Fig.~\ref{fig:desi}. There we see that similarly to our 3D analysis,
the linear theory gives highly sub-optimal constraints 
with $\sigma(\sigma_8/\sigma_{8,\mathrm{fid}})\simeq 13\%$,
which improve by about a factor of two when the model
is upgraded to one-loop. The two-loop model improves over the 
one-loop one by $\approx 20\%$ in the conservative analysis,
and by a factor of two in the optimistic analysis. Our conservative
results are somewhat worse than the results of the above 3D forecast, but nevertheless the improvement
is quite sizeable. 
At the same time, the optimistic 
forecast with the narrow priors on EFT parameters achieves the same significant $\simeq 50\%$ 
improvements as the 3D case. All in all, 
we see that the two-loop
computation returns a better measurement of $\sigma_8$
than the current analysis pipelines for both considered configurations, which motivates its application to data.}

An even more important point than a better measurement of $\sigma_8$
in $\Lambda$CDM is that the two-loop EFT calculation provides a robust 
theoretical control over a much larger range of scales than the linear theory does. 
This is especially
important for the new physics models, which in principle, 
can be analytically implemented in the two-loop EFT computations.
For instance, the previous BOSS results on the ultralight axions showed
that the masses $m_a= 10^{-25}$ eV can be probed with the one-loop
EFT model thanks to an extension of $\kmax$ to $0.4~\hMpc$~\cite{Rogers:2023ezo}. 
Our two-loop EFT calculations can be pushed to 
$\kmax=0.85~\hMpc$, which 
allows us to probe larger masses $m_a\sim 
10^{-24}$ eV, which are currently only weakly constrained with 
the DES weak lensing data~\cite{Dentler:2021zij,Preston:2025tyl} (see also 
the recent limits from CMB lensing~\cite{Lague:2026sbd}). We leave the analysis
of this and other new physics models for future work.

\section{Conclusions}
\label{sec:conclusions}

We have computed the real 
space galaxy power spectrum
in cosmological perturbation theory 
(effective field theory)
at two-loop order. We have derived 
fifth order bias operators using a hybrid scheme
which classifies all operators into Eulerian 
local evolution operators and the Lagrangian non-local evolution operators. We have shown that there are 29 distinct bias operators 
relevant at {fifth order}, in agreement
with other fifth order bias studies~\cite{Bakx:2025cvu}.
We have derived 
the fifth order bias and cubic order 
{higher-derivative counterterm}
renormalization
conditions, 
which are mathematically equivalent to the 
renormalization conditions for the one-loop galaxy trispectrum. 
In passing, we 
recovered the one-loop power spectrum and one-loop bispectrum
renormalization conditions, which is an important consistency check of our new two-loop computation. 

We have shown that only 17 operators produce non-redundant contributions
to the galaxy power spectrum at the two-loop order. 
We have computed the relevant two-loop shapes numerically using the 
Bogoliubov-Parasiuk-Hepp-Zimmermann renormalization scheme, 
which allows us to connect our two-loop results with the common one-loop 
computations such as the ones used in the \texttt{CLASS-PT} code.
We have then derived the relevant two-loop counterterms and stochastic 
contributions. 

We have shown that the two-loop galaxy power spectrum 
allows one to model the $z=0.61$ 
galaxy clustering data from the PT Challenge
N-body simulation to $\kmax=0.85~\hMpc$. 
Assuming conservative physically 
motivated priors for EFT
parameters, this leads to a
significantly more
precise inference of the mass fluctuation amplitude $\sigma_8$,
which is a factor of three better than the linear theory result.
We stress that this improvement happens when all 37 relevant EFT parameters of 
dark matter and galaxies
are varied within conservative physically-motivated
priors. {We have also carried out a Fisher
forecast for the realistic $2\times 2$pt analysis 
of the projected DESI galaxy clustering and CMB lensing cross-correlations, and found $(20-50)\%$ improvements there as well, depending on the assumed priors on EFT parameters.}
Extrapolating this to the projected galaxy clustering and galaxy--galaxy lensing data suggests a factor of three improvement
of limits on the lensing parameter $S_8$, and also novel opportunities to 
test new physics that modifies the shape of the matter power
spectrum on scales $k\lesssim (0.4-0.8)~\hMpc$, such as the 
ultralight axion dark matter sub-component with masses $m_a\sim 10^{-24}$~eV. At face value, this implies exciting new opportunities 
for galaxy clustering and weak lensing data from DES, LSST and
Roman surveys.
It will be important to study the implications of the two-loop
computation both for $\Lambda$CDM and beyond it in more
realistic analyses of 
the projected clustering data.

Our analysis can be extended in multiple other ways. First, one can 
implement a fast evaluation of the two-loop integrals, e.g. by means 
of the COBRA technique~\cite{Bakx:2024zgu,Bakx:2025cvu,Bakx:2025jwa}, which will allow us to carry out inference of cosmological
parameters beyond $\sigma_8$.
Second, our computation can be extended to redshift space,
though the field-level results of Refs.~\cite{Schmittfull:2020trd,Ivanov:2024xgb,Sullivan:2025eei} 
suggest that the 
actual bottleneck there is the modeling of the stochastic
(one-halo) contribution, which becomes non-perturbative
for wavenumbers $\sim 0.3~\hMpc$.
Since the EFT is really 
useful only to model the deterministic part, it remains 
unclear if the two-loop computation in redshift space will be useful
for the actual data analysis.  
The two-loop computations, however, may improve the modeling 
of the Lyman-$\alpha$ forest, where the EFT stochasticity 
is smaller than that of galaxies~\cite{Ivanov:2023yla,Ivanov:2024jtl,deBelsunce:2024rvv,deBelsunce:2025bqc,deBelsunce:2026tks}.
Third, the fifth order EFT computation we carried out here 
can be used to compute the one-loop galaxy trispectrum in EFT, 
extending the previous matter clustering results~\cite{Bertolini:2016bmt}. {As a first step in this direction, we have computed here the one-loop galaxy trispectrum renormalization conditions 
for the galaxy bias and higher derivative operators.}

We leave these and other research directions for future investigation. 

\acknowledgements
\noindent
We thank Takahiro Nishimichi for providing the PT Challenge galaxy and matter power spectrum measurements. This work was completed at the Aspen
Center for Physics, which is supported by National 
Science Foundation grant PHY-2210452. 

\appendix

\section{Matter Power Spectrum at Two-Loop Order}\label{sec:2lmatter}

In this Appendix we review the two-loop matter power spectrum
and show that its counterterm contribution 
can be expressed in terms of the master integrals
presented in Section~\ref{sec:HD}.
Let us start with the general 2-loop matter power spectrum expression from~\cite{Anastasiou:2025jsy}:
\be\label{eq:P2loop_full}
P(k) = \PL(k) + \underbrace{P_{13}+P_{22}+P^{\rm ctr}_{13}}_{\text{one loop}}
+ \underbrace{P_{15}+P_{24}+P_{33}
+ P^{\rm ctr}_{15}+P^{\rm ctr}_{24}+P^{\rm ctr}_{33}}_{\text{two loops}}
\;+\;\text{stoch.}\,,
\ee
with the standard SPT diagrams
($P_{13}=6\PL(k)\int_{\q}F_3(\k,\q,-\q)\PL(q)$,
$P_{22}=2\int_{\q}F_2^2(\q,\k{-}\q)\PL(q)\PL(|\k{-}\q|)$, etc.) and the counterterm diagrams
\be\label{eq:anast_ctr_diagrams}
\begin{split}
P^{\rm ctr}_{13}(k) &= 2\,F_{{\rm ct},1}(k)\,\PL(k)\,,\\
P^{\rm ctr}_{15}(k) &= 6\,\PL(k)\!\int_{\q} F_{{\rm ct},3}(\k,\q,-\q)\,\PL(q)
 \;+\; 2\,F_{\partial^2{\rm ct},1}(k)\,\PL(k)\,,\\
P^{\rm ctr}_{33}(k) &= 6\,F_{{\rm ct},1}(k)\,\PL(k)\!\int_{\q}
 F_{3}(\k,\q,-\q)\,\PL(q) \;+\; \bigl[F_{{\rm ct},1}(k)\bigr]^2\PL(k)\,,\\
P^{\rm ctr}_{24}(k) &= 4\!\int_{\q} F_{{\rm ct},2}(\k{-}\q,\q)\,
 F_{2}(\k{-}\q,\q)\,\PL(q)\,\PL(|\k{-}\q|)\,.
\end{split}
\ee
The counterterm kernels are defined as
\be\label{eq:anast_ctr_kernels}
F_{{\rm ct},1}(k) = -\,c_{\delta,1}\frac{k^2}{9}\,,\qquad
F_{\partial^2{\rm ct},1}(k) = c_{\partial^2\delta,1}\frac{k^4}{26}\,,
\ee
where we absorbed all factors of $k_{\rm NL}^2$ into the counterterms making them dimensionful, and 
\be
4\, F_{{\rm ct},2}(\k{-}\q,\q) = \sum_{i=1}^{4}
\mathcal{B}_i\,
 e^{(i)}_{{\rm ct},2}(\k,\q)\,,\qquad
6\, F_{{\rm ct},3}(\k,\q,-\q) = \sum_{i=1}^{7}
\mathcal{A}_i\,
 e^{(i)}_{{\rm ct},3}(\k,\q)\,,
\ee
with $(\mathcal{B}_1,\dots,\mathcal{B}_4)=(c_{\delta,1},\,c_{r\delta,1},\,
c_{\delta^2,1},\,c_{\delta,2})$,
$(\mathcal{A}_1,\dots,\mathcal{A}_7)=(c_{\delta,1},\,c_{\delta,2},\,c_{\delta,3},\,
c_{r\delta,1},\,c_{r\delta,2},\,c_{\delta^2,1},\,c_{r\delta^2,1})$ in the
gauge $c_{r,1}=c_{r,2}=c_{r^2,1}=0$.  
The four
quadratic kernels are
\be\label{eq:ect2_kernels}
\begin{split}
e^{(1)}_{{\rm ct},2} &= \frac{-24(\k\cdot\q)^3
 + 8(\k\cdot\q)^2\,(7k^2+3q^2) - 2(\k\cdot\q)\,(11k^4+32k^2q^2)
 + 10\,k^2q^2(k^2+2q^2)}{99\,q^2\,|\k-\q|^2}\,,\\[4pt]
e^{(2)}_{{\rm ct},2} &= \frac{8(\k\cdot\q)^3 - 4(\k\cdot\q)^2(k^2+2q^2)
 + 8(\k\cdot\q)\,k^2q^2 - 4k^4q^2}{33\,q^2\,|\k-\q|^2}\,,\\[4pt]
e^{(3)}_{{\rm ct},2} &= -\frac{8k^2}{33}\,,\qquad
e^{(4)}_{{\rm ct},2} = -\frac{8k^2\bigl(2(\k\cdot\q)^2
 - 14(\k\cdot\q)q^2 + 5k^2q^2 + 7q^4\bigr)}{231\,q^2\,|\k-\q|^2}\,,
\end{split}
\ee
while the seven cubic kernels
$e^{(i)}_{{\rm ct},3}$ 
are given by:
\begin{subequations}\label{eq:ect3_kernels}
\begin{align}
e^{(1)}_{{\rm ct},3} &= \frac{1}{9009\,k^{2}\,q^{4}\,|\k-\q|^{2}\,|\k+\q|^{2}}
  \Bigl[\bigl(-3976 k^{4} - 7144 k^{2} q^{2} - 880 q^{4}\bigr)(\k\cdot\q)^{4}
  \nonumber\\
&\quad
  + \bigl(1001 k^{8} + 4039 k^{6} q^{2} + 4135 k^{4} q^{4} + 2813 k^{2} q^{6}\bigr)(\k\cdot\q)^{2}
  - 63 k^{8} q^{4} + 6 k^{6} q^{6} + 69 k^{4} q^{8}\Bigr]\,,\\[6pt]
e^{(2)}_{{\rm ct},3} &= \frac{88\,(\k\cdot\q)^{4} + 340 k^{2} q^{2}\,(\k\cdot\q)^{2} + 300 k^{4} q^{4}}{3003\,k^{2}\,q^{4}}\,,\\[6pt]
e^{(3)}_{{\rm ct},3} &= \frac{\bigl(4 k^{2} + 100 q^{2}\bigr)(\k\cdot\q)^{4} - \bigl( 32 k^{4} q^{2} + 100 k^{2} q^{4} - 24 q^{6}\bigr)(\k\cdot\q)^{2} -  35 k^{6} q^{4} - 74 k^{4} q^{6} - 39 k^{2} q^{8}}{273\,q^{4}\,|\k-\q|^{2}\,|\k+\q|^{2}}\,,\\[6pt]
e^{(4)}_{{\rm ct},3} &= \frac{\bigl(- 4 k^{4} - 906 k^{2} q^{2} + 22 q^{4}\bigr)(\k\cdot\q)^{4} + \bigl(226 k^{6} q^{2} + 102 k^{4} q^{4} + 200 k^{2} q^{6}\bigr)(\k\cdot\q)^{2} + 90 k^{8} q^{4} + 180 k^{6} q^{6} + 90 k^{4} q^{8}}{1287\,k^{2}\,q^{4}\,|\k-\q|^{2}\,|\k+\q|^{2}}\,,\\[6pt]
e^{(5)}_{{\rm ct},3} &= \frac{16\,(\k\cdot\q)^{6} - \bigl( 4 k^{4} + 86 k^{2} q^{2} + 6 q^{4}\bigr)(\k\cdot\q)^{4} - \bigl( 21 k^{6} q^{2} + 6 k^{4} q^{4} - 41 k^{2} q^{6}\bigr)(\k\cdot\q)^{2} -  17 k^{8} q^{4} - 24 k^{6} q^{6} - 7 k^{4} q^{8}}{91\,k^{2}\,q^{4}\,|\k-\q|^{2}\,|\k+\q|^{2}}\,,\\[6pt]
e^{(6)}_{{\rm ct},3} &= \frac{44\,(\k\cdot\q)^{2} + 60 k^{2} q^{2}}{429\,q^{2}}\,,\\[6pt]
e^{(7)}_{{\rm ct},3} &= \frac{-2\,(\k\cdot\q)^{2} -  k^{2} q^{2}}{13\,q^{2}}\,.
\end{align}
\end{subequations}
Matching the single-hard limits of $P_{24}$ and $P_{15}$ produces the 
following renormalization conditions for the ``infinite'' parts of the counterterms:
\be\label{eq:F3ctr_conditions}
\begin{split}
&c_{\delta,1} = -\frac{61}{70}\, \sigma^2_{-2}\,,\qquad
c_{\delta,2} = -\frac{12409}{5880}\, \sigma^2_{-2}\,,\qquad
c_{\delta,3} = -\frac{167099}{64680}\, \sigma^2_{-2}\,,\\
&c_{r\delta,1} = -\frac{6997}{3430}\, \sigma^2_{-2}\,,\qquad
c_{r\delta,2} = -\frac{27241}{44100}\, \sigma^2_{-2}\,,\qquad
c_{\delta^2,1} = \frac{63149}{41160}\, \sigma^2_{-2}\,,\\
&c_{r\delta^2,1} = \frac{60145999}{169785000}\,\sigma^2_{-2}\,,
\qquad\qquad
\sigma^2_{-2} \,\equiv\, \!\int_{\q}\frac{\PL(q)}{q^{2}}\,.
\end{split}
\ee
Note that the sign of $c_{\delta^2,1}$ corrects the one printed
in~\cite{Anastasiou:2025jsy}, which is additionally verified
by the trispectrum renormalization conditions.

The kernels above can be mapped onto ours.
Using Eq.~\eqref{eq:Endef}, we find
\be\label{eq:ect2_decomposition}
\begin{split}
e^{(1)}_{{\rm ct},2} &= -\tfrac{4}{9}\,k^2F_2
 + \tfrac{32}{99}\,E_1 + \tfrac{8}{63}\,E_2 + \tfrac{8}{33}\,E_3\,,\qquad
e^{(2)}_{{\rm ct},2} = -\tfrac{4}{33}\,E_1 - \tfrac{8}{33}\,E_3\,,\\
e^{(3)}_{{\rm ct},2} &= -\tfrac{8}{33}\,E_1\,,\qquad\qquad\qquad\qquad\quad\;\;
e^{(4)}_{{\rm ct},2} = -\tfrac{8}{33}\,E_1 - \tfrac{16}{231}\,E_2\,,
\end{split}
\ee
allowing us to rewrite $P^{\rm ctr}_{24}$ as
\be\label{eq:P24ctr_general_note}
P^{\rm ctr}_{24}(k)
= -\frac{2c_{\delta,1}}{9}\,k^2P_{22}^{\rm SPT}
+ Q_{\delta^2}\,k^2\Ids + Q_{\Gal_2}\,k^2\IG
+ Q_{E_3}\,\mathcal J_{F_2}^{E_3}\,,
\ee
with the identification
\begin{align}
Q_{\delta^2}
&=
\frac{16}{99}c_{\delta,1}
-\frac{2}{33}c_{r\delta,1}
-\frac{4}{33}c_{\delta^2,1}
-\frac{4}{33}c_{\delta,2},
\label{eq:Qd2}\\
Q_{\Gal_2}
&=
\frac{4}{63}c_{\delta,1}
-\frac{8}{231}c_{\delta,2},
\label{eq:QG2}\\
Q_{E_3}
&=
\frac{4}{33}
\left(c_{\delta,1}-c_{r\delta,1}\right)
\label{eq:QE3}\,.
\end{align}
The $P_{13}$-like integral simplifies upon the angular integration of
the cubic kernels.  
However, there is one conceptual point that emerges there. 
We expect that some of these kernels can be recast into $k^2P^{\rm SPT}_{13}$.
However, bare $k^2P^{\rm SPT}_{13}$
template is IR-singular, which should produce a $k^4\sigma_v^2\PL$ term (with
$\sigma_v^2\equiv\int_{|\q|\leq \Lambda_{\rm IR}} \PL(q)/q^2$) being the 
equivalence-principle-violating IR variance of the displacement field. 
$\sigma_v^2$ should not be confused with $\sigma^2_{-2}$
because the former is integrated over the IR modes, while the latter is an 
integral over the UV modes.
As we shall see, the two-loop counterterm contribution 
will involve both $k^2P^{\rm SPT}_{13}$ and its IR-finite
part.   
We therefore define the IR-subtracted template
\be\label{eq:Pbar13}
k^2\bar P_{13}(k) \equiv \PL(k)\!\int_{\q}
\Bigl[6k^2F_3(\k,\q,-\q) + \frac{k^4}{3q^2}\Bigr]\PL(q)
 \;=\; k^2P^{\rm SPT}_{13}(k) + \frac{k^4}{3}\,\sigma_v^2\,\PL(k)\,,
\ee
which isolates the offending IR contribution. We will show momentarily
that it is precisely cancelled by the $k^2P^{\rm SPT}_{22}$
term from the quadratic counterterm.
Defining
\be\label{eq:Ii_def}
\mathcal{I}_i(k)\equiv
\PL(k)\int_{\vq}
e^{(i)}_{{\rm ct},3}(\k,\vq)\,\PL(q)\,,
\ee
the seven $P_{13}$-like integrals decompose exactly as:
\begin{align}
\mathcal{I}_1={}&
-\frac{40}{297}\,k^2\bar P_{13}
+\frac{2957}{57915}\,k^2\FG
-\frac{40}{819}\,\mathcal{F}^{E_4}_{\Gal_2}
+\frac{5308}{27027}\,\sigma_0^2\,k^2\PL
+\frac{1}{27}\,\sigma_v^2\,k^4\PL\,,
\label{eq:I1}\\
\mathcal{I}_3={}&
-\frac{1}{195}\,k^2\FG
-\frac{47}{273}\,\sigma_0^2\,k^2\PL\,,
\label{eq:I3}\\
\mathcal{I}_4={}&
\frac{7}{297}\,k^2\bar P_{13}
-\frac{931}{115830}\,k^2\FG
+\frac{1}{117}\,\mathcal{F}^{E_4}_{\Gal_2}
+\frac{688}{6435}\,\sigma_0^2\,k^2\PL\,,
\label{eq:I4}\\
\mathcal{I}_5={}&
-\frac{1}{26}\,k^2\FG
+\frac{5}{91}\,\mathcal{F}^{E_4}_{\Gal_2}
-\frac{452}{1365}\,\sigma_0^2\,k^2\PL\,,
\label{eq:I5}\\
\bigl(\mathcal{I}_2,\,\mathcal{I}_6,\,\mathcal{I}_7\bigr)
={}&
\Bigl(\tfrac{6464}{45045},\;\tfrac{224}{1287},\;-\tfrac{5}{39}\Bigr)\,
\sigma_0^2\,k^2\PL\,,
\label{eq:I267}
\end{align}
with $\sigma_0^2 \equiv \sigma^2=\int_{\q}\PL(q)$. 
The $k^2P_{13}$ kernel is recovered by the exact relation
\be\label{eq:k2F3_relation}
6\,k^2F_3(\k,\q,-\q) = -9\,e^{(1)}_{{\rm ct},3}
 - \tfrac{99}{10}\,e^{(2)}_{{\rm ct},3} - 26\,e^{(3)}_{{\rm ct},3}
 - 9\,e^{(4)}_{{\rm ct},3} - \tfrac{33}{5}\,e^{(5)}_{{\rm ct},3}
 - \tfrac{19293}{665}\,e^{(6)}_{{\rm ct},3}
 - \tfrac{13166}{665}\,e^{(7)}_{{\rm ct},3}\,.
\ee
The $k^2\PL$ pieces are absorbed into
the free finite part of $c_{\delta,1}$
in much the same way  as $\sigma^2$-dependent pieces
in the one-loop galaxy power spectrum
are absorbed into $b_1$. We stress that just like 
in the $b_1$ case this does not make the  $c_{\delta,1}$
parameter entering the tree-level and one-loop counterterms different. 
The difference between the ``old'' and ``new'' $c_{\delta,1}$
is a three-loop effect, which can be consistently 
neglected at the precision of our computation.
Also note that exactly one $k^4\sigma_v^2\PL$ term survives,
and it multiplies the advected
linear counterterm parameter $c_{\delta,1}$,
precisely as required to 
cancel the IR singularities
in the counterterm diagram sum 
($\tfrac{c_{\delta,1}}{27}$ from
$P^{\rm ctr}_{15}$, $\tfrac{c_{\delta,1}}{27}$ from $P^{\rm ctr}_{33}$,
$-\tfrac{2c_{\delta,1}}{27}$ from $P^{\rm ctr}_{24}$). 
It follows that the cubic matter
counterterm contribution can be expressed exactly in terms of
three non-trivial shapes plus two counterterm-degenerate local terms:
\begin{align}
6\PL(k)\int_{\vq}
F_{{\rm ct},3}(\k,\vq,-\vq)\,\PL(q)
={}&
C_{13}\,k^2\bar P_{13}
+C_{\FG}\,k^2\FG
+C_{E_4}\,\mathcal{F}^{E_4}_{\Gal_2}
+C_{\alpha}\,\sigma_0^2\,k^2\PL
+\frac{c_{\delta,1}}{27}\,\sigma_v^2\,k^4\PL\,,
\label{eq:F3ct_reduced}
\end{align}
where
\begin{align}
C_{13}={}&
\frac{-40c_{\delta,1}+7c_{r\delta,1}}{297}\,,
\label{eq:C13}\\
C_{\FG}={}&
\frac{
5914c_{\delta,1}
-594c_{\delta,3}
-931c_{r\delta,1}
-4455c_{r\delta,2}
}{115830}\,,
\quad \quad \quad 
C_{E_4}={}
\frac{
-40c_{\delta,1}
+7c_{r\delta,1}
+45c_{r\delta,2}
}{819}\,,
\label{eq:CE4}
\end{align}
and the unobservable renormalization of $c_{\delta,1}$
depends on
\begin{align}
C_\alpha 
= \frac{5308}{27027}c_{\delta,1}
+ \frac{6464}{45045}c_{\delta,2}
- \frac{47}{273}c_{\delta,3}
+ \frac{688}{6435}c_{r\delta,1}
- \frac{452}{1365}c_{r\delta,2}
+ \frac{224}{1287}c_{\delta^2,1}
- \frac{5}{39}c_{r\delta^2,1}\,.
\label{eq:Calpha}
\end{align}
The contributions proportional
to $c_{\delta,2}$, $c_{\delta^2,1}$, and $c_{r\delta^2,1}$ enter
only through $C_\alpha$: they are scaleless and vanish in
dimensional regularization.

Combining Eq.~\eqref{eq:P24ctr_general_note} above with the cubic
reduction, the sum
$P^{\rm ctr}_{15}+P^{\rm ctr}_{24}+P^{\rm ctr}_{33}$ takes the template
form, for arbitrary EFT coefficients,
\begin{align}
P_{\rm mm}^{\rm ct,\,2\text{-}loop}(k)
={}&
-\frac{2c_{\delta,1}}{9}\,
k^2P_{\rm 1\text{-}loop}^{\rm SPT}(k)
+\frac{7\,(c_{r\delta,1}-c_{\delta,1})}{297}\,k^2\bar P_{13}(k)
+Q_{\delta^2}\,k^2\Ids
+Q_{\Gal_2}\,k^2\IG
\nonumber\\
&\hspace{20mm}
+Q_{E_3}\,\mathcal J_{F_2}^{E_3}
+C_{\FG}\,k^2\FG
+C_{E_4}\,\mathcal F_{\Gal_2}^{E_4}
+
k^4\PL(k)
\left(
\frac{c_{\partial^2\delta,1}}{13}
+\frac{c_{\delta,1}^{\,2}}{81}
\right),
\label{eq:Pmm_ct_general}
\end{align}
where $Q_{\delta^2},Q_{\Gal_2},Q_{E_3}$
are given by 
Eqs.~(\ref{eq:Qd2},\ref{eq:QG2},\ref{eq:QE3}),
respectively. 
Using $c_{r\delta,1}-c_{\delta,1}
 = -\tfrac{33}{4}\,Q_{E_3}$, the complete matter counterterm 
can be recast as 
\be\label{eq:our_ctr_model}
\begin{split}
P^{\rm ctr}_{mm}(k) ={}& -2c_s\,k^2\bigl[\PL
 + P^{\rm SPT}_{\rm 1\text{-}loop}\bigr]
 + \bigl(c_s^2 + 2\tilde c_s\bigr)\,k^4\PL
 + 2e_1\,k^2\Ids + 2e_2\,k^2\IG\\
&+ 2e_3\Bigl[\mathcal{J}_{F_2}^{E_3}
 - \tfrac{7}{36}\,k^2\bar P_{13}\Bigr]
 + e_4\,k^2\FG + e_5\,\mathcal{F}^{E_4}_{\Gal_2}\,,
\end{split}
\ee
where $-2c_s k^2\PL$ is the one-loop counterterm and every other
term is two-loop order.
The $\bar P_{13}$ term descends
from a field-level cubic counterterm that can be rewritten as
\be\label{eq:FPbar13}
\delta_m \supset -\tfrac{7}{18}\,e_3\!
\int_{\vq_1\vq_2\vq_3}\!\!(2\pi)^3\delta_D(\k-\vq_{123})\,
F_{\bar P_{13}}(\vq_1,\vq_2,\vq_3)\,
\delta_L(\vq_1)\delta_L(\vq_2)\delta_L(\vq_3)\,,
\ee
whose power-spectrum configuration is $1/6$ of the integrand
of Eq.~\eqref{eq:Pbar13},
\be\label{eq:FPbar13_config}
F_{\bar P_{13}}(\k,\q,-\q)
 = \frac16\Bigl[6k^2F_3(\k,\q,-\q) + \frac{k^4}{3q^2}\Bigr]
 = k^2F_3(\k,\q,-\q) + \frac{k^4}{18\,q^2}\,,
\ee
which ensures $6\,\PL(k)\!\int_{\q}F_{\bar P_{13}}(\k,\q,-\q)\,\PL(q)
 = k^2\bar P_{13}(k)$.
Our analysis thus shows that the two-loop
counterterm can be viewed as originating from the field-level
expression~\eqref{eq:matter_ctr2}.

This field-level form fixes how the term propagates to the tracer
spectra.  Since $\delta_g \supset b_1\delta_m$ carries the matter
counterterms with coefficients fixed by the dark matter
renormalization conditions, and since $k^2\bar P_{13}$ is
not in the galaxy shape basis $P_i^{\rm HD}$, the galaxy-side
copy of this operator cannot be reabsorbed into the free galaxy
parameters (unlike the $E_n$, $\FG$, $\mathcal{F}^{E_4}_{\Gal_2}$ and
$k^4\PL$ pieces). 
Thus an independent $e_3$ template survives in every
power and cross
spectrum:
\be\label{eq:Pbar13_ladder}
\Delta P^{\rm ctr}_{mm} = -\tfrac{7}{18}\,e_3\,k^2\bar P_{13}\,,\qquad
\Delta P^{\rm ctr}_{gm} = -\tfrac{7}{18}\,e_3\,b_1\,k^2\bar P_{13}\,,\qquad
\Delta P^{\rm ctr}_{gg} = -\tfrac{7}{18}\,e_3\,b_1^2\,k^2\bar P_{13}\,,
\ee
where we also copied its contribution 
to $\Delta P^{\rm ctr}_{mm}$ for completeness.
The galaxy--matter counterterm, with no free parameters beyond those of
$P_{gg}$ and $P_{mm}$, gives eq.~\eqref{eq:Pgm_ctr_final}.

\section{Quadratic Displacement Counterterm}\label{app:Psi2ctr}

Here we derive 
explicitly the 
second order density, velocity,
and displacement counterterm
corrections. 
The continuity and Euler equations sourced by the EFT stress tensor
$\tau^{ij}$ are given by
\be\label{eq:fluid}
\partial_\eta\delta+\nabla_i\pi^i=0\,,
\qquad
\partial_\eta\pi^i+\tfrac12\,\pi^i+\tfrac32\,A^i
=-\,\nabla_j S^{ij}\,,
\ee
where $A^i\equiv\frac{\nabla_i}{\nabla^2}\,\delta$, 
$u^i= v^i_{\rm pec}/(aHf)$ is the normalized peculiar velocity,
and the momentum 
is $\pi^i=(1+\delta)u^i$. The total source is then:
\be\label{eq:Sdef}
S^{ij}\;\equiv\;\tfrac32\Bigl(A^iA^j-\tfrac12\delta^{K}_{ij}\,
A\!\cdot\!A\Bigr)\;+\;\frac{\pi^i\pi^j}{1+\delta}
\;+\;{\tau^{ij}}\,.
\ee
As shown in~\cite{Anastasiou:2025jsy}, the relevant solutions of the fluid equations for density and momentum counterterms 
scaling as $X^{(n)}_{\rm ct}\propto D^n$
can be written as
\be\label{eq:ctr_greens}
\delta^{(n)}_{\rm ct}=\frac{\alpha^{(n)}_S}{n+2}\,
\nabla_i\nabla_j S^{ij}_{(n)}\,,
\qquad
\pi^{i,(n)}_{\rm ct}=
-\,\alpha^{(n)}_S\,\frac{\nabla_i\nabla_j\nabla_k}{\nabla^2}S^{jk}_{(n)}
\;-\;\alpha^{(n)}_V\Bigl[\nabla_jS^{ij}_{(n)}
-\frac{\nabla_i\nabla_j\nabla_k}{\nabla^2}S^{jk}_{(n)}\Bigr]\,,
\ee
where the scalar and vector coefficients are given by
\be 
\alpha^{(n)}_S=\frac{2(n+2)}{(n+1)(2n+7)}\,,
\qquad
\alpha^{(n)}_V=\frac{2}{2n+5}\,.
\ee
If $\alpha^{(n)}_S=\alpha^{(n)}_V$, the r.h.s. in $\pi^i_{\rm ct}$ would collapse to a local gradient source.
If $\alpha^{(n)}_S-\alpha^{(n)}_V\neq 0$,
$\pi^i_{\rm ct}$ picks up a non-local
contribution.
Once $\delta_{\rm ct}$ and $\pi^i_{\rm ct}$ are known, one can reconstruct 
$u^i_{\rm ct}$ and then the displacement 
field through the identity:
\be\label{eq:Psi_exact}
\partial_\eta\vPsi(\vec q,\eta)=\vec{u}\bigl(\vec q+\vPsi(\vec q,\eta),\,\eta\bigr)
=\sum_{p\ge0}\frac{1}{p!}\,
\Psi^{j_1}\!\cdots\Psi^{j_p}\,
\nabla_{j_1}\!\cdots\nabla_{j_p}u^i(\vec q,\eta)\,,
\ee
where $\vec{q}$ is the Lagrangian coordinate. Indeed, the same equation can be used to relate the LPT and SPT kernels as 
\be\label{eq:Psi_recursion}
n\,\Psi^{i,(n)}=\sum_{p\ge0}\frac{1}{p!}
\Bigl[\Psi^{j_1}\!\cdots\Psi^{j_p}\nabla_{j_1}\!\cdots\nabla_{j_p}u^i
\Bigr]^{(n)}\,.
\ee
However, the counterterms of 
order $m$
have time-dependence $\propto D^{m+2}$,
so we get
\be 
(m+2)\,\Psi^{i,(m)}_{\rm ct}=\sum_{p\ge0}\frac{1}{p!}
\Bigl[\Psi^{j_1}\!\cdots\Psi^{j_p}\nabla_{j_1}\!\cdots\nabla_{j_p}u^i\Bigr]^{(m)}_{\tau}\,,
\ee
where the $\tau$ subscript means that
one insertion above should be that of a counterterm. Let us see how this works in practice. At first order we get
\be 
S^{ij}_{(1)}=\tau^{ij,(1)}_{\rm ct}
=\bigl[c_{\delta,1}\,\delta^{K}_{ij}
+c_{r,1}\,\nabla_i\nabla_j/\nabla^2\bigr]\delta^{(1)}\,,
\ee 
which gives:
\be\label{eq:ctr_first_order}
\begin{split}
& \delta^{(1)}_{\rm ct}=\frac{c_{\delta,1}+c_{r,1}}{9}\nabla^2\delta^{(1)}\,,
\quad
u^{i,(1)}_{\rm ct}=\pi^{i,(1)}_{\rm ct}
=-\frac{c_{\delta,1}+c_{r,1}}{3}\nabla_i\delta^{(1)}\,,\\
&
\Psi^{i,(1)}_{\rm ct}=\tfrac13 u^{i,(1)}_{\rm ct}\,,\, \quad 
\Phi^{(1)}_{v,\rm ct}=\frac{c_{\delta,1}+c_{r,1}}{3}\,\nabla^2\Phi^{(1)}_{v}\,.
\end{split}
\ee
At second order we find a non-zero
transverse part~\cite{Carrasco:2013mua,Mercolli:2013bsa} that stems 
from the source
\be\label{eq:Sij2}
\begin{split}
S^{ij}_{(2)}=\;&\tfrac32\Bigl[A^{i,(1)}_{\rm ct}A^{j,(1)}
+A^{i,(1)}A^{j,(1)}_{\rm ct}
-\delta^K_{ij}\,A^{(1)}_{\rm ct}\!\cdot\!A^{(1)}\Bigr]
\;+\;\pi^{i,(1)}_{\rm ct}\pi^{j,(1)}+\pi^{i,(1)}\pi^{j,(1)}_{\rm ct}\\
&+c_{\delta,2}\,\delta^K_{ij}\delta^{(2)}
+c_{r,2}\,t^{(2)}_{ij}
+c_{r^2,1}\,t^{(1)}_{il}t^{(1)}_{lj}
+c_{r\delta,1}\,t^{(1)}_{ij}\delta^{(1)}
+c_{\delta^2,1}\,\delta^K_{ij}(\delta^{(1)})^2\,,
\end{split}
\ee
with $A^{i,(1)}_{\rm ct}=\tfrac19(c_{\delta,1}+c_{r,1})
\nabla_i\delta^{(1)}$ from eq.~\eqref{eq:ctr_first_order} and
$t^{(n)}_{ij}$ the $n$-th order piece of $t_{ij}=\nabla_i\nabla_j\Phi=\nabla_i\nabla_j\delta/\nabla^2$. We obtain then:
\be\label{eq:u2ct}
{u}^i_{\rm ct}=\Bigl[\frac{{\pi}^i}{1+\delta}\Bigr]_{\tau}
=\frac{{\pi}^i_{\rm ct}}{1+\delta}
-\frac{\pi^i\delta_{\rm ct}}{(1+\delta)^2}
\;\;\Longrightarrow\;\;
u^{i,(1)}_{\rm ct}=\pi^{i,(1)}_{\rm ct}\,,\quad
u^{i,(2)}_{\rm ct}=\pi^{i,(2)}_{\rm ct}
-\delta^{(1)}u^{i,(1)}_{\rm ct}-\delta^{(1)}_{\rm ct}u^{i,(1)}\,,
\ee
with $\Phi^{(2)}_{v,\rm ct}=(\nabla_i/\nabla^2)\,u^{i,(2)}_{\rm ct}$.
Of the five quadratic stress-tensor structures in eq.~\eqref{eq:Sij2} only $c_{r^2,1}t^{(1)}_{il}t^{(1)}_{lj}$ 
survives the $\alpha_S-\alpha_V$ projector. Together with the 
non-transverse pieces from the linear counterterms they form
the non-local counterterm (NLC)~\cite{DAmico:2022ukl}:
\be\label{eq:u2NLC}
u^{i,(2)}_{\rm ct}\Big|_{\rm NLC}=-c_5^{\pi}\,e_5^i\,,\quad  
e_5^i\equiv\frac{\nabla_i\nabla_j\nabla_k}{\nabla^2}
\left(
\frac{\nabla_j\nabla_l}{\nabla^2}\delta 
\frac{\nabla_l\nabla_k}{\nabla^2}\delta\right)\,,
\ee
where $c_5^{\pi}$ is a free counterterm coefficient. Its value required for the trispectrum renormalization (discussed shortly) is
\be\label{eq:c5formula}
c_5^{\pi} \;=\; \frac{\alpha^{(2)}_S-\alpha^{(2)}_V}{\knl^{2}}
\bigl(2c_{r,1}-c_{r,2}+c_{\delta,1}-c_{r\delta,1}-c_{r^2,1}\bigr)\Big|_{c^{*}}
\;=\;\frac{668}{56595}\times2\sigma^2_{-2}\,.
\ee
The full displacement reads
\be\label{eq:Psict2}
4\,\vPsi^{(2)}_{\rm ct}=\vec{u}^{(2)}_{\rm ct}
+\bigl(\vPsi^{(1)}_{\rm ct}\!\cdot\!\boldsymbol{\nabla}\bigr)\vec{u}^{(1)}
+\bigl(\vPsi^{(1)}\!\cdot\!\boldsymbol{\nabla}\bigr)\vec{u}^{(1)}_{\rm ct}\,.
\ee
Collecting
eqs.~\eqref{eq:Sij2}, \eqref{eq:ctr_greens}, \eqref{eq:u2ct} and
\eqref{eq:Psict2} gives $\vPsi^{(2)}_{\rm ct}$ in closed form:
\be\label{eq:L2ct_def}
\vPsi^{(2)}_{\rm ct}=i\,D^{4}\!\int_{\q_1\q_2}\!(2\pi)^3\delta^{(3)}_D(\k-\q_{12})\,
\Vec{L}^{\rm ct}_2(\q_1,\q_2)\,\delta_0(\q_1)\delta_0(\q_2)\,,\quad \Vec{L}^{\rm ct}_2(\q_1,\q_2)
=\ell^{\rm ct}_2(\q_1,\q_2)\,\q_1+\ell^{\rm ct}_2(\q_2,\q_1)\,\q_2\,,
\ee
where the $\ell^{\rm ct}_2$ kernel is given by ($\k=\q_1+\q_2$):
\be\label{eq:Ascalar}
\begin{split}
\ell^{\rm ct}_2(\q_1,\q_2)=\;&
\frac{c_{\delta,1}+c_{r,1}}{9}
\Bigg[
\frac{3}{8}
+\frac{(\k\cdot\q_2)}{8q_1^2}
+\frac{3\,(\q_1\cdot\q_2)}{8q_2^2}
-\frac{q_1^2+q_2^2}{88q_1^2q_2^2}
\left(
11\,(\k\cdot\q_2)
-4\,(\q_1\cdot\q_2)
+\frac{2(\k\cdot\q_1)(\k\cdot\q_2)}{k^2}
\right)
\Bigg]
\\[2pt]
&-\frac{2}{33}
\Bigg[
\bigl(c_{\delta,2}+c_{r,2}\bigr)
F_2(\q_1,\q_2)
+c_{\delta^2,1}
\Bigg]
-c_{r^2,1}\,
\frac{(\q_1\cdot\q_2)(\k\cdot\q_2)}
     {36q_1^2q_2^2}
\left(
1+\frac{2\,(\k\cdot\q_1)}{11k^2}
\right) \\[2pt]
&
-c_{r\delta,1}
\Bigg[
\frac{(\k\cdot\q_1)}{36q_1^2}
+\frac{1}{396k^2}
\left(
\frac{(\k\cdot\q_1)^2}{q_1^2}
+\frac{(\k\cdot\q_2)^2}{q_2^2}
\right)
\Bigg]\,.
\end{split}
\ee
Inserting the trispectrum renormalization
conditions (the trispectrum point $c^{*}$ of
eq.~\eqref{eq:cstar_canonical} derived in the next section) for which
$c_{r,1}=c_{r,2}=0$, only four stress coefficients survive. 
They give the kernel that all the numerical renormalization conditions of
Sec.~\ref{sec:hdbias} are built from. In the $2\sigma^2_{-2}$ units used
for the $\beta$'s, 
\be\label{eq:Ascalar_cstar}
\begin{split}
\frac{\ell^{*}_2(\q_1,\q_2)}{2\sigma^2_{-2}}
=\;&
-\frac{61}{1260}
\left[
\frac{3}{8}
+\frac{(\k\cdot\q_2)}{8q_1^2}
+\frac{3\,(\q_1\cdot\q_2)}{8q_2^2}
\right]
+\frac{61}{110880}\,
\frac{q_1^2+q_2^2}{q_1^2q_2^2}
\left[
11\,(\k\cdot\q_2)
-4\,(\q_1\cdot\q_2)
+\frac{2(\k\cdot\q_1)(\k\cdot\q_2)}{k^2}
\right]
\\[2pt]
&+\frac{146978057}{1581620040}
F_2(\q_1,\q_2)
-\frac{167110717}{2214268056}
-\frac{22916149}{3019456440}\,
\frac{(\q_1\cdot\q_2)(\k\cdot\q_2)}
     {q_1^2q_2^2}
\left[
1+\frac{2\,(\k\cdot\q_1)}{11k^2}
\right]
\\[2pt]
&+\frac{216929939}{6038912880}
\left[
\frac{(\k\cdot\q_1)}{q_1^2}
+\frac{1}{11k^2}
\left(
\frac{(\k\cdot\q_1)^2}{q_1^2}
+\frac{(\k\cdot\q_2)^2}{q_2^2}
\right)
\right]\,.
\end{split}
\ee
The renormalized displacement ``inherits'' the non-local structure 
from $u^{(2)}_{\rm ct}$ in Eq.~\eqref{eq:u2NLC}: 
\be\label{eq:Psi2NLC}
\vPsi^{(2)}_{\rm ct}\Big|_{\rm NLC}=\tfrac14\,\vec{u}^{(2)}_{\rm ct}
\Big|_{\rm NLC}=-\tfrac14\,c_5^{\pi}\,\vec{e}_5\,.
\ee

\section{Trispectrum-level UV-matching of higher-derivative operators 
for matter and galaxies}
\label{app:trispec}

We start off by generalizing the dark matter renormalization
conditions to the trispectrum case. 
The power-spectrum conditions of Eq.~\eqref{eq:F3ctr_conditions} are
derived for the $(\k,\q,-\q)$-configuration only. 
At the level of general external momenta, the cubic counterterm kernel
is fixed by the single-hard limit of the fifth-order kernel,
\be
\label{eq:P15renormGen}
\delta K_3(\vk_1,\vk_2,\vk_3) \;=\; 10 \int_{\q}
    K_5(\vk_1,\vk_2,\vk_3,\q,-\q)\,\PL(q)\Big|_{\rm q~hard}
    \;=\; \sum_{O,\alpha} c_{O,\alpha}\, F^{(O,\alpha)}_{{\rm ct},3}(\vk_1,\vk_2,\vk_3)\,.
\ee
The prescription here is the one already set out in
Sec.~\ref{sec:bias}: ``$q$ hard'' is the pure single-hard limit, 
i.e. the leading Laurent coefficient of the $q\to\infty$ expansion, 
and the BPHZ prescription 
subtracts the two single-hard limits and restores the double-hard
overlap.  
The matter sector inherits the same structure. 
The single-hard identity fixes the cubic vertex $F_{{\rm ct},3}$, but
it does not renormalize the double-hard region of the two-loop
diagrams built from that vertex: closing the matched vertex
into the remaining loop we get
\be\label{eq:Fct3_hard_tail}
\Bigl\langle \sum_{O,\alpha} c_{O,\alpha}\,
F^{(O,\alpha)}_{{\rm ct},3}(\k,\q,-\q)\Bigr\rangle_{\hat\q}
\;\xrightarrow[\;q\to\infty\;]{}\;
\left(F^{{\rm fin},k^2}_{{\rm ct},3}\,k^2
\;+\,F^{{\rm fin},k^4}_{{\rm ct},3}\,\frac{k^4}{q^2}\right)
\,,
\ee
which generates the double-hard divergences of $P^{\rm ctr}_{15}$.
Just like $\delta b_1|_{\sigma^4}$ is absorbed by the linear
bias, these are absorbed by the linear counterterms
$F_{{\rm ct},1}$ and $F_{\partial^2{\rm ct},1}$ as
\be\label{eq:dh_conditions}
\delta c_{\delta,1}
 = 27\,F^{{\rm fin},k^2}_{{\rm ct},3}\;\sigma^2_{0}\,,
\qquad
\delta c_{\partial^2\delta,1}
 = -\,78\,F^{{\rm fin},k^4}_{{\rm ct},3}\;\sigma^2_{-2} \,.
\ee
The opposite signs follow from the counterterm kernels,
$F_{1,\partial^2{\rm ct}} =c_{\partial^2\delta,1}k^4/26$ and
$F_{1,{\rm ct}} = -c_{\delta,1}k^2/9$. The double-hard subtraction 
enters both contributions 
with the same sign, 
giving
\be\label{eq:dh_conditions_explicit}
\begin{split}
\delta c_{\delta,1} = \Bigl[
&\tfrac{178}{429}\,c_{\delta,1}
+\tfrac{3232}{5005}\,c_{\delta,2}
-\tfrac{141}{182}\,c_{\delta,3}
+\tfrac{1208}{2145}\,c_{r\delta,1}-\tfrac{438}{455}\,c_{r\delta,2}
+\tfrac{112}{143}\,c_{\delta^2,1}
-\tfrac{15}{26}\,c_{r\delta^2,1}
\Bigr]\sigma^2_{0}\,,\\[4pt]
\delta c_{\partial^2\delta,1} = \Bigl[
&-\tfrac{4681}{24255}\,c_{\delta,1}
-\tfrac{32}{315}\,c_{\delta,3}
-\tfrac{68}{2079}\,c_{r\delta,1}
+\tfrac{16}{49}\,c_{r\delta,2}
\Bigr]\sigma^2_{-2}\,.
\end{split}
\ee

We now generalize the power-spectrum matching of~\cite{Anastasiou:2025jsy}
to the one-loop
bispectrum and trispectrum level.  These correspond to the renormalization
conditions for the $B_{411}, T_{4211}$ and $T_{5111}$ SPT diagrams.
There is a single matching principle:
at every order $n$ the counterterm vertex must absorb the single-hard
tadpole integral of the corresponding SPT kernel,
\be\label{eq:master_matching}
\sum_{O,\alpha} c_{O,\alpha}\, F^{(O,\alpha)}_{{\rm ct},n}(\vk_1,\ldots,\vk_n)
\;=\; -\,W_n\, \hat{V}_n(\vk_1,\ldots,\vk_n)\,,\qquad
\hat{V}_n = \int_{\q} F_{n+2}(\vk_1,\ldots,\vk_n,\q,-\q)\,\PL(q)\Big|_{\rm q\ hard},
\ee
with the Wick contraction coefficient $W_n = \binom{n+2}{2} = 3,\,6,\,10$ for
$n = 1, 2, 3$. As before, the coefficients will be in units of
$\sigma_{-2}^2=\int_{\q} \PL(q)/q^2$.  The
one-loop power spectrum renormalization corresponds 
to $n = 1$ at a single external momentum. 
$P_{24}$,
$B_{411}$ and $T_{4211}$ are simultaneously 
renormalized by the quadratic vertex ($n = 2$). The momentum configurations
in these integrals are general, and hence no degeneracy would be present if we analyzed e.g. 
only $P_{24}$.
Finally,  
$P_{15}$ matches the cubic vertex ($n = 3$) at the tadpole
configuration $(\k,\q,-\q)$ after the angular average. 
This is a reduced configuration that misses the general-momentum information. 
Only
$T_{5111}$ is sensitive to the cubic counterterm 
vertex at general external momenta.  
The full $P/B/T$-level matching is therefore a single nested linear system on
the seventeen stress-tensor coefficients $c_{O,\alpha}$. Our task here is to
solve this system in a consistent manner while paying attention to the
degeneracies that appear in the $P$-only matching.

The first important fact is that the
bispectrum adds no new renormalization conditions beyond the power
spectrum: the 
$n \leq 2$ vertices are already fixed by $P_{13}$ and
$P_{24}$, so $B_{411}$ and $T_{4211}$ 
divergences should simply reproduce the known
combinations. As we see below, this nontrivial consistency test is passed exactly by our computation.  
Second, all genuinely new information enters through
$T_{5111}$: the cubic-vertex system at general momenta has rank 12,
containing the seven $P$-level combinations, which we re-derive from
$T_{5111}$ alone, plus five new conditions
invisible to any power-spectrum or bispectrum measurement.

We fit the set of angle-averaged single-hard SPT kernels for the kinematic 
momentum configurations~\eqref{eq:master_matching} to the general 
counterterm kernel. 
The fitted conditions below are written in manifestly gauge-invariant
form. For $n=1,2$ we obtain (in units $\sigma^2_{-2}$):
\begin{align}
& c_{\delta,1} + c_{r,1} = -\tfrac{61}{70} \,, \notag\\
& c_{\delta,2} + \tfrac{7}{4}\,c_{r^2,1} - \tfrac{3}{4}\,c_{r,2} + \tfrac{7}{4}\,c_{r,1} = -\tfrac{12409}{5880} \,, \notag\\
& c_{r\delta,1} + c_{r^2,1} + c_{r,2} - c_{r,1} = -\tfrac{6997}{3430} \,, \notag\\
& c_{\delta^2,1} - \tfrac{7}{4}\,c_{r^2,1} + \tfrac{3}{4}\,c_{r,2} - \tfrac{3}{4}\,c_{r,1} = \tfrac{63149}{41160} \,.
\label{eq:level2conds}
\end{align}
The four combinations on the left are exactly the gauge-invariant
functionals that the power spectrum
fixes through $P_{13}$ and $P_{24}$, and the values on the right are
exactly those of Eq.~\eqref{eq:F3ctr_conditions}.  This is an
independent, bispectrum-level confirmation of the power-spectrum
conditions, including the sign of $c_{\delta^2,1}$ mentioned earlier.  

At the cubic level all seventeen coefficients enter, and the
general-momentum system has rank 12:
\begin{align}
& c_{\delta,1} + c_{r,1} = -\tfrac{61}{70} \,, \notag\\
& c_{\delta,2} + \tfrac{1225}{494}\,c_{r^2,2} - \tfrac{395}{494}\,c_{r,2} + \tfrac{889}{494}\,c_{r,1} - \tfrac{63}{247}\,c_{\delta^3,1} + \tfrac{1085}{988}\,c_{\delta r^2,1} = -\tfrac{146978057}{47927880} \,, \notag\\
& c_{\delta,3} + \tfrac{1225}{494}\,c_{r^2,2} - \tfrac{395}{494}\,c_{r,2} + \tfrac{889}{494}\,c_{r,1} - \tfrac{310}{247}\,c_{\delta^3,1} + \tfrac{1085}{988}\,c_{\delta r^2,1} = -\tfrac{1645826839}{503242740} \,, \notag\\
& c_{r\delta,1} + \tfrac{350}{247}\,c_{r^2,2} + \tfrac{240}{247}\,c_{r,2} - \tfrac{240}{247}\,c_{r,1} - \tfrac{36}{247}\,c_{\delta^3,1} + \tfrac{155}{247}\,c_{\delta r^2,1} = -\tfrac{216929939}{83873790} \,, \notag\\
& c_{r\delta,2} + \tfrac{35}{19}\,c_{r^2,2} + \tfrac{5}{19}\,c_{r,2} - \tfrac{5}{19}\,c_{r,1} + \tfrac{4}{19}\,c_{\delta^3,1} - \tfrac{7}{38}\,c_{\delta r^2,1} = -\tfrac{4862834}{9677745} \,, \notag\\
& c_{\delta^2,1} - \tfrac{1225}{494}\,c_{r^2,2} + \tfrac{395}{494}\,c_{r,2} - \tfrac{395}{494}\,c_{r,1} + \tfrac{63}{247}\,c_{\delta^3,1} - \tfrac{1085}{988}\,c_{\delta r^2,1} = \tfrac{167110717}{67099032} \,, \notag\\
& c_{r\delta^2,1} - \tfrac{105}{494}\,c_{r^2,2} + \tfrac{175}{494}\,c_{r,2} - \tfrac{175}{494}\,c_{r,1} - \tfrac{44}{247}\,c_{\delta^3,1} + \tfrac{895}{988}\,c_{\delta r^2,1} = -\tfrac{260049833}{71891820} \,, \notag\\
& c_{r^3,1} + \tfrac{9}{247}\,c_{r^2,2} - \tfrac{15}{247}\,c_{r,2} + \tfrac{15}{247}\,c_{r,1} + \tfrac{64}{247}\,c_{\delta^3,1} + \tfrac{135}{494}\,c_{\delta r^2,1} = -\tfrac{166884661}{251621370} \,, \notag\\
& c_{r,3} - \tfrac{320}{247}\,c_{r^2,2} - \tfrac{290}{247}\,c_{r,2} + \tfrac{43}{247}\,c_{r,1} - \tfrac{80}{247}\,c_{\delta^3,1} + \tfrac{70}{247}\,c_{\delta r^2,1} = -\tfrac{22304069}{125810685} \,, \notag\\
& c_{\delta^2,2} - \tfrac{1225}{988}\,c_{r^2,2} + \tfrac{395}{988}\,c_{r,2} - \tfrac{395}{988}\,c_{r,1} + \tfrac{557}{494}\,c_{\delta^3,1} - \tfrac{1085}{1976}\,c_{\delta r^2,1} = \tfrac{1594799603}{1006485480} \,, \notag\\
& c_{r^2r,1} - \tfrac{105}{494}\,c_{r^2,2} + \tfrac{175}{494}\,c_{r,2} - \tfrac{175}{494}\,c_{r,1} - \tfrac{44}{247}\,c_{\delta^3,1} - \tfrac{93}{988}\,c_{\delta r^2,1} = \tfrac{110949109}{503242740} \,, \notag\\
& c_{r^2,1} - \tfrac{350}{247}\,c_{r^2,2} + \tfrac{7}{247}\,c_{r,2} - \tfrac{7}{247}\,c_{r,1} + \tfrac{36}{247}\,c_{\delta^3,1} - \tfrac{155}{247}\,c_{\delta r^2,1} = \tfrac{22916149}{41936895} \,.
\label{eq:level3conds}
\end{align}
As we shall see, five coefficients ($c_{r^2,2}$, $c_{r,2}$, $c_{r,1}$,
$c_{\delta^3,1}$, $c_{\delta r^2,1}$) 
can be consistently set to zero: they
parametrize the five-dimensional null space of the cubic vertex at
general momenta ($17 - 12 = 5$), i.e. they correspond to five exact kernel
identities of the seventeen $F^{(O,\alpha)}_{{\rm ct},3}$. Let us describe now the 
principled way to solve the above equations while preserving the 
power spectrum renormalization conditions.

As shown in~\cite{Anastasiou:2025jsy}, the expansion of $\delta K_3$ over the seventeen kernels
$F^{(O,\alpha)}_{{\rm ct},3}$ is not unique at the power-spectrum
configuration, where only seven combinations of the $c_{O,\alpha}$ affect the
result, whilst ten independent shifts leave it unchanged: eight
annihilate the $(\k,\q,-\q)$ integrand before the angular average, and
two more (the ``angular'' directions) only after it.  Taking the
power-spectrum solution $c^{P}_{O,\alpha}$ of
Eq.~\eqref{eq:F3ctr_conditions} as a reference point, the general
power-spectrum-consistent solution is the ten-parameter family
\be\label{eq:cstar_family}
c_{O,\alpha} \;=\; c^{P}_{O,\alpha}
\;+\; \sum_{i} t_i\, v^{(i)}_{O,\alpha}\,,
\ee
with $i$ running over the ten labels of Eqs.~\eqref{eq:flatdirs} and
$t_i$ denoting the ten amplitudes the power spectrum cannot determine.  
We call the $v^{(i)}$ the $P$-flat directions and write them as
$v^{(i)} = \sum_{O,\alpha} v^{(i)}_{O,\alpha}\,\hat e_{O,\alpha}$,
where $\hat e_{O,\alpha}$ is the elementary shift that raises
$c_{O,\alpha}$ by one and leaves the other sixteen coefficients
unchanged; each $v^{(i)}$ is thus a correlated deformation of several
coefficients at once, with purely numerical (momentum-independent)
components. Following~\cite{Anastasiou:2025jsy}, 
eight of the ten are normalized to unit component along
the coefficient that names them; the two angular directions have no
such natural pivot and are normalized to unit component on
$\hat e_{r\delta,2}$ and $\hat e_{r\delta^2,1}$, respectively:
\begin{subequations}\label{eq:flatdirs}
\begin{align}
v^{(\delta^2,2)} &= -\tfrac{775}{187}\,\hat e_{\delta,3} + \tfrac{130}{187}\,\hat e_{r\delta,2} + \tfrac{401}{187}\,\hat e_{r\delta^2,1} -\tfrac{60}{187}\,\hat e_{r^3,1} -\tfrac{200}{187}\,\hat e_{r,3} + \hat e_{\delta^2,2}\,,\\
v^{(r^2r,1)} &= -\tfrac{1085}{374}\,\hat e_{\delta,3} + \tfrac{91}{187}\,\hat e_{r\delta,2} + \tfrac{1459}{374}\,\hat e_{r\delta^2,1} -\tfrac{416}{187}\,\hat e_{r^3,1} -\tfrac{140}{187}\,\hat e_{r,3} + \hat e_{r^2 r,1}\,,\\
v^{(r^2,1)} &= -\tfrac{7}{4}\,\hat e_{\delta,2} -\hat e_{r\delta,1} + \tfrac{7}{4}\,\hat e_{\delta^2,1} + \hat e_{r^2,1}\,,\\
v^{(r^2,2)} &= \tfrac{1225}{374}\,\hat e_{\delta,3} -\tfrac{525}{187}\,\hat e_{r\delta,2} -\tfrac{1225}{374}\,\hat e_{r\delta^2,1} + \tfrac{156}{187}\,\hat e_{r^3,1} + \tfrac{520}{187}\,\hat e_{r,3} + \hat e_{r^2,2}\,,\\
v^{(r,2)} &= \tfrac{3}{4}\,\hat e_{\delta,2} -\tfrac{705}{374}\,\hat e_{\delta,3} -\hat e_{r\delta,1} + \tfrac{35}{187}\,\hat e_{r\delta,2} -\tfrac{3}{4}\,\hat e_{\delta^2,1} + \tfrac{705}{374}\,\hat e_{r\delta^2,1} -\tfrac{160}{187}\,\hat e_{r^3,1} + \tfrac{90}{187}\,\hat e_{r,3} + \hat e_{r,2}\,,\\
v^{(r,1)} &= -\hat e_{\delta,1} -\tfrac{7}{4}\,\hat e_{\delta,2} + \tfrac{331}{374}\,\hat e_{\delta,3} + \hat e_{r\delta,1} -\tfrac{35}{187}\,\hat e_{r\delta,2} + \tfrac{3}{4}\,\hat e_{\delta^2,1} -\tfrac{705}{374}\,\hat e_{r\delta^2,1} + \tfrac{160}{187}\,\hat e_{r^3,1} + \tfrac{97}{187}\,\hat e_{r,3} + \hat e_{r,1}\,,\\
v^{(\delta^3,1)} &= -\tfrac{1085}{374}\,\hat e_{\delta,3} + \tfrac{91}{187}\,\hat e_{r\delta,2} + \tfrac{711}{374}\,\hat e_{r\delta^2,1} -\tfrac{42}{187}\,\hat e_{r^3,1} -\tfrac{140}{187}\,\hat e_{r,3} + \hat e_{\delta^3,1}\,,\\
v^{(\delta r^2,1)} &= \tfrac{1085}{748}\,\hat e_{\delta,3} -\tfrac{91}{374}\,\hat e_{r\delta,2} -\tfrac{1833}{748}\,\hat e_{r\delta^2,1} + \tfrac{21}{187}\,\hat e_{r^3,1} + \tfrac{70}{187}\,\hat e_{r,3} + \hat e_{\delta r^2,1}\,,\\
v^{({\rm ang},1)} &= -\tfrac{155}{26}\,\hat e_{\delta,3} + \hat e_{r\delta,2} + \tfrac{33513}{2366}\,\hat e_{r^3,1} -\tfrac{20}{13}\,\hat e_{r,3}\,,\\
v^{({\rm ang},2)} &= \hat e_{r\delta^2,1} -\tfrac{25}{13}\,\hat e_{r^3,1}\,.
\end{align}
\end{subequations}
All ten $P$-flat directions annihilate
the $n = 1$ and $n = 2$ vertices identically at general
momenta since their quadratic parts are exact kernel identities of
$F_{{\rm ct},n\leq 2}$.  Plugging Eq.~\eqref{eq:cstar_family} into Eq.~\eqref{eq:level3conds} 
we find exactly five independent conditions:
\begin{subequations}\label{eq:T5conditions}
\begin{align}
t_{\delta^2,2} -\tfrac{1225}{988}\,t_{r^2,2} + \tfrac{395}{988}\,t_{r,2} -\tfrac{395}{988}\,t_{r,1} + \tfrac{557}{494}\,t_{\delta^3,1} -\tfrac{1085}{1976}\,t_{\delta r^2,1} &= \tfrac{1594799603}{1006485480}\,,\\
t_{r^2r,1} -\tfrac{105}{494}\,t_{r^2,2} + \tfrac{175}{494}\,t_{r,2} -\tfrac{175}{494}\,t_{r,1} -\tfrac{44}{247}\,t_{\delta^3,1} -\tfrac{93}{988}\,t_{\delta r^2,1} &= \tfrac{110949109}{503242740}\,,\\
t_{r^2,1} -\tfrac{350}{247}\,t_{r^2,2} + \tfrac{7}{247}\,t_{r,2} -\tfrac{7}{247}\,t_{r,1} + \tfrac{36}{247}\,t_{\delta^3,1} -\tfrac{155}{247}\,t_{\delta r^2,1} &= \tfrac{22916149}{41936895}\,,\\
t_{{\rm ang},1} &= -\tfrac{17506567}{16008300}\,,\\
t_{{\rm ang},2} &= -\tfrac{5123157163}{622545000}\,.
\end{align}
\end{subequations}
$v^{({\rm ang},1)}$ and
$v^{({\rm ang},2)}$ above are precisely the angular-degeneracy directions
along which Ref.~\cite{Anastasiou:2025jsy} removes $c_{r^3,1}$ and
$c_{r,3}$ from the operator list as a gauge choice for the
power spectrum (which works for the bispectrum as well), but
the general-momentum matching forces both amplitudes to specific
nonzero values.  

The five combinations of the $t_i$ orthogonal to
Eq.~\eqref{eq:T5conditions} remain exactly kernel-flat (as opposed to $P$-flat) at one loop because 
the residual five-parameter family moves the
coefficients only along the null space of the cubic vertex at general
momenta, i.e. along exact identities of the seventeen kernels
$F^{(O,\alpha)}_{{\rm ct},n}$ themselves.  The ten $P$-flat directions
thus split into five kernel-flat that are pure reparametrizations of the
stress-tensor (no observable at this order depends on them)
and five that are physical but invisible at the $P$ level.  
Since nothing depends
on the flat five, we gauge them to zero,
$t_{r^2,2}=t_{r,2}=t_{r,1}=t_{\delta^3,1}=t_{\delta r^2,1}=0$.
Imposing the five conditions of Eq.~\eqref{eq:T5conditions} and this
gauge defines the ``trispectrum point''  $c^{*}$; its nonzero entries are (in
units of $\sigma^{2}_{-2}$),
\be\label{eq:cstar_canonical}
\begin{split}
&c^{*}_{\delta,1} = -\tfrac{61}{70}\,,\quad
c^{*}_{\delta,2} = -\tfrac{146978057}{47927880}\,,\quad
c^{*}_{\delta,3} = -\tfrac{1645826839}{503242740}\,,\quad
c^{*}_{r\delta,1} = -\tfrac{216929939}{83873790}\,,\\
&c^{*}_{r\delta,2} = -\tfrac{4862834}{9677745}\,,\quad
c^{*}_{\delta^2,1} = \tfrac{167110717}{67099032}\,,\quad
c^{*}_{\delta^2,2} = \tfrac{1594799603}{1006485480}\,,\quad
c^{*}_{r\delta^2,1} = -\tfrac{260049833}{71891820}\,,\\
&c^{*}_{r,3} = -\tfrac{22304069}{125810685}\,,\quad
c^{*}_{r^2,1} = \tfrac{22916149}{41936895}\,,\quad
c^{*}_{r^3,1} = -\tfrac{166884661}{251621370}\,,\quad
c^{*}_{r^2r,1} = \tfrac{110949109}{503242740}\,,
\end{split}
\ee
with $c^{*}_{r,1}=c^{*}_{r,2}=c^{*}_{r^2,2}=c^{*}_{\delta^3,1}
=c^{*}_{\delta r^2,1}=0$.  Only $c^{*}_{\delta,1}$ coincides with its
power-spectrum-gauge counterpart $c^{P}_{\delta,1}$; every other entry
differs from Eq.~\eqref{eq:F3ctr_conditions}, because the trispectrum
matching is not consistent with the power-spectrum gauge choice
$c_{r^2,1} = 0$ (indeed $c^{*}_{r^2,1} = \tfrac{22916149}{41936895}$).  Consistency with the power-spectrum and bispectrum
conditions nevertheless holds by construction: 
$c^{*} - c^{P}$ lies in the span of the ten flat directions,
with expansion amplitudes exactly the five values of
Eq.~\eqref{eq:T5conditions} and zero along the kernel-flat five, so every
$P$-visible combination of $c^{*}$ reproduces
Eq.~\eqref{eq:F3ctr_conditions}. In particular the two points give
identical renormalization conditions for $e_{1,2,3}$, $C_{13}$,
$C_{\FG}$ and $C_{E_4}$ of Eqs.~\eqref{eq:C13}--\eqref{eq:CE4}.  At
the power-spectrum configuration the fixed $t$-shifts act only as an
unobservable renormalization of the $k^2\PL$ and $k^4\PL$
counterterms; at general external momenta they are physical (the
fixed directions are $P$-flat, not kernel-flat).  
They do not, however, modify the double-hard conditions of
Eq.~\eqref{eq:dh_conditions}.  
Eight of the ten $P$-flat directions are exact
identities among the $(\k,\q,-\q)$ kernels, 
while the two angular directions vanish upon 
angular integration for any $q$.  
The angle-averaged
cubic kernel is therefore invariant along the whole ten-parameter family,
and so is its large-$q$ expansion: $F^{{\rm fin},k^2}_{{\rm ct},3}$ and
$F^{{\rm fin},k^4}_{{\rm ct},3}$, and with them
Eqs.~\eqref{eq:dh_conditions}
and~\eqref{eq:dh_conditions_explicit}, take the same values 
at $c^{*}$ and $c^{P}$.

For a biased tracer the same matching applies to the fifth-order
galaxy kernel,
\be
\label{eq:P15renormTracer}
\delta K^{g}_3(\vk_1,\vk_2,\vk_3) \;=\; 10 \int_{\q}
    K^{g}_5(\vk_1,\vk_2,\vk_3,\q,-\q)\,\PL(q)\Big|_{\rm q~hard}\,.
\ee
There are three ingredients needed for its renormalization at the level of higher-derivative operators: 
(a) the dark matter renormalization discussed above, 
(b) the renormalized velocity potentials which have to be inserted in the usual one-loop Galileon bias expansion as 
$\Phi_v\to \Phi_v + \Phi_{v,\mathrm{ct}}$,
and (c) the 
higher-derivative bias kernels 
$\beta_{a}\, K^{(a)}(\vk_1,\vk_2,\vk_3)$
from Eq.~\eqref{eq:k2d3ops}.

The first two ingredients are quite important: they fix the renormalization conditions dictated by the dark matter field. 
While for the most part this produces the operators degenerate with the usual higher-derivative bias expansion, 
there are unique structures 
that appear in $b_1(\delta+\delta_{\rm ct})$ and 
$\mathcal{G}_2(\Phi_v + \Phi_{v,\mathrm{ct}})$, which cannot
be generated by the usual higher-derivative bias expansion. 
With these three ingredients, we have found that the 
trispectrum renormalization condition takes the form:
\be
\label{eq:dK3tracer}
\delta K^{g}_3
 = b_1\,\mathcal{M}_3(\vk_1,\vk_2,\vk_3)
 + \sum_{a=1}^{17} \beta_a\, K^{(a)}(\vk_1,\vk_2,\vk_3)
 + 2\gamma_2 \Gal_2(\Phi^{(2)}_{v,\mathrm{ct}},\Phi^{(1)}_v)\,,
\ee
where 
$\mathcal{M}_3 = \sum_{O,\alpha}c^{*}_{O,\alpha}\sigma_{-2}^2F^{(O,\alpha)}_{{\rm ct},3}$ is the dark-matter cubic counterterm at the
trispectrum renormalization point of Eq.~\eqref{eq:cstar_canonical}: 
it enters with
the coefficient $b_1$ without any extra free parameters, since any shift
of $c^*$ would
break the renormalization conditions for the matter sector. 
Likewise, the operator $2\gamma_2 \Gal_2(\Phi^{(2)}_{v,\mathrm{ct}},\Phi^{(1)}_v)$ is produced by the renormalization
of the dark matter velocity potential 
in the quadratic Galileon and its coefficient is therefore completely locked.
For consistency with the bispectrum, one may also add 
$2\gamma_2 \Gal_2(\Phi^{(1)}_{v,\mathrm{ct}},\Phi^{(1)}_v)$
at the quadratic order, even though it is fully degenerate with 
the $\beta_{2,3}$ and $\beta_{2,4}$ contributions,
\be\label{eq:g2op_quadratic}
2\,\Gal_2\bigl(\Phi^{(1)}_{v,\rm ct},\Phi^{(1)}_v\bigr)
= 
-\frac{c_{\delta,1}+c_{r,1}}{3}
(q_1^2+q_2^2)\,\Kcal(\vq_1,\vq_2)
= 
\frac{c_{\delta,1}+c_{r,1}}{3}
\Bigl[-k^2+2\,(\vq_1\!\cdot\!\vq_2)\Bigr]
\Kcal(\vq_1,\vq_2)
\,.
\ee
We proceed without these contributions 
noting that our 
$\beta_{2,3}$ and $\beta_{2,4}$ 
are different from those obtained in
the scheme with explicit $2\gamma_2 \Gal_2(\Phi^{(1)}_{v,\mathrm{ct}},\Phi^{(1)}_v)$.
Matching eq.~(\ref{eq:P15renormTracer}) at general external momenta then
fixes every $\beta_a$ as an exact function of the bias parameters 
(quoted in units
$2\,\sigma^2_{-2}$):
\begin{subequations}\label{eq:betaB_conditions}
\begin{align}
\beta_{1} &= -\tfrac{16}{35}\,\gamma_{2} +\tfrac{8}{15}\,\gamma_{21}\,,\\
\beta_{2,1} &= -\tfrac{47}{3528}\,b_2 -\tfrac{8}{1715}\,\gamma_{2} -\tfrac{76}{105}\,\gamma_{21} +\tfrac{16}{105}\,\gamma_{21}^{\times} -\tfrac{32}{35}\,\gamma_{22} -\tfrac{32}{245}\,\gamma_{2}^{\times} -\tfrac{46}{735}\,\gamma_{31}\,,\\
\beta_{2,2} &= \tfrac{64}{2205}\,b_2 -\tfrac{32}{105}\,\gamma_{21}^{\times} +\tfrac{64}{245}\,\gamma_{2}^{\times}\,,\\
\beta_{2,3} &= \tfrac{4}{245}\,b_2 +\tfrac{491}{2058}\,\gamma_{2} -\tfrac{4}{7}\,\gamma_{21} +\tfrac{4}{15}\,\gamma_{211} -\tfrac{6}{35}\,\gamma_{21}^{\times} -\tfrac{32}{105}\,\gamma_{22} +\tfrac{36}{245}\,\gamma_{2}^{\times} -\tfrac{12}{35}\,\gamma_{3} +\tfrac{8}{735}\,\gamma_{31}\,,\\
\beta_{2,4} &= -\tfrac{8}{245}\,b_2 -\tfrac{673}{2058}\,\gamma_{2} +\tfrac{24}{35}\,\gamma_{21} -\tfrac{8}{15}\,\gamma_{211} +\tfrac{12}{35}\,\gamma_{21}^{\times} +\tfrac{32}{105}\,\gamma_{22} -\tfrac{72}{245}\,\gamma_{2}^{\times} +\tfrac{24}{35}\,\gamma_{3} -\tfrac{52}{2205}\,\gamma_{31}\,,\\
\beta_{3,1} &= -\tfrac{2528}{33957}\,b_2 -\tfrac{43}{17640}\,b_3 +\tfrac{24064}{169785}\,\gamma_{2} +\tfrac{144}{49}\,\gamma_{21} +\tfrac{32}{63}\,\gamma_{212} -\tfrac{16}{15}\,\gamma_{21}^{\times} +\tfrac{4}{105}\,\gamma_{21}^{\times\times} +\tfrac{256}{35}\,\gamma_{22} -\tfrac{64}{35}\,\gamma_{22}^{\times} \notag \\ &  +\tfrac{736}{2205}\,\gamma_{23} -\tfrac{64}{147}\,\gamma_{2sq} -\tfrac{9104}{46305}\,\gamma_{2}^{\times} -\tfrac{8}{245}\,\gamma_{2}^{\times\times} +\tfrac{1012}{2205}\,\gamma_{31} -\tfrac{412}{6615}\,\gamma_{31}^{\times} +\tfrac{21856}{8085}\,\gamma_{41}\,,\\
\beta_{3,2} &= -\tfrac{9956}{169785}\,b_2 +\tfrac{1}{420}\,b_3 +\tfrac{11789}{56595}\,\gamma_{2} +\tfrac{2888}{735}\,\gamma_{21} +\tfrac{16}{15}\,\gamma_{211} -\tfrac{4}{315}\,\gamma_{212} -\tfrac{176}{105}\,\gamma_{21}^{\times} +\tfrac{1088}{105}\,\gamma_{22} +\tfrac{32}{35}\,\gamma_{221} \notag \\ &  -\tfrac{64}{35}\,\gamma_{22}^{\times} +\tfrac{1192}{2205}\,\gamma_{23} +\tfrac{8}{735}\,\gamma_{2sq} -\tfrac{1324}{46305}\,\gamma_{2}^{\times} +\tfrac{24}{1715}\,\gamma_{3} +\tfrac{592}{735}\,\gamma_{31} +\tfrac{24}{245}\,\gamma_{311} -\tfrac{962}{6615}\,\gamma_{31}^{\times} +\tfrac{449942}{56595}\,\gamma_{41}\,,\\
\beta_{3,3} &= -\tfrac{492}{18865}\,b_2 +\tfrac{682697}{2376990}\,\gamma_{2} +\tfrac{272}{735}\,\gamma_{21} -\tfrac{4}{21}\,\gamma_{211} -\tfrac{4}{35}\,\gamma_{211}^{\times} +\tfrac{248}{315}\,\gamma_{212} +\tfrac{62}{735}\,\gamma_{21}^{\times} +\tfrac{832}{735}\,\gamma_{22} -\tfrac{16}{35}\,\gamma_{22}^{\times} \notag \\ &  +\tfrac{16}{441}\,\gamma_{23} -\tfrac{496}{735}\,\gamma_{2sq} -\tfrac{52231}{185220}\,\gamma_{2}^{\times} +\tfrac{482}{15435}\,\gamma_{31} -\tfrac{2}{147}\,\gamma_{311} +\tfrac{19}{6615}\,\gamma_{31}^{\times} +\tfrac{36}{245}\,\gamma_{3}^{\times} -\tfrac{20161}{18865}\,\gamma_{41}\,,\\
\beta_{3,4} &= -\tfrac{1847}{75460}\,b_2 +\tfrac{4}{245}\,b_3 +\tfrac{4241}{43218}\,\gamma_{2} +\tfrac{964}{735}\,\gamma_{21} -\tfrac{8}{15}\,\gamma_{211} +\tfrac{4}{15}\,\gamma_{211}^{\times} +\tfrac{88}{315}\,\gamma_{212} +\tfrac{16}{245}\,\gamma_{21}^{\times} -\tfrac{6}{35}\,\gamma_{21}^{\times\times} \notag \\ &  +\tfrac{2176}{735}\,\gamma_{22} -\tfrac{16}{35}\,\gamma_{221} -\tfrac{32}{105}\,\gamma_{22}^{\times} +\tfrac{4}{45}\,\gamma_{23} -\tfrac{176}{735}\,\gamma_{2sq} +\tfrac{4777}{185220}\,\gamma_{2}^{\times} +\tfrac{36}{245}\,\gamma_{2}^{\times\times} -\tfrac{12}{1715}\,\gamma_{3} +\tfrac{1546}{15435}\,\gamma_{31} \notag \\ &  -\tfrac{12}{245}\,\gamma_{311} +\tfrac{76}{1323}\,\gamma_{31}^{\times} -\tfrac{12}{35}\,\gamma_{3}^{\times} -\tfrac{179}{147}\,\gamma_{41}\,,\\
\beta_{3,5} &= \tfrac{4784}{169785}\,b_2 -\tfrac{8}{245}\,b_3 -\tfrac{1823}{24255}\,\gamma_{2} -\tfrac{1432}{735}\,\gamma_{21} +\tfrac{16}{15}\,\gamma_{211} -\tfrac{8}{15}\,\gamma_{211}^{\times} -\tfrac{58}{105}\,\gamma_{212} -\tfrac{16}{35}\,\gamma_{21}^{\times} +\tfrac{12}{35}\,\gamma_{21}^{\times\times} \notag \\ &  -\tfrac{64}{15}\,\gamma_{22} +\tfrac{32}{35}\,\gamma_{221} +\tfrac{32}{105}\,\gamma_{22}^{\times} -\tfrac{8}{63}\,\gamma_{23} +\tfrac{116}{245}\,\gamma_{2sq} -\tfrac{58}{15435}\,\gamma_{2}^{\times} -\tfrac{72}{245}\,\gamma_{2}^{\times\times} +\tfrac{24}{1715}\,\gamma_{3} -\tfrac{248}{2205}\,\gamma_{31} \notag \\ &  +\tfrac{24}{245}\,\gamma_{311} -\tfrac{253}{2205}\,\gamma_{31}^{\times} +\tfrac{24}{35}\,\gamma_{3}^{\times} +\tfrac{47986}{18865}\,\gamma_{41}\,,
\\
\beta_{3,6} &= -\tfrac{22391}{679140}\,b_2 +\tfrac{165325}{475398}\,\gamma_{2} +\tfrac{104}{147}\,\gamma_{21} -\tfrac{4}{21}\,\gamma_{211} -\tfrac{2}{315}\,\gamma_{212} +\tfrac{2}{21}\,\gamma_{21}^{\times} +\tfrac{96}{49}\,\gamma_{22} +\tfrac{136}{2205}\,\gamma_{23} +\tfrac{4}{735}\,\gamma_{2sq} \notag \\ &  -\tfrac{15067}{92610}\,\gamma_{2}^{\times} +\tfrac{232}{3087}\,\gamma_{31} -\tfrac{2}{147}\,\gamma_{311} +\tfrac{2}{6615}\,\gamma_{31}^{\times} -\tfrac{57419}{56595}\,\gamma_{41}\,,\\
\beta_{3,7} &= -\tfrac{188}{18865}\,b_2 +\tfrac{36271}{1018710}\,\gamma_{2} +\tfrac{1936}{6615}\,\gamma_{21} -\tfrac{4}{21}\,\gamma_{211} -\tfrac{4}{35}\,\gamma_{211}^{\times} +\tfrac{8}{189}\,\gamma_{212} +\tfrac{6}{35}\,\gamma_{21}^{\times} +\tfrac{32}{105}\,\gamma_{22} -\tfrac{16}{105}\,\gamma_{221} \notag \\ &  +\tfrac{4}{441}\,\gamma_{23} -\tfrac{16}{441}\,\gamma_{2sq} -\tfrac{5192}{138915}\,\gamma_{2}^{\times} +\tfrac{3137}{20580}\,\gamma_{3} -\tfrac{11}{1323}\,\gamma_{31} +\tfrac{82}{6615}\,\gamma_{311} +\tfrac{1}{3969}\,\gamma_{31}^{\times} +\tfrac{36}{245}\,\gamma_{3}^{\times} -\tfrac{60271}{169785}\,\gamma_{41}\,,\\
\beta_{3,8} &= \tfrac{3428}{56595}\,b_2 -\tfrac{16739}{33957}\,\gamma_{2} -\tfrac{544}{735}\,\gamma_{21} +\tfrac{88}{105}\,\gamma_{211} +\tfrac{24}{35}\,\gamma_{211}^{\times} -\tfrac{82}{315}\,\gamma_{212} -\tfrac{36}{35}\,\gamma_{21}^{\times} -\tfrac{32}{35}\,\gamma_{22} +\tfrac{16}{21}\,\gamma_{221} \notag \\ &  -\tfrac{4}{315}\,\gamma_{23} +\tfrac{164}{735}\,\gamma_{2sq} +\tfrac{10946}{46305}\,\gamma_{2}^{\times} -\tfrac{1744}{1715}\,\gamma_{3} +\tfrac{52}{735}\,\gamma_{31} -\tfrac{136}{2205}\,\gamma_{311} +\tfrac{1}{6615}\,\gamma_{31}^{\times} -\tfrac{216}{245}\,\gamma_{3}^{\times} +\tfrac{37054}{18865}\,\gamma_{41}\,,\\
\beta_{3,9} &= \tfrac{24}{3773}\,b_2 +\tfrac{117752}{1188495}\,\gamma_{2} -\tfrac{583}{1764}\,\gamma_{21} +\tfrac{8}{35}\,\gamma_{211} -\tfrac{36}{245}\,\gamma_{21}^{\times} +\tfrac{48}{245}\,\gamma_{22} +\tfrac{8}{15}\,\gamma_{221} -\tfrac{24}{35}\,\gamma_{22}^{\times} +\tfrac{118}{2205}\,\gamma_{23} \notag \\ &  +\tfrac{16}{1715}\,\gamma_{31} -\tfrac{2264}{56595}\,\gamma_{41}\,,\\
\beta_{3,10} &= -\tfrac{48}{3773}\,b_2 -\tfrac{296957}{2376990}\,\gamma_{2} +\tfrac{1}{5}\,\gamma_{21} -\tfrac{16}{35}\,\gamma_{211} +\tfrac{72}{245}\,\gamma_{21}^{\times} -\tfrac{704}{735}\,\gamma_{22} -\tfrac{16}{15}\,\gamma_{221} +\tfrac{48}{35}\,\gamma_{22}^{\times} -\tfrac{92}{735}\,\gamma_{23} \notag \\ &  -\tfrac{2207}{61740}\,\gamma_{31} +\tfrac{6758}{56595}\,\gamma_{41}\,,\\
\beta_{3,11} &= -\tfrac{468}{18865}\,b_2 -\tfrac{1193}{9702}\,\gamma_{2} +\tfrac{437}{735}\,\gamma_{21} -\tfrac{8}{15}\,\gamma_{211} +\tfrac{2}{315}\,\gamma_{212} +\tfrac{12}{35}\,\gamma_{21}^{\times} +\tfrac{16}{105}\,\gamma_{22} -\tfrac{4}{7}\,\gamma_{221} +\tfrac{24}{35}\,\gamma_{22}^{\times} \notag \\ &  -\tfrac{113}{2205}\,\gamma_{23} -\tfrac{4}{735}\,\gamma_{2sq} -\tfrac{7582}{46305}\,\gamma_{2}^{\times} +\tfrac{573}{1715}\,\gamma_{3} -\tfrac{269}{8820}\,\gamma_{31} -\tfrac{1}{245}\,\gamma_{311} +\tfrac{1}{945}\,\gamma_{31}^{\times} -\tfrac{24151}{56595}\,\gamma_{41}\,,\\
\beta_{3,12} &= -\tfrac{5116}{33957}\,b_2 +\tfrac{1823}{24255}\,\gamma_{2} +\tfrac{1432}{735}\,\gamma_{21} -\tfrac{16}{15}\,\gamma_{211} +\tfrac{4}{315}\,\gamma_{212} +\tfrac{16}{15}\,\gamma_{21}^{\times} +\tfrac{64}{15}\,\gamma_{22} -\tfrac{32}{35}\,\gamma_{221} +\tfrac{8}{63}\,\gamma_{23} \notag \\ &  -\tfrac{8}{735}\,\gamma_{2sq} -\tfrac{5668}{9261}\,\gamma_{2}^{\times} -\tfrac{24}{1715}\,\gamma_{3} +\tfrac{248}{2205}\,\gamma_{31} -\tfrac{24}{245}\,\gamma_{311} +\tfrac{82}{1323}\,\gamma_{31}^{\times} -\tfrac{47986}{18865}\,\gamma_{41}\,,\\
\beta_{3,13} &= 24\,\gamma_{41}\,.
\end{align}
\end{subequations}
The quadratic and linear entries ($\beta_1$, $\beta_{2,i}$) agree
with the values fixed at the lower orders: the trispectrum-level
matching recovers them through the SPT-evolution terms.  
 $\gamma_{41}$ is the only operator that 
contributes to 
$\beta_{3,13}$.

Just like in the
matter discussion around Eq.~\eqref{eq:Fct3_hard_tail}, we need to supplement the 
single-hard renormalization of $\delta K^{g}_3$ with a double-hard subtraction.
Upon the subtraction of the usual bias contributions, it takes the form:
\be\label{eq:tracer_op_tails}
\bigl\langle \delta K_3^g(\k,\q,-\q)\bigr\rangle_{\hat\q}
\;\xrightarrow[\;q\to\infty\;]{}\;
x \,q^2 \;+\; \varphi \,k^2 \;+\; \psi \,\frac{k^4}{q^2}
\;+\;O(q^{-4})\,,
\ee
where the new $x\,q^2$ kernel,
forbidden for matter,
is present for several higher-derivative bias operators.  
This new kernel produces a pure renormalization of the linear
bias $b_1$.  Contracting the exact tails against the
decomposition~\eqref{eq:dK3tracer}, the three channels of the full vertex
are
\begin{align}
x \;=\;
&\tfrac{82}{63}\,\beta_{2,2}
 -\tfrac{2}{3}\,\beta_{3,1} +\tfrac{1}{3}\,\beta_{3,2}
 +\tfrac{4}{9}\,\beta_{3,3} +\tfrac{4}{9}\,\beta_{3,4}
 -\tfrac{4}{9}\,\beta_{3,6} +\tfrac{1}{9}\,\beta_{3,12}\,,
\label{eq:tracer_op_chi}\\[2pt]
\varphi\;=\;&\ -\tfrac{68}{63}\beta_{2,1}+\tfrac29\beta_{2,2}
-\tfrac{124}{315}\beta_{2,4}-\tfrac13\beta_{3,1}+\tfrac49\beta_{3,4}
+\tfrac{16}{45}\beta_{3,10}\\
& \quad -\tfrac{16}{45}\beta_{3,11} -\tfrac{964132}{7640325}\hat{\g}_2+\tfrac{156487}{15280650}\beta_{3,13}
+\tfrac{120424}{9029475}(2\sigma_{-2}^2)b_1,
\label{eq:tracer_op_phi}\\[2pt]
\psi \;=\;
&\tfrac{61}{1890}\beta_1+\tfrac{32}{105}\beta_{2,3}
+\tfrac{32}{147}\beta_{2,4}-\tfrac{16}{45}\beta_{3,9}
-\tfrac{16}{63}\beta_{3,10} +\tfrac{380936}{17827425}\hat{\g}_2-\tfrac{5344}{17827425}\beta_{3,13}
-\tfrac{10541261}{5562156600}(2\sigma_{-2}^2)b_1\,.
\label{eq:tracer_op_psi}
\end{align}
The operators $K^{(3,5)}$, $K^{(3,7)}$ and $K^{(3,8)}$ vanish identically in all three channels due to their Galileon structure.  The $b_1$
entries above stem from the dark matter
counterterm $\mathcal{M}_3$.
The full vertex renormalization is
\be\label{eq:tracer_dh_master}
\bigl\langle \delta K^{g}_3(\k,\q,-\q)\bigr\rangle_{\hat\q}
\;\xrightarrow[\;q\to\infty\;]{}\;
X\,q^2 \;+\; \Phi\,k^2 \;+\; \Psi\,\frac{k^4}{q^2}\,,
\ee
with 
\begin{align}
X \;=\;
&\big(\tfrac{6016}{138915}\,b_{2} +\tfrac{64}{6615}\,b_{3}
 +\tfrac{64}{945}\,\gamma_{211}^{\times} +\tfrac{128}{945}\,\gamma_{212}
 -\tfrac{32}{315}\,\gamma_{21}^{\times}
 -\tfrac{32}{315}\,\gamma_{21}^{\times\times}\nonumber\\
&+\tfrac{256}{945}\,\gamma_{22}^{\times}
 -\tfrac{256}{2205}\,\gamma_{2sq}
 +\tfrac{48896}{138915}\,\gamma_{2}^{\times}
 +\tfrac{64}{735}\,\gamma_{2}^{\times\times}
 +\tfrac{176}{6615}\,\gamma_{31}^{\times}
 -\tfrac{64}{735}\,\gamma_{3}^{\times}\big)\times 2\sigma^2_{-2}\,,
\label{eq:tracer_dh_X}\\[2pt]
\Phi \;=\;
&\big(\tfrac{120424}{9029475}\,b_{1} +\tfrac{52889}{1018710}\,b_{2}
 +\tfrac{61}{7560}\,b_{3} +\tfrac{992}{305613}\,\gamma_{2}
 -\tfrac{848}{33075}\,\gamma_{21}\nonumber\\
&+\tfrac{16}{135}\,\gamma_{211}^{\times}
 -\tfrac{32}{675}\,\gamma_{212}
 -\tfrac{4}{45}\,\gamma_{21}^{\times\times}
 -\tfrac{1024}{1575}\,\gamma_{22} -\tfrac{256}{675}\,\gamma_{221}
 +\tfrac{3392}{4725}\,\gamma_{22}^{\times}\nonumber\\
&-\tfrac{9728}{99225}\,\gamma_{23} +\tfrac{64}{1575}\,\gamma_{2sq}
 +\tfrac{312511}{694575}\,\gamma_{2}^{\times}
 +\tfrac{8}{105}\,\gamma_{2}^{\times\times} -\tfrac{96}{245}\,\gamma_{3}
 -\tfrac{1108}{33075}\,\gamma_{31}\nonumber\\
&-\tfrac{32}{1575}\,\gamma_{311}
 +\tfrac{4556}{99225}\,\gamma_{31}^{\times}
 -\tfrac{16}{105}\,\gamma_{3}^{\times}
 -\tfrac{94544}{94325}\,\gamma_{41}\big)\times 2\sigma^2_{-2}\,,
\label{eq:tracer_dh_Phi}\\[2pt]
\Psi \;=\;
&\big(-\tfrac{10541261}{5562156600}\,b_{1} -\tfrac{768}{660275}\,b_{2}
 +\tfrac{5512}{1188495}\,\gamma_{2} +\tfrac{13672}{231525}\,\gamma_{21}\nonumber\\
&+\tfrac{4864}{33075}\,\gamma_{22} +\tfrac{128}{1575}\,\gamma_{221}
 -\tfrac{128}{1225}\,\gamma_{22}^{\times}
 +\tfrac{8864}{694575}\,\gamma_{23}
 -\tfrac{1152}{60025}\,\gamma_{2}^{\times}\nonumber\\
&+\tfrac{384}{8575}\,\gamma_{3}
 +\tfrac{548}{138915}\,\gamma_{31}
 -\tfrac{415328}{17827425}\,\gamma_{41}\big)\times 2\sigma^2_{-2}\,.
\label{eq:tracer_dh_Psi}
\end{align}
The biases $b_4$, $b_5$, $b_{\Gal_2\Gal_3}$, $\gamma_{211}$,
$\gamma_{2sq}^{\times}$, $\gamma_{2}^{\times\times\times}$ and
$\gamma_{3}^{\times\times}$ drop from the double-hard limit entirely.  
To proceed, we isolate the pure dark matter parts from these expressions,
$\Phi=\tilde{\Phi}+b_1\Phi_{\mathcal{M}_3}$, $\Psi=\tilde{\Psi}+b_1\Psi_{\mathcal{M}_3}$, with
\be
\Phi_{\mathcal{M}_3} = \tfrac{120424}{9029475}\times 2\sigma^2_{-2}\,,\qquad
\Psi_{\mathcal{M}_3} = -\tfrac{10541261}{5562156600}\times 2\sigma^2_{-2}\,.
\ee
Closing the vertex into the remaining loop with the $P_{13}$-type
contraction gives
\be\label{eq:dKg3_closure}
\Delta P_{gg}(k)\Big|_{\delta K^{g}_3}
= 6b_1\,\PL(k)\int_{\q}
\bigl\langle \delta K^{g}_3(\k,\q,-\q)\bigr\rangle_{\hat\q}\,\PL(q)
= 6b_1\,\PL(k)\Bigl[X\,\sigma^2_{2} + \Phi\,k^2\,\sigma^2_{0}
+ \Psi\,k^4\,\sigma^2_{-2}\Bigr]\,,
\ee
with $\sigma^2_{n}\equiv\int_{\q}q^{n}\,\PL(q)$.  
Matching
this against the tree-level $b_1^2\,\PL$ and against the
$-2b_1\beta_1\,k^2\PL$ and $\bigl(\beta_1^2+2b_1\tilde\beta_1\bigr)k^4\PL$
counterterms, whose coefficients are $c_0=\beta_1$ and
$c_7=\beta_1^2+2b_1\tilde\beta_1$ in eq.~\eqref{eq:master}, the double-hard limit is absorbed via
\be\label{eq:tracer_dh_shifts}
\delta b_1 = -3X\,\sigma^2_{2}\,,\qquad
\delta\beta_1 = 3\tilde{\Phi}\,\sigma^2_{0}\,,\qquad
\delta\tilde\beta_1 = -3\tilde{\Psi}\,\sigma^2_{-2}\,,\qquad
\delta c_s = 3\Phi_{\mathcal{M}_3}\,\sigma^2_{0}\,,
\qquad \delta \tilde{c}_s = -3\Psi_{\mathcal{M}_3}\,\sigma^2_{-2}\,.
\ee
The $q^2$ channel is a 
renormalization of the linear bias, and the $k^2$ and $k^4$ channels are
renormalizations of the $\nabla^2\delta$ and $\nabla^4\delta$ biases
$\beta_1$ and $\tilde\beta_1$.  
This is the exact galaxy version
of the matter assignment~\eqref{eq:dh_conditions}.

Finally, let us mention that the consistent renormalization
of the dark matter field and the use of the renormalized dark matter displacement to build the basis of Lagrangian operators
are crucial in order to obtain a minimal complete basis of higher-derivative operators. This is important for the discussion
of Ref.~\cite{Bakx:2026rmd}, which pointed out that the renormalization of the one-loop galaxy trispectrum 
requires structures beyond the simplistic Eulerian 
local gradient expansion. One can check that their
$\hat\Pi^{(3)}$ and $\Pi_{B1}-\Pi_{B2}$ structures 
exactly emerge
from the dark matter density and  
velocity renormalization, respectively. 
The $\hat\Pi^{(3)}$ operator
is supplied by the dark matter density counterterm
contained in our $\mathcal{M}_3$, 
and corresponds
to our $e_3$ counterterm.
The $\Pi_{B1}-\Pi_{B2}$
combination emerges as
\begin{equation}
\;\Gal_2\!\bigl(\vec{e}_5,\vPsi^{(1)}\bigr)
=-\tfrac57\,\Gal_2(\nabla^2\varphi_2,\varphi_1)-\bigl(\Pi_{B1}-\Pi_{B2}\bigr)\,,
\label{eq:e5beta}
\end{equation}
where $e_5^i$ is the non-local velocity counterterm 
contribution from Eq.~\eqref{eq:u2NLC}. In our basis this is included in 
$\Gal_2\!\bigl(\vPsi^{(2)}_{\rm ct},\vPsi^{(1)}\bigr)$ because 
$\vec{e}_5$ appears in the renormalized dark matter displacement~\eqref{eq:Psict2},
and also in $\Gal_2(\Phi_{v,\mathrm{ct}}^{(2)},\Phi^{(1)})$, which 
provides an extra direction in operator space.
Had we not made the $\Phi_{v}\to \Phi_{v}+\Phi_{v,\mathrm{ct}}$ substitution, 
$K_{3,13}$ would have received an extra contribution from $\gamma_2$.
A compelling feature of our Lagrangian basis is that it explains why $\Pi_{B1},\Pi_{B2}$
do not appear separately: the renormalized dark matter
velocity only supplies the $\Pi_{B1}-\Pi_{B2}$ combination,
which is indeed the combination that appears in the UV limit of $T_{5111}$.

Finally, let us speculate that the simplicity of 
the renormalization
condition $\beta_{3,13}=24\gamma_{41}$ for the non-local 
operator
$\Gal_2\!\bigl(\vPsi^{(2)}_{\rm ct},\vPsi^{(1)}\bigr)$
suggests that there might exist a field redefinition,
similar to $\vPsi \to \vPsi +\vPsi_{\rm ct}$,
which could absorb this operator altogether,
just like the field redefinition $\Phi_{v}\to \Phi_{v}+\Phi_{v,\mathrm{ct}}$
absorbs the non-local content of the $\gamma_2$ operator.
We hope to revisit this point in the future.

\section{Redundancy of the mixed stochastic-deterministic 
power spectrum contributions to all orders}\label{sec:mixed}

Let us consider the following non-perturbative expression
for the galaxy density field in real space:
\be 
\delta_g=\delta_{\rm det}+\epsilon + \delta_{\rm mix}\epsilon\,,
\ee 
where $\delta_{\rm det}, \delta_{\rm mix}$ are most general 
composite operators built from the deterministic fields, 
such as the linear matter density, gravitational potential, 
LPT vector potential, etc. The only approximation we have made
so far is to assume a local coupling between stochastic 
and deterministic fields in $\delta_{\rm mix}\epsilon$.
By definition, $\langle \delta_{\rm mix}\epsilon\rangle =0$, 
$\langle \delta_{\rm det}\epsilon\rangle =0$. The fully non-linear power spectrum reads:
\be \label{eq:stochNP}\begin{split}
\langle |\delta_g |^2\rangle'=\langle |\delta_{\rm det} |^2\rangle'+
P_\epsilon(k)+P_{\rm mixed,~tot}\,, \quad 
P_{\rm mixed,~tot}\equiv \int_{\q}P_{\rm NL, mix}(q)P_\epsilon(|\k-\q|)\,,
\end{split}
\ee 
where $P_{\rm NL, mix}$ and $P_\epsilon(k)$ are the non-perturbative power spectra
of $\delta_{\rm mix}$ and $\epsilon$, respectively.
In EFT $P_\epsilon(k)$ assumes a series expansion 
in $k^2$ in the $k\ll k_{\rm stoch}$ limit:
\be \label{eq:stoch_P}
P_\epsilon(k)=\sum_{n=0}c_n k^{2n}\,.
\ee 
Plugging this into Eq.~\eqref{eq:stochNP} we find
that the $P_{\rm mixed,~tot}$ integral produces 
a similar power series
\be \label{eq:mixed_final}
P_{\rm mixed,~tot}=
\sum_{n=0}\int_{\q}P_{\rm NL, mix}(q)
c_n (q^2 -2(\k\cdot\q)+k^2)^{2n}=
\sum_{n'=0}
c'_{n'} k^{2n'}\,,
\ee 
which implies that all $c'_n$ can be absorbed by a re-definition
of the stochastic counterterms $c_n$. The key property that we used
is that $P_{\rm NL, mix}(q)$ does not depend on the on-shell momentum $k$;
the $k$-dependence is entirely determined by the stochastic power 
spectrum~\eqref{eq:stoch_P}. 
The contributions from modes with $q\gtrsim k_{\rm stoch}$
are also renormalized by the stochastic counterterms $c_n$.
Therefore, the mixed stochastic-deterministic 
couplings are completely redundant at the level of the galaxy power spectrum. We stress, however, that they are not redundant
at the level of higher-order $n$-point functions, see 
e.g.~\cite{Ivanov:2021kcd,Bakx:2025pop} for detailed 
discussions in the context of galaxy bispectrum. 

It is trivial now to generalize this argument to the case of 
non-local mixed couplings, e.g. $\nabla_i\nabla_jF_{\rm mix}[\delta,\Phi,\varphi_2,\Vec{A}_3,...]\nabla^i\nabla^j\epsilon$, where $F_{\rm mix}[\delta,\Phi,\varphi_2,\Vec{A}_3,...]$ is a general function of the deterministic fields and their potentials. The non-trivial derivative structure will collapse into extra integral kernels, 
such as $((\k\cdot\q)-q^2)^2$, which will simply shift 
$c'_{n'}$ coefficients in Eq.~\eqref{eq:mixed_final}, without 
affecting the conclusions.
This completes the proof that the mixed 
stochastic deterministic contributions 
are completely redundant in EFT at the level of the 
galaxy power spectrum to all orders. 

\section{Forecast for DESI Projected Clustering and CMB Lensing}
\label{app:desi_forecast}

In this Appendix we describe the Fisher forecast 
for the $2\times 2$pt analysis of the DESI galaxy catalogs and 
their cross-correlations with CMB lensing from \textit{Planck} PR4
and ACT DR6. Our analysis follows the projected clustering
analysis of Ref.~\cite{Ivanov:2026dvl}. We use the 2D auto 
spectra $C_\ell^{g_ig_i}$ and cross-spectra $C_\ell^{g_i\kappa_a}$
with $g_i=\{\mathrm{BGS, LRG1, LRG2, LRG3}\}$ samples of DESI DR1
and $\kappa_a=\{\kappa_{\rm PR4},\kappa_{\rm DR6}\}$.
We use the same Gaussian covariance matrices for our Fisher forecast 
as in the actual data analysis of~\cite{Ivanov:2026dvl}. 
The effective redshifts of BGS, LRG1, LRG2, LRG3
are $z=0.296,0.509,0.706,0.922$.

For the fiducial cosmology, we have used two choices: \textit{Planck} 
2018 cosmology and PT Challenge cosmology. 
In the former case we rescale the two-loop power spectrum
templates to have the same $\sigma_8(z)$ as \textit{Planck} 
2018. 
We have found consistent
results between the two analyses, implying that they are not 
sensitive to the fiducial cosmology choice. In what follows we report the PT Challenge cosmology results. We use the same fiducial cosmology
in all four analyses. 
For the linear and one-loop analyses, we use the 
fiducial EFT parameters from the best-fits to the actual 
DESI DR1 data.
For the two-loop EFT parameters, 
we use the fiducial nuisance parameters from the best-fit 
to PT Challenge data. This way the key EFT parameters (shot noise,
number density, and the linear bias) are the same across all four 
analyses. 
 We propagate the redshift dependence of the two-loop
 bias templates and counterterms by rescaling each appearance
 of the linear matter power spectrum with the growth factor,
 giving $\Delta P^{\rm 2-loop}_{gg,gm}\propto D^6(z)$, $P^{\rm HD}_{i=0,\dots,6}\propto D^4(z)$, and $P^{\rm HD}_7\propto D^2(z)$. We do not rescale the fiducial stochastic parameter template
 with redshift.

Assuming flat $\kmax/(\hMpc)=0.08,0.3,0.85$ in the linear, 1-loop, and two-loop analyses respectively, we compute the angular  
scale cuts through $\ell_{\rm max}=\chi(z_{\rm eff})\kmax-1/2$, finding 
\be 
\begin{split}
& \text{Linear}~(\kmax=0.08~\hMpc):~\ell_{\rm max}=\{65, 106, 140,171\}\,,\\
& \text{1-loop}~(\kmax=0.3~\hMpc):~\ell_{\rm max}=\{256, 400, 525,646\}\,,\\
& \text{2-loop}~(\kmax=0.85~\hMpc):~\ell_{\rm max}=\{700, 1133, 1488,1830\}\,,
\end{split}
\ee 
for BGS, LRG1, LRG2, LRG3. In our Fisher forecast we vary only  
$A\equiv\sigma_8/\sigma_{8,\mathrm{fid}}$ and keep all other cosmological parameters
fixed.

\bibliographystyle{apsrev4-1}
\bibliography{short}

\end{document}